\documentclass[%
    aps,
    prd,
    letterpaper,
    superscriptaddress,
    preprintnumbers,
    floatfix,
    twocolumn,
    nofootinbib,
    showpacs,
    hyphens
]{revtex4-2}  

\usepackage{amsmath}
\usepackage{amsfonts}
\usepackage{mathtools}
\usepackage{cases}
\usepackage{lmodern}
\usepackage[T1]{fontenc}
\usepackage{upgreek}
\usepackage{microtype}
\usepackage{graphicx} 
\usepackage{comment}    
\usepackage{dcolumn}   
\usepackage{bm}       
\usepackage[caption=false,justification=justified]{subfig}
\usepackage{ragged2e} 
\DeclareCaptionJustification{justified}{\justifying}
\usepackage{multirow}
\usepackage{slashed}
\usepackage{placeins}
\usepackage{epstopdf}
\usepackage{color}
\usepackage{tikz,pgfplots}
\pgfplotsset{compat=newest}
\usepackage{enumerate,comment}
\usepackage{setspace}
\usepackage{dsfont}
\usepackage{nicefrac}
\usepackage{siunitx}
\usepackage{physics}
\usepackage[normalem]{ulem}
\usepackage{tensor}
\usepackage{tabularx}
\usepackage{listings}
\usepackage{adjustbox}
\usepackage{hyperref}  
\hypersetup{breaklinks=true}
\usepackage[capitalize]{cleveref}
\usepackage{duckuments}

\definecolor{c1}{RGB}{206,0,0}
\definecolor{c2}{RGB}{249,149,0}
\definecolor{c3}{RGB}{153,0,210}
\definecolor{c4}{RGB}{0,109,219}
\definecolor{c5}{RGB}{0,146,146}
\definecolor{c6}{RGB}{255,109,182}
\pgfplotscreateplotcyclelist{cycle1}{
{c1}, {c2}, {c3}, {c4}, {c5}, {c6}
}

\begin{document}
\title{Determination of the Collins-Soper kernel from Lattice QCD}
\author{Artur Avkhadiev}
\affiliation{Center for Theoretical Physics, Massachusetts Institute of Technology, Cambridge, MA 02139, U.S.A.}
\author{Phiala E. Shanahan}
\affiliation{Center for Theoretical Physics, Massachusetts Institute of Technology, Cambridge, MA 02139, U.S.A.}
\author{Michael L. Wagman}
 \affiliation{Fermi National Accelerator Laboratory, Batavia, IL 60510, USA}  
\author{Yong Zhao}
\affiliation{Physics Division, Argonne National Laboratory, Lemont, IL 60439, USA}

\preprint{FERMILAB-PUB-24-0037-T, MIT-CTP/5677}

\newcommand{\Min}{{\small \mathrm{M}}} 
\newcommand{\Euc}{{\small \mathrm{Euc.}}} 
\renewcommand{\vec}[1]{\boldsymbol{#1}}
\newcommand{\mom}[1]{\mathbf{#1}}
\newcommand{\pos}[1]{\mathbf{#1}}
\newcommand{\tran}{T} 
\newcommand{\wf}{\tilde{\phi}}
\newcommand{\symm}{{\small \textrm{symm.}}}
\newcommand{\qcntr}{{\small q\textrm{-cent.}}}
\newcommand{\unsubtr}{{\small \mathrm{unsub.}}} 
\newcommand{\bare}{{\small \mathrm{bare}}} 
\newcommand{\unexp}{{\mathrm{u}}} 
\newcommand{\ren}{{\small \mathrm{ren.}}} 
\newcommand{\nn}{\nonumber} 
\DeclareRobustCommand{\Eq}[1]{Eq.~\eqref{eq:#1}}
\DeclareRobustCommand{\Eqs}[2]{Eqs.~\eqref{eq:#1} and \eqref{eq:#2}}
\DeclareRobustCommand{\fig}[1]{Fig.~\ref{fig:#1}}
\DeclareRobustCommand{\figs}[2]{Figs.~\ref{fig:#1} and \ref{fig:#2}}
\DeclareRobustCommand{\app}[1]{App.~\ref{app:#1}}
\DeclareRobustCommand{\sec}[1]{Sec.~\ref{sec:#1}}
\DeclareRobustCommand{\secs}[2]{Secs.~\ref{sec:#1} and \ref{sec:#2}}
\DeclareRobustCommand{\tbl}[1]{Table~\ref{tbl:#1}}
\DeclareRobustCommand{\refcite}[1]{Ref.~\cite{#1}}
\DeclareRobustCommand{\refcites}[1]{Refs.~\cite{#1}}
\newcommand{\MSbar}{
\overline{\rm MS}    
} 
 \newcommand{\xMOM}{ 
 \rm RI/xMOM
 } 
\newcommand{\MOM}{
    \rm RI^\prime/MOM
 }

\newcommand{\DNP}{\mathcal{D}_{\text{NP}}}
\newcommand{\DP}{\mathcal{D}_\text{res}}

\newcommand{\arthur}[1]{{\color{red} \textbf{Arthur:}    #1}}
\newcommand{\phiala}[2]{{{\color[rgb]{0.5,0.3,0} 
\sout{#1}}{\color[rgb]{0.0,0.5,0.8} \textbf{ Phiala:}    #2}}}
\newcommand{\yong}[2]{{{\color[rgb]{0.0,0.0,1} \sout{#1}}{\color[rgb]{0.66,0.0,0.3} \textbf{ Yong:}    #2}}}
\newcommand{\mike}[2]{{{\color[rgb]{0.5,0.3,0} 
\sout{#1}}{\color[rgb]{0.7,0.5,0.8} \textbf{ Mike:}    #2}}}

\begin{abstract}
  This work presents a determination of the quark Collins-Soper kernel, which relates transverse-momentum-dependent parton distributions (TMDs) at different rapidity scales, using lattice quantum chromodynamics (QCD). 
    This is the first such determination with systematic control of quark mass, operator mixing, and discretization effects.
    Next-to-next-to-leading logarithmic matching is used to match lattice-calculable distributions to the corresponding TMDs. 
    The continuum-extrapolated lattice QCD results are consistent with several recent phenomenological parameterizations of the Collins-Soper kernel and are precise enough to disfavor other parameterizations.
\end{abstract}

\maketitle

Elucidating the three-dimensional structure of the proton is a key target of current and future experimental programs worldwide, including the COMPASS~\cite{Chiosso:2013ila,COMPASS:2021bws,COMPASS:2022xig,Alexeev:2022wgr,COMPASS:2023cgk,COMPASS:2023vqt,COMPASS:2023vhr} experiment at CERN, RHIC~\cite{Aschenauer:2015eha,RHICSPIN:2023zxx} at BNL, the 12~GeV program~\cite{Dudek:2012vr,CLAS:2021opg,CLAS:2021ovm,CLAS:2021lky,CLAS:2023wda,CLAS:2023gja} at the TJNAF, and future experiments at the planned Electron-Ion Collider~\cite{Boer:2010zf,Boer:2011fh,Accardi:2012qut,Zheng:2018ssm,AbdulKhalek:2022hcn,Burkert:2022hjz}. In recent years, considerable developments have been made to constrain the proton's transverse structure in particular, as parameterized by transverse-momentum-dependent parton distributions (TMDs)~\cite{Collins:1981uk,Collins:1981va,Collins:1984kg}. 

In this context, a quantity of particular importance is the Collins-Soper (CS) kernel: a fundamental nonperturbative function that appears as the universal rapidity evolution kernel for TMDs, which can be considered to characterise the QCD vacuum~\cite{Collins:1981uk,Collins:1981va,Collins:1984kg}. The CS kernel is not only a fundamental proton and nuclear structure observable of importance in its own right, it is also needed to compare TMDs measured at different scales and is required as input into measurements of electroweak observables including the W-boson mass~\cite{Bozzi:2019vnl} and in various nuclear structure studies~\cite{AbdulKhalek:2022hcn}.

Phenomenological extractions of the CS kernel from global fits of experimental data from Drell-Yan and Semi-Inclusive Deep-Inelastic Scattering (SIDIS) processes, however, are largely unconstrained in the nonperturbative region, specifically for kinematics with small transverse momentum scale $q_T \lesssim 0.3~\text{GeV}$. First constraints of the CS kernel from lattice QCD calculations~\cite{Shanahan:2019zcq,Shanahan:2020zxr,LatticeParton:2020uhz,Li:2021wvl,Schlemmer:2021aij,Shanahan:2021tst,LPC:2022ibr,Shu:2023cot,Chu:2023flm,Avkhadiev:2023poz} demonstrate the potential of this approach to provide first-principles information with sufficient precision to distinguish between different phenomenological models in this regime. Nevertheless, key systematic uncertainties, in particular discretization effects, remain to be controlled in all such calculations to date.

This work presents a new lattice QCD determination of the CS kernel which includes systematic control of quark mass, operator renormalization, and discretization effects, and uses next-to-next-to-leading logarithmic matching to TMDs from the corresponding lattice-calculable distributions. This allows a parameterization of the CS kernel to be constrained entirely by first-principles calculations for the first time. 

{\it \bf The Collins-Soper Kernel:}
The transverse momentum of a parton of flavor $i$ in a given hadron state is encoded in the TMDs $f_i^\text{TMD}(x,b_T,\mu,\zeta)$, which are functions of the longitudinal momentum fraction $x$ carried by the parton, the transverse displacement $b_T$ (the Fourier conjugate of $q_T$), the virtuality scale $\mu$, and the hadron momentum through the rapidity scale $\zeta$. Unlike the $\mu$-evolution of the TMDs, which is perturbative for perturbative scales $\mu$ and $\zeta$, the $\zeta$ evolution of TMDs is nonperturbative and is encoded in the CS kernel~\cite{Collins:1981uk,Collins:1981va}: 
\begin{equation}
    \gamma_i(b_T,\mu) = 2 \frac{d}{d \ln \zeta} \ln f_i^\text{TMD}(x,b_T,\mu,\zeta).
\end{equation}
The quark CS kernel $\gamma_q(b_T,\mu)$ is independent of flavor.
The kinematic regime of particular interest is for $b_T \gtrsim 0.6$~fm, where there is some tension between different phenomenological parameterizations of the kernel~\cite{Landry:2002ix,Scimemi:2019cmh,Bacchetta:2019sam,Bacchetta:2022awv,Moos:2023yfa,Isaacson:2023iui}. 

{\it \bf Lattice QCD calculation:} 
Constraints on the quark CS kernel are extracted from lattice QCD calculations on 3 ensembles of gauge fields produced by the MILC collaboration~\cite{MILC:2012znn} with 2+1+1 dynamical quark flavors, the one-loop Symanzik improved gauge action~\cite{Symanzik:1983dc,Curci:1983an,Luscher:1984xn,Luscher:1985zq}, and the highly improved staggered quark action with sea quark masses tuned to reproduce the physical pion mass~\cite{Follana:2003fe,Follana:2003ji,Follana:2006rc}. The calculations are performed as detailed in Ref.~\cite{Avkhadiev:2023poz}, which presented results on one ensemble of lattice gauge fields with four-volume $L^3\times T = (48a)^3 \times 64a$ with $a = 0.12$~fm. This work adds calculations on an additional two ensembles of gauge fields with four-volumes $L^3\times T = (32a)^3 \times 48a$ with $a = 0.15$~fm and $L^3\times T = (64a)^3 \times 96a$ with $a = 0.09$~fm, enabling systematic investigation of discretization effects for the first time. A summary of the computation and analysis are included below, with further details and figures provided in the Supplementary Material.

Within the Large-Momentum Effective Theory (LaMET)~\cite{Ji:2013dva,Ji:2014gla,Ji:2020ect} framework, the lightlike-separated operators that define physical TMDs are related to lattice-calculable `quasi-distributions' defined by matrix elements of purely space-like separated operators at large hadron momentum $\lvert \mom{P} \rvert \gg \Lambda_{\mathrm{QCD}}$~\cite{Ji:2014hxa,Ji:2018hvs,Ebert:2018gzl,Ebert:2019okf,Ebert:2019tvc,Ji:2019sxk,Ji:2019ewn,Ebert:2020gxr,Vladimirov:2020ofp,Ji:2020jeb,Ji:2021znw,Ebert:2022fmh,Schindler:2022eva,Zhu:2022bja,Rodini:2022wic} with matching coefficients computed perturbatively.
Using this approach, the quark CS kernel may be extracted from ratios of matrix elements of staple-shaped Wilson line operators in hadron states at different boost momenta $P_1^z$, $P_2^z$~\cite{Ji:2014hxa,Ebert:2018gzl,Ji:2019sxk}:
\begin{widetext}
    \begin{equation}
    \begin{aligned}
    \label{eq:kernel-wf-lattice}
    \gamma_q^{\MSbar}(b_{\tran}, \mu)
                  &=
                    \lim_{a\to 0}
                    \lim_{\ell \to \infty}
                    \frac{1}{\ln(P_{1}^{z}/P_{2}^{z})} 
                    \ln\frac
                    {\displaystyle
                        \int_{-\infty}^\infty \frac{\dd{b^z}}{2\pi}
                        e^{i \left(x-\frac{1}{2}\right) P_1^z b^z}
                        P_1^{z}
                        N_\Gamma(P_1^z)
                        \sum_{\Gamma^\prime} Z^{\MSbar}_{\Gamma \Gamma^\prime}(\mu,a)
                        W^{(0)}_{\Gamma^\prime}(b_\tran, b^{z}, P_1^{z}, \ell,a)
                    }
                    {\displaystyle
                        \int_{-\infty}^\infty \frac{\dd{b^z}}{2\pi}
                        e^{i \left(x-\frac{1}{2}\right) P_{2}^z b^z}
                        P_2^{z}
                        N_\Gamma(P_2^z)
                        \sum_{\Gamma^\prime} Z^{\MSbar}_{\Gamma \Gamma^\prime}(\mu,a)
                        W^{(0)}_{\Gamma^\prime}(b_\tran, b^z, P_2^{z}, \ell,a)
                    } 
                    \\ &\quad\quad+
                    {\delta \gamma}^{\MSbar}_q(\mu, x, P_1^z, P_2^z)
                     + \mathrm{p.c.}
    \end{aligned}
    \end{equation}
\end{widetext}
Here, the perturbative matching correction is denoted $\delta \gamma^{\MSbar}_q(\mu, x, P_1^z, P_2^z)$, $\mathrm{p.c.}$ denotes ($x$-dependent) power corrections proportional to powers of $1/(b_T P^z)^2$ and $(\Lambda/P^z)^2$ where $\Lambda$ is a generic hadronic scale, $N_{\Gamma}(P^z)$ are normalization factors corresponding to $N_{\gamma_3 \gamma_5}(P^z) = -im_h/P^z$ and $N_{\gamma_4 \gamma_5}(P^z) = m_h/E_h(P^z)$, where $m_h$ and $E_h(P^z)$ are the hadron mass and energy respectively, $Z^{\MSbar}_{\Gamma \Gamma^\prime}(\mu)$ are $16\times16$ renormalization matrices in Dirac space, and $W^{(0)}_\Gamma(b_\tran, b^z, P^{z}, \ell,a)$ denotes ratios of bare quark quasi-TMD wavefunctions (quasi-TMD WFs):
 \begin{align}
    \label{eq:quasi-wf-ratio}
        W^{(0)}_{\Gamma}(b_\tran, b^z, P^{z}, \ell,a) 
            &= \frac
                {\wf_{\Gamma}(b_\tran, b^{z}, P^z, \ell,a)}
                {\wf_{\gamma_4\gamma_5}(b_\tran, 0, 0, \ell,a)}.
    \end{align}
Here $\wf_{\Gamma}(b_\tran, b^z, P^{z}, \ell,a)$ are defined as matrix elements of nonlocal quark bilinear operators ${\mathcal{O}^{\Gamma}_{u\bar{d}}(b_\tran, b^z, \ell)}$ with the $u$ and $\bar{d}$ quarks separated by 4-vector $b = (\pos{b}_{\tran}, b^{z}, 0)$ and connected by a staple-shaped Wilson line of total length $\ell + b_\tran$, between the QCD vacuum and a hadron state at boost $P^{z}$:
\begin{equation}
    \label{eq:wf-bare}
    \begin{aligned}
    \wf_{\Gamma}(b_\tran, b^z, P^{z}, \ell,a)
                &= \mel{0} {\mathcal{O}^{\Gamma}_{u\bar{d}}(b_\tran, b^z, \ell)}{h(P^z)},
    \end{aligned} 
    \end{equation}
computed with lattice discretization scale $a$.
Because the kernel is independent of the choice of hadron state, pion states are used here for simplicity.
Similar approaches have been used in previous lattice QCD studies to constrain the quark CS kernel~\cite{Shanahan:2019zcq,Shanahan:2020zxr,LatticeParton:2020uhz,Li:2021wvl,Schlemmer:2021aij,Shanahan:2021tst,LPC:2022ibr,Shu:2023cot,Chu:2023flm} and other TMD quantities~\cite{LPC:2022zci,Chu:2023jia,LatticeParton:2020uhz,Li:2021wvl,Chu:2023flm}.

\begin{ruledtabular}
        \begin{table}[t]
            \begin{tabular}{cccc}
            $n^z$ & $P^z$ [GeV] & $\ell/a$ & $N_\text{cfg}$\\\hline
            \multicolumn{4}{c}{$L^3\times T = (32a)^3 \times 48a$, $a = 0.15$~fm}\\\hline
                0  & 0    & \{7,\,10,\,13,\,14,\,17,\,21,\,25\}  & 229     \\
                3  & 0.77 & \{21,\,25\} & 1105 \\
                5  & 1.29 & \{14,\,17\} & 1105 \\
                7  & 1.81 & \{10,\,13\} & 1105 \\
                9 & 2.32 & \{7,\,10\} & 1105 \\\hline
            \multicolumn{4}{c}{$L^3\times T = (48a)^3 \times 64a$, $a = 0.12$~fm} \\\hline
                0  & 0    & \{11,\,14,\,17,\,20,\,26,\,32\}  & 79     \\
                4  & 0.86 & \{26,\,32\} & 469 \\
                6  & 1.29 & \{17,\,20\} & 472 \\
                8  & 1.72 & \{14,\,17\} & 523 \\
                10 & 2.15 & \{11,\,14\} & 481 \\\hline
            \multicolumn{4}{c}{$L^3\times T = (64a)^3 \times 96a$, $a = 0.09$~fm} \\\hline
                0  & 0    & \{12,\,17,\,22,\,27,\,32,\,35,\,43\}  & 47     \\
                4  & 0.86 & \{35,\,43\} & 303 \\
                6  & 1.29 & \{27,\,32\} & 472 \\
                8  & 1.72 & \{17,\,22\} & 269 \\
                10 & 2.15 & \{12,\,17\} & 270 
            \end{tabular}
            \caption{\label{tab:measurements}
            Details of the parameters used for calculation on each ensemble of lattice gauge fields. Lattice momenta are specified as $n^z$, with $P^z = \frac{2\pi}{L} n^z$. For operators with staple extent $\ell/a$, all geometries with $-\ell/a \leq b^z \leq \ell/a$ and $0 \leq b_\tran/a \leq 7$ along $\hat{n}_T \in \lbrace \pm \hat{x}, \pm \hat{y} \rbrace$ are computed, for all of the 16 Dirac structures $\Gamma$. The number of configurations used for each measurement is denoted $N_\text{cfg}$; on each configuration, sources on a $2^4$ grid bisecting the lattice along each dimension are used.
            } 
        \end{table}
        \end{ruledtabular}

\begin{figure}[t]
                \centering
                \includegraphics[width=0.49\textwidth]{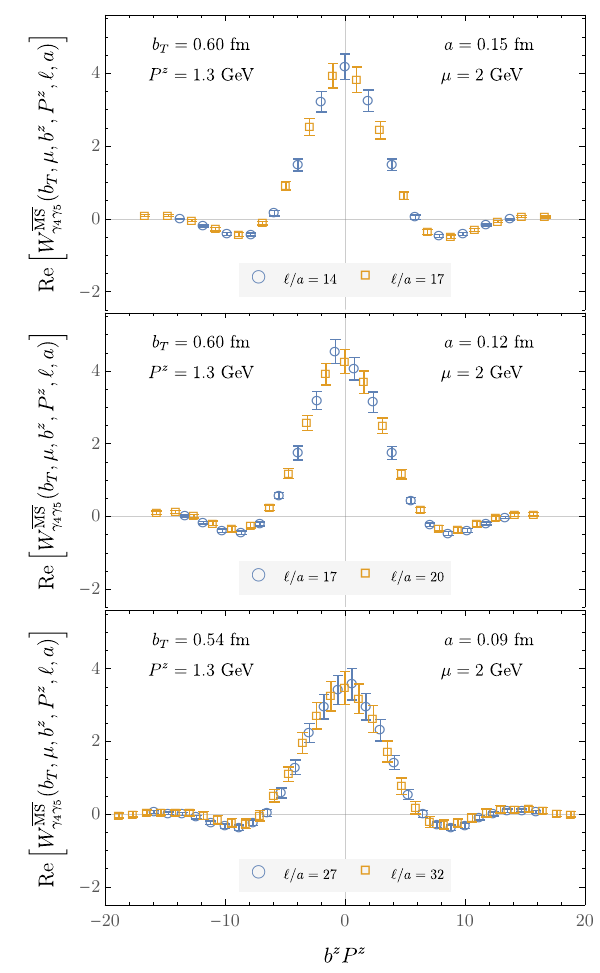}
            \caption{\label{fig:MSbarQuasiTMDWFratios}
                The real parts of the $\MSbar$-renormalized quasi-TMD WF ratios, $W^{\MSbar}_{\Gamma}(b_\tran, \mu, b^z, P^z, \ell,a)$, computed on each ensemble. Panels show results on ensembles with  $a\in\{0.15,0.12,0.09\}$~fm from top to bottom, as functions of $b^z$, for $\Gamma = \gamma_4\gamma_5$, $P^z = 1.3$ GeV, and similar though not identical $b_T$ values as indicated in each panel.
                }
        \end{figure}

The numerical calculation proceeds as detailed in Ref.~\cite{Avkhadiev:2023poz}:

{\it Computation of bare quasi-TMD WF ratios:}
    Bare quasi-TMD WFs $\wf_{\Gamma}(b_\tran, b^z, P^{z}, \ell,a)$ are extracted from bootstrap-level fits to the Euclidean time-dependence of two-point correlation functions both with and without staple-shaped operators. 
    Pion states are created with momentum-smeared interpolating fields~\cite{Bali:2016lva}, and the tree-level $\mathcal{O}(a)$-improved Wilson clover fermion action~\cite{Sheikholeslami:1985ij,Luscher:1996ug,Jansen:1998mx} is used for propagator computation, with clover term coefficient $c_{sw} = 1.0$. Hopping parameters are $\kappa \in\{0.12575, 0.12547, 0.1252\}$ for calculations on the ensembles with  $a\in\{0.15,0.12,0.09\}$~fm respectively, and field configurations are treated with Wilson flow to flow-time $\mathfrak{t} = 1.0$~\cite{Luscher:2010iy} and gauge-fixed to Landau gauge\footnote{Gauge-fixing is necessary for the computation of the renormalization factors discussed below.} before measurements are made.
    
    Calculations are performed on each ensemble for the operator choices (defined by $\Gamma$ and the staple geometry specified by $\ell$, $b^z$, $b_T$), choices of momenta $P^z$, and the numbers of configurations specified in Table~\ref{tab:measurements}. For the geometries with odd $\ell/a$ where no $b^z=0$ matrix elements are available (as used in the denominator of Eq.~\eqref{eq:quasi-wf-ratio} to form ratios), the average of those with $b^z/a = \pm 1$ are used. 

{\it Determination of renormalization factors $Z^{\MSbar}_{\Gamma\Gamma^\prime}(\mu, a)$:}
The $16\times 16$ matrices of renormalization factors $Z^{\MSbar}_{\Gamma\Gamma^\prime}(\mu)$ are computed using the $\xMOM$ renormalization scheme~\cite{Ji:2017oey,Green:2017xeu,Green:2020xco} and converted to $\MSbar$ using a conversion coefficient computed in continuum perturbation theory~\cite{Green:2020xco}. Calculations use $N_\text{cfg}\in\{120,32,30\}$ gauge field configurations on the ensembles with $a\in\{0.15,0.12,0.09\}$~fm, and statistical uncertainties are estimated using bootstrap resampling. As in Ref.~\cite{Avkhadiev:2023poz}, a range of renormalization scales defined by Wilson line lengths $\xi_\mathrm{R}/a \in \lbrace \{3,4,5\},\{2,3,4\},\{2,3,4\} \rbrace$ and off-shell quark momenta $p^\mu_\mathrm{R} = \frac{2\pi}{L}(0,0,n^z,0)$, with $n^z \in \lbrace \{8,10,12\},\{8,10,12\},\{6,8,10\} \rbrace$ are used on the ensembles with  $a\in\{0.15,0.12,0.09\}$~fm respectively. Central values are defined as those with $\xi_\mathrm{R}/a \in \lbrace 4,3,3 \rbrace$ and $n^z \in \lbrace 10,10,8 \rbrace$ for calculations with $a\in\{0.15,0.12,0.09\}$~fm, and a systematic uncertainty, added in quadrature with the statistical uncertainty, is defined as half the difference between the maximum and minimum $\xMOM$ renormalization factor over the scales studied. Examples of the $\MSbar$-renormalized quasi-TMD WF ratios, $W^{\MSbar}_{\Gamma}(b_\tran, \mu, b^z, P^z, \ell,a) = \sum_{\Gamma^\prime} Z^{\MSbar}_{\Gamma \Gamma^\prime}(\mu,a) W^{(0)}_{\Gamma^\prime}(b_\tran, b^z, P^z, \ell,a)$, computed on each ensemble, are shown in Fig.~\ref{fig:MSbarQuasiTMDWFratios}. 

{\it Fourier transformation:}
After renormalization and multiplication by $N_\Gamma(P^z)$, a Discrete Fourier Transform (DFT) is used to realize the Fourier transforms in the numerator and denominator of Eq.~\eqref{eq:kernel-wf-lattice}, where the quasi-TMD WF ratios for each $P^z$ are first averaged over $\pm b^z$ and the relevant values of $\ell(P^z)$. This yields $x$-space renormalized quasi-TMD WF ratios $W_\Gamma^{\MSbar}(b_\tran,  \mu, x, P^{z},a)$. 
Because results are computed for a finite range of $b^z$, the DFT is effectively truncated to a finite range.
The effects of this truncation are studied by comparing results using subsets of the data with $b^z < b^z_{\text{max}}$ and varying $b^z_{\text{max}}$, as well as by comparing results where an analytical model used to extrapolate to $b^z > b^z_{\text{max}}$.
Truncation effects are found in this way to lead to negligible systematic uncertainties, as detailed for the $a = 0.12$ fm ensemble in Ref.~\cite{Avkhadiev:2023poz}.

{\it Perturbative matching:} The perturbative matching correction $\delta\gamma^{\MSbar}_q(\mu, x,P_1^z, P_2^z)$ appearing in \cref{eq:kernel-wf-lattice} is taken to be the `$b_T$-unexpanded resummed next-to-next-to-leading order' (uNNLL) correction detailed in Ref.~\cite{Avkhadiev:2023poz}. It was found in the analysis of Ref.~\cite{Avkhadiev:2023poz}, for the same $a=0.12$~fm ensemble also studied here, that this choice reduces the effect of $b_T$-dependent power corrections and offers the best convergence compared with other currently-available matching prescriptions, i.e., fixed-order~\cite{Ji:2021znw,Deng:2022gzi,delRio:2023pse,Ji:2023pba} and resummed~\cite{Ji:2019ewn,Ebert:2022fmh,Stewart:2010qs} corrections up to next-to-next-to-leading order.

{\it Extraction of the CS kernel:} Each choice of $x$, $P^z$, and $a$ defines an estimator for the CS kernel (corresponding to the right-hand side of Eq.~\eqref{eq:kernel-wf-lattice} with the Fourier Transform implemented via DFT, neglecting the p.c.\ term, and without the limit $a\rightarrow 0$ being taken): 
\begin{equation}
    \begin{aligned}
    \label{eq:kernel-wf-estimator}
    \hat{\gamma}&_{\Gamma}^{\MSbar}(b_{\tran}, \mu, x, P_1^z, P_2^z, a) \\
    &= \dfrac{1}{\ln(P_{1}^{z}/P_{2}^{z})}
                  \ln\Bigg\lbrack
                   \dfrac
                    {\displaystyle
                        W_\Gamma^{\MSbar}(b_\tran,  \mu, x, P^{z}_1, a)
                    }
                    {\displaystyle
                        W_\Gamma^{\MSbar}(b_\tran,  \mu, x, P^{z}_2, a)
                    } 
                \Bigg\rbrack
        \\&\quad+ \delta \gamma_q^{\MSbar}(  \mu, x, P_1^z, P_2^z).
    \end{aligned}
    \end{equation}
In principle, it would be desirable to exploit the multiple lattice ensembles available in this calculation to perform a continuum extrapolation of the CS kernel estimator for individual $b_\tran$ values; this would require matched $b\tran$ across the ensembles. Alternatively, one might aim to disentangle power corrections and discretization effects by fitting all results for $\mathrm{Re}\big\lbrack\hat{\gamma}_{\Gamma}^{\MSbar}(b_{\tran},  \mu, x, P_1^z, P_2^z,a)\big\rbrack$ to a parameterization of the CS kernel plus $P^z$, $a$, and $b_T$-dependent terms\footnote{Specifically such corrections would be proportional to terms such as $a/b_T$, $a^2/b_T^2$, 
$\frac{1}{\ln(P_1^z/P_2^z)}  \left[ \frac{1}{b_T^2 (P_1^z)^2} - \frac{1}{b_T^2 (P_2^z)^2} \right]$, $\frac{ \Lambda^2}{\ln(P_1^z/P_2^z)}  \left[ \frac{1}{(P_1^z)^2} - \frac{1}{(P_2^z)^2} \right]$, $\frac{1}{\ln(P_1^z/P_2^z)}\left[ a(P_1^z - P_2^z) \right]$, $\frac{1}{\ln(P_1^z/P_2^z)}\left[ a^2(P_1^z)^2 - a^2 (P_2^z) \right]$, \ldots}. In practice, estimators for different $\{P_1^z, P_2^z\}$ are largely consistent, and as such the data is insufficient to constrain momentum-dependent power corrections.

\begin{figure}[t]
                \centering
                \includegraphics[width=0.48\textwidth]{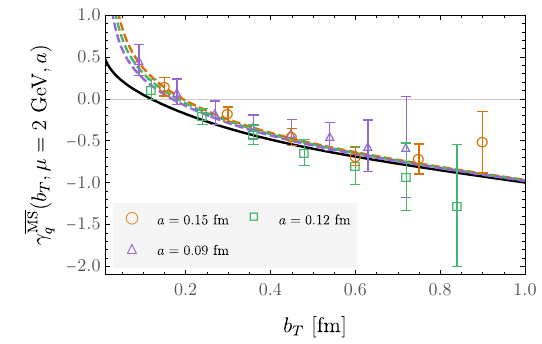}\\[2pt]
                \includegraphics[width=0.48\textwidth]{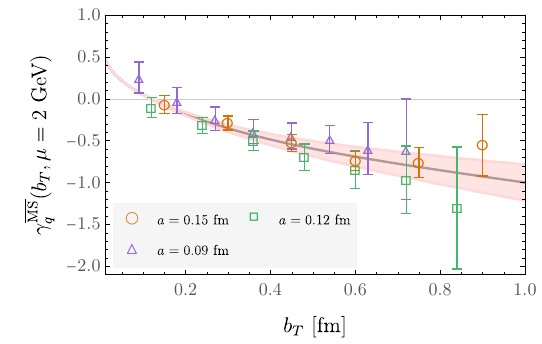}
            \caption{\label{fig:CSkerneldata}
                [Upper] Averaged CS kernel estimators computed on each ensemble, including a fit to a parameterization of the CS kernel plus $\mathcal{O}(a/b_T)$ discretization effects, as described in the text. The colored dashed curves correspond to $\gamma^{\text{param}}_q(b_{\tran}, \mu, a)$, with the best-fit values of $(B_\text{NP},c_0,c_1,k_1,k_2)$ as described in the text, at each corresponding value of $a$, while the solid black curve shows the result at $a=0$. [Lower] Lattice QCD constraints on the CS kernel, with $\mathcal{O}(a/b_T)$ artefacts subtracted as defined in the text, and the best-fit parameterization of the CS kernel fit to the lattice results shown as a solid black curve with the 1$\sigma$ uncertainty band shown as a shaded red region.
                }
\end{figure}

Instead, the CS kernel on each ensemble is determined as a bootstrap-level weighted average of $\mathrm{Re}\big\lbrack\hat{\gamma}_{\Gamma}^{\MSbar}(b_{\tran},  \mu, x, P_1^z, P_2^z,a)\big\rbrack$ over $\Gamma \in \{\gamma_4 \gamma_5, \gamma_3\gamma_5\}$, all available combinations of $\{P_1^z, P_2^z\}$, and $x \in [0.3, 0.7]$, with weights proportional to the inverse variance, just as done in Ref.~\cite{Avkhadiev:2023poz}. These $P_z$-, $\Gamma$-, and $x$-averaged CS kernel constraints, denoted $\gamma^{\MSbar}_q(b_{\tran}, \mu, a)$, should agree with the CS kernel up to discretization effects.
The results (including a fit to a parameterization of the CS kernel and the leading $a/b_T$ discretization effects, as discussed in the next section), are shown in Fig.~\ref{fig:CSkerneldata}.

Additionally, an analogous analysis of $\mathrm{Im}\big\lbrack\hat{\gamma}_{\Gamma}^{\MSbar}(b_{\tran},  \mu, x, P_1^z, P_2^z,a)\big\rbrack$ can be performed. As the CS kernel is purely real, significant deviation of the resulting numerical results from zero would provide an indication of systematic uncertainties beyond those that are quantified in this calculation. This analysis is presented in the Supplementary Material. Including the uNNLL matching of Ref.~\cite{Avkhadiev:2023poz} and recent devlopments~\cite{Liu:2023onm} accounting for a linear infrared renormalon in the imaginary part of the matching coefficient for the quasi-TMD WF, there is no evidence in the numerical data for significant additional unconstrained systematic uncertainties.

{\it \bf Parameterization:} The Lattice QCD constraints on the CS kernel are fit to the parameterization of Ref.~\cite{Moos:2023yfa}, with the addition of terms accounting for lattice discretization effects proportional to $a/b_T$, $a^2/b_T^2$:
\begin{equation}\label{eq:CSkernelparam}
\begin{aligned}
    \gamma_q^\text{param.}&(b_T,\mu,a; B_\text{NP},c_0,c_1;k_1,k_2) =  -2\DP(b^*,\mu)\\
    & -2\DNP(b_T;B_\text{NP},c_0,c_1) + k_1 \frac{a}{b_T} + k_2 \frac{a^2}{b_T^2},
    \end{aligned}
\end{equation}
where\footnote{$\DP$ is given explicitly in the Supplementary Material.} $\DP$ is the resummed leading power expression for the CS kernel computed in the operator product expansion, evolved to scale $\mu$, and the parameterization of the remaining nonperturbative piece is
\begin{equation} \label{eq:DNP}
    \DNP(b_T;B_{\text{NP}}) = b_T b^* \left[ c_0+c_1\ln\left({\frac{b^*}{B_\text{NP}}}\right)\right],
\end{equation}
and
\begin{equation}
    b^*(b_T; B_{\text{NP}}) = \frac{b_T}{\sqrt{1+\frac{b_T^2}{B^2_\text{NP}}}}.
\end{equation}
The expression of Eq.~\eqref{eq:CSkernelparam} is thus a three-parameter $(B_\text{NP},c_0,c_1)$ parameterization of the CS kernel, with an additional two parameters $(k_1,k_2)$ modeling lattice discretization effects. 

The lattice QCD constraints on the CS kernel, for each of the three values of $a$ used in the numerical calculations, are fit simultaneously to Eq.~\eqref{eq:CSkernelparam} to yield $(B_\text{NP},c_0,c_1,k_1,k_2$).
To diagnose overfitting, additional fits are performed in which subsets of the model parameters are held fixed at reference values, namely $c_1 = k_1 = k_2 = 0$ and $B_\text{NP} = 2$ GeV, while others are optimized.
The Akaike Information Criterion (AIC)~\cite{AkaikeAIC} is used to quantify the relative goodness-of-fit for models including different parameter subsets.
The minimum AIC model is found to be $(c_0,k_1)$ with $c_1=k_2=0$ and $B_\text{NP} = 2$ GeV.
The corresponding fit results are
\begin{equation}
    c_0 = 0.032(12),\ k_1 = 0.22(8),
\end{equation}
with a $\chi^2 / \text{dof} = 0.39$.
These fit results, and the resulting parameterization of the CS kernel, are shown in Fig.~\ref{fig:CSkerneldata}.
Overall fit quality is illustrated through the comparison of $\gamma^{\text{param}}_q(b_{\tran}, \mu, a=0)$ with best-fit values for $(B_\text{NP},c_0,c_1,k_1,k_2)$ with the lattice QCD results where discretization effects have been subtracted, i.e., $\gamma^{\MSbar}_q(b_{\tran}, \mu) \equiv \gamma^{\MSbar}_q(b_{\tran}, \mu, a) - k_1 (a/b_T)$ using the best-fit results for $k_1$.

\begin{figure}[t]
                \centering
                \includegraphics[width=0.48\textwidth]{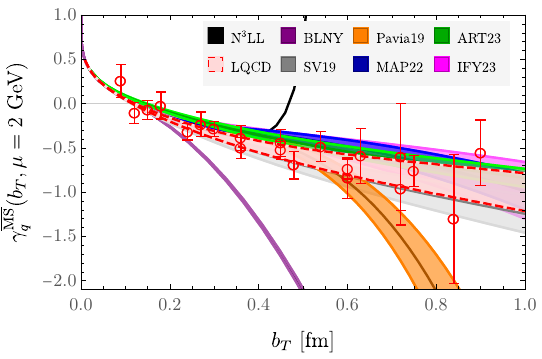}
            \caption{\label{fig:phenocomparison}
                Comparison of lattice QCD parameterization of the CS kernel compared with phenomenological parameterizations~\cite{Landry:2002ix,Scimemi:2019cmh,Bacchetta:2019sam,Bacchetta:2022awv,Moos:2023yfa,Isaacson:2023iui} of experimental data (BLNY, SV19, Pavia19, MAP22, ART23, IFY23), and perturbative results from Refs.~\cite{Collins:2014jpa,Li:2016ctv,Vladimirov:2016dll} ($\mathrm{N}^3\mathrm{LL}$).
                }
\end{figure}

These continuum-limit results are compared with phenomenological parameterizations of experimental data in Fig.~\ref{fig:phenocomparison}.
In particular, the parameterization used in Ref.~\cite{Scimemi:2019cmh} corresponds to the AIC-preferred parameterization used here and leads to a consistent result $c_0^{\text{SV19}} = 0.043(11)$ with $B_{\text{NP}}^{\text{SV19}} = 1.9(2)$ GeV. The global fits performed in Ref.~\cite{Moos:2023yfa} also give a consistent result, $c_0^\text{ART23} = 0.037(6)$, though in that work $c_1$ is also included as a fit parameter.

Fits to other parameter subsets $(c_0,k_2)$ and $(c_0,k_1,k_2)$ give consistent results for $c_0$ at $1\sigma$ with uncertainties that differ by $\lesssim 10\%$.
The magnitudes of $k_1$ and $k_2$ range from 0.1 - 0.3 in all cases, which suggests that the size of discretization effects is consistent with naive dimensional analysis.
Fits including $B_{\text{NP}}$ or $c_1$ as free parameters give consistent results for $c_0$ with larger uncertainties.

Other parameterizations for the nonperturbative function $\DNP(b)$ have been used in fits to experimental data~\cite{Landry:2002ix,Sun:2014dqm}, for example the BLNY parameterization $\DNP^{\text{BLNY}}(b) = g_2 b^2$ with free parameters $g_2$ and $B_{\text{NP}}$ (which enters $\DP$).
Fits to this parameterization with $B_{\text{NP}} = 1.5$ GeV lead to the result $g_2 = 0.085(26)$ with comparable goodness-of-fit, $\chi^2 / \text{dof} = 0.58$, to the fits using the parametrization of Eq.~\eqref{eq:DNP} described above.
This is consistent with the phenomenological fit results of Ref.~\cite{Isaacson:2023iui}, which use the same value of $B_{\text{NP}}$ and find $g_2 = 0.053(24)$.
These lattice QCD constraints on the CS kernel are therefore not sufficient to establish a clear preference between functional forms for $\DNP$;
however they do provide a significant preference for the recent fit results from Refs.~\cite{Scimemi:2019cmh,Bacchetta:2022awv,Moos:2023yfa,Isaacson:2023iui}
in comparison with Ref.~\cite{Bacchetta:2019sam} and especially with older BLNY fit results~\cite{Landry:2002ix} at large $b_T$.

{\it \bf Summary:} This work presents the first lattice QCD calculation of the CS kernel with systematic control of quark mass, operator renormalization, and discretization effects. The results are used to constrain a `pure-theory' parameterization of the CS kernel through a direct fit to lattice QCD results for the first time. These lattice QCD results for the CS kernel are consistent with the most recent phenomenological results. This opens the door for future first-principles QCD predictions of the CS kernel beyond the region constrained by current experiments, as well as joint fits to experimental data and lattice QCD results. As more precise lattice QCD results are achieved at larger values of $b_T$ in future calculations, this promises to be increasingly valuable.

\begin{acknowledgments}
We thank Yang Fu for helpful discussions and Johannes Michel for valuable comments on the manuscript.
This work is supported in part by the U.S. Department of Energy, Office of Science, Office of Nuclear Physics, under grant Contract Numbers DE-SC0011090, DE-AC02-06CH11357, and by Early Career Award DE-SC0021006. PES is supported in part by Simons Foundation grant 994314 (Simons Collaboration on Confinement and QCD Strings). YZ is also supported in part by the 2023 Physical Sciences and Engineering (PSE) Early Investigator Named Award program at Argonne National Laboratory.
This manuscript has been authored by Fermi Research Alliance, LLC under Contract No.\ DE-AC02-07CH11359 with the U.S.\ Department of Energy, Office of Science, Office of High Energy Physics.

This research used resources of the National Energy Research Scientific Computing Center (NERSC), a U.S. Department of Energy Office of Science User Facility operated under Contract No. DE-AC02-05CH11231, the Extreme Science and Engineering Discovery Environment (XSEDE) Bridges-2 at the Pittsburgh Supercomputing Center (PSC) through allocation TG-PHY200036, which is supported by National Science Foundation grant number ACI-1548562, facilities of the USQCD Collaboration, which are funded by the Office of Science of the U.S. Department of Energy. 
We gratefully acknowledge the computing resources provided on Bebop, a high-performance computing cluster operated by the Laboratory Computing Resource Center at Argonne National Laboratory, and the computing resources at the MIT SuperCloud and Lincoln Laboratory Supercomputing Center~\cite{reuther2018interactive}.
The Chroma~\cite{Edwards:2004sx}, QLua \cite{qlua},  QUDA \cite{Clark:2009wm,Babich:2011np,Clark:2016rdz}, QDP-JIT \cite{6877336}, and QPhiX~\cite{10.1007/978-3-319-46079-6_30} software libraries were used in this work.
Data analysis used NumPy~\cite{harris2020array} and Julia~\cite{Julia-2017,mogensen2018optim}, and figures were produced using Mathematica \cite{Mathematica}.
\end{acknowledgments}

\bibliography{refs}

\begin{figure*}[!t]
            \centering
                \includegraphics[width=0.32\textwidth]{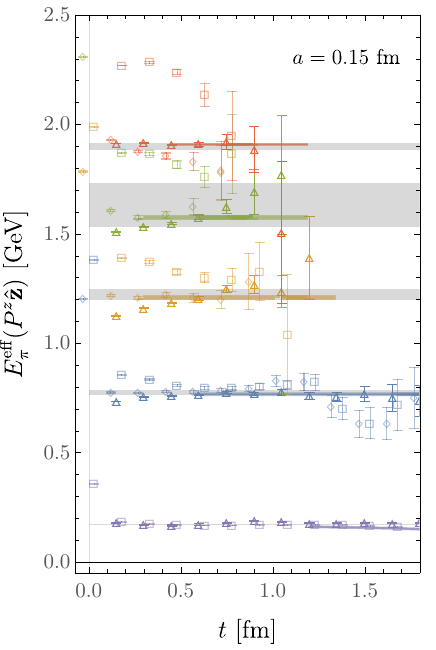}
                \includegraphics[width=0.32\textwidth]{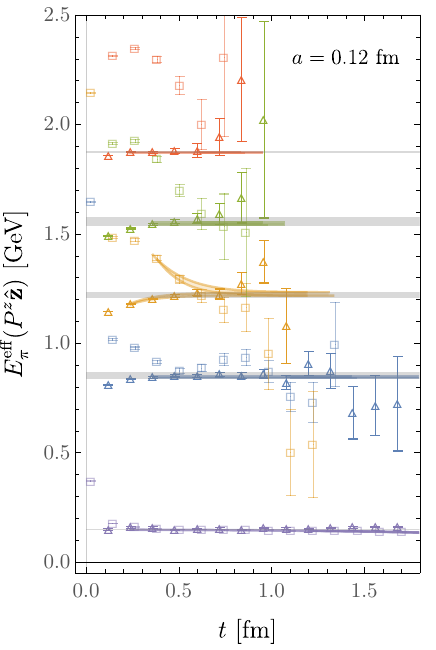}   
                \includegraphics[width=0.32\textwidth]{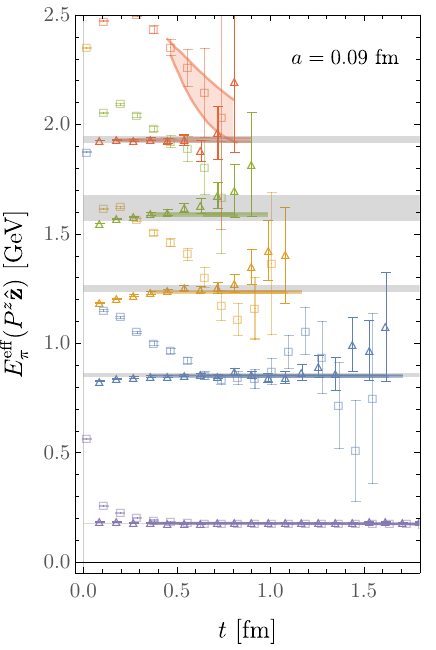}   
                \caption{\label{fig:EMP} Effective energies (\cref{eq:eff-energy}), constructed using $C^{\pi}_{2\mathrm{pt}}$ (squares, offset slightly on the horizontal axis) and the most statistically precise $C^{\Gamma }_{2\mathrm{pt}}$ (triangles) for each choice of $P^z$ on each ensemble.
                The gray bands show fit results obtained as detailed in Ref.~\cite{Avkhadiev:2023poz} as weighted averages over fits with different fit ranges and numbers of excited states, while the colored bands show the single highest-weight fit included in each average.
                }
            \end{figure*}

\newpage
\section*{Supplementary material}

This Supplementary Material collects additional results of the intermediate stages of analysis in the numerical calculation of the CS kernel. 

\subsection{Two-point correlation functions and effective energies}\label{app:2pt}

For each of the ensembles detailed in Table~\ref{tab:measurements}, two-point correlation functions both with and without staple-shaped Wilson line operators are constructed, and used to extract the bare quasi-TMD WF ratios ratios $W_\Gamma^{(0)}(b_\tran, b^{z}, P^{z}, \ell)$ defined in \cref{eq:quasi-wf-ratio}, as described in the text. Pion interpolating operators are defined with momentum-smeared quark fields constructed with a Gaussian momentum-smearing kernel $F_{\mom{K}}$ with $\mom{K}=\pm \mom{P}/2$, realized with $n_{\mathrm{smear}} = \num{50}$ smearing steps and width $\varepsilon = \num{0.2}$~\cite{Bali:2016lva}:
        \begin{equation}
            \chi^\dagger_{\mom{P}}(x) = \bar{u}_{F_{\mom{P}/2}}(x) \gamma_5 d_{ F_{-\mom{P}/2}}(x).
        \end{equation}
The two-point correlation functions are constructed as
        \begin{equation}
        \begin{aligned}
                C&_{\mathrm{2pt}}^{\pi}(t, \mom{P})  \equiv a^6 \sum_{\mathclap{\pos{y}}}     
                    e^{i \mom{P}\cdot \pos{y}} 
                \left<
                            \chi_{\mathbf{P}}(y)
                            \chi^\dagger_{\mathbf{P}}(0) 
                \right>,\\
            C&^{\Gamma}_{\mathrm{2pt}}(t, b_\tran, b^z, \mom{P}, \ell)  \\
                &\equiv a^6 \sum_{\mathclap{\pos{y}}}    
                e^{i \mom{P}\cdot \pos{y}} 
                \left<
                    \mathcal{O}^{\Gamma}_{u\bar{d}}(b_\tran, b^z, y, \ell)
                        \chi^\dagger_{\mom{P}}(0)
                \right>,
        \end{aligned}
        \end{equation}
        where labels $a$ denoting the discretization scales of different ensembles are suppressed, and $\mom{P} = P^{z} \hat{\mom{z}}$, $t= y_4$.
        
            \begin{figure}[!t]
            \centering
         \includegraphics[width=0.46\textwidth]{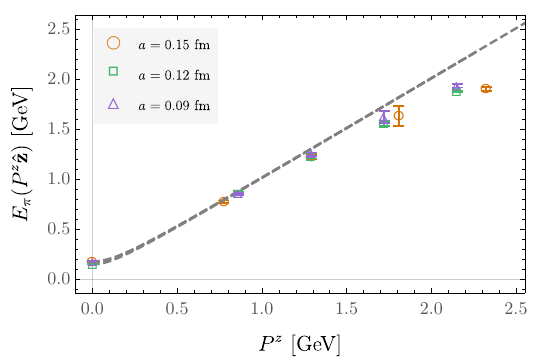}
                \caption{
                Extracted energies $E_\pi(P^z\hat{\mom{z}})$ on each ensemble as functions of $P^z$.
                The energy determined from the continuum dispersion relation is shown as a grey dashed line.
                \label{fig:energy-disp}}
            \end{figure}

Energies $E_\pi(P^z \hat{\mom{z}})$ on each ensemble, for each considered value of momentum $P^z$, are extracted from effective energies constructed using both $C^{\pi}_{2\mathrm{pt}}$ and the most statistically precise $C^{\Gamma }_{2\mathrm{pt}}$ for each choice of $P^z$, as detailed in Ref.~\cite{Avkhadiev:2023poz}:        
\begin{equation}
            \label{eq:eff-energy}
            \begin{aligned}
                a E_{\pi}^{\mathrm{eff}}\left(t + \frac{1}{2}a, \mom{P}=P^z\hat{\mom{z}}\right)
                        &=  \mathrm{log}
                                \left[ \frac{
                                     C^{\pi/\Gamma}_{2\mathrm{pt}}(t, \mom{P})
                                }
                                {C^{\pi/\Gamma}_{2\mathrm{pt}}(t+a, \mom{P})} \right] \\
                        &\xrightarrow{T \gg t \gg 0} a E_{\pi}(\mom{P}) + \ldots.
                    \end{aligned}
            \end{equation}
Effective energies on each ensemble are shown in Fig.~\ref{fig:EMP}, and the extracted values of $E_\pi(P^z \hat{\mom{z}})$, compared with the continuum dispersion relation $E_{\pi}(P^z \hat{\mom{z}}) = \sqrt{E_\pi(0)^2 + \left\lvert P^{z} \hat{\mom{z}} \right\rvert^2}$, are shown in Fig.~\ref{fig:energy-disp}.
Deviations between $E_\pi(P^z \hat{\mom{z}})$ and the continuum dispersion relation are observed to increase approximately linearly with $a P^z$, as expected since tree-level $O(a)$ improvement only partially mitigates $O(a)$ discretization effects.
The deviations from the continuum dispersion relation on each ensemble are $\lesssim 19\%$ with $a=0.15$ fm, $\lesssim 15\%$ with $a=0.12$ fm, and $\lesssim 10\%$ with $a = 0.09$ fm.

\subsection{Bare and renormalized quasi-TMD WF ratios}\label{app:WF}

\begin{figure*}[t]
    \centering
        \includegraphics[width=0.98\textwidth]{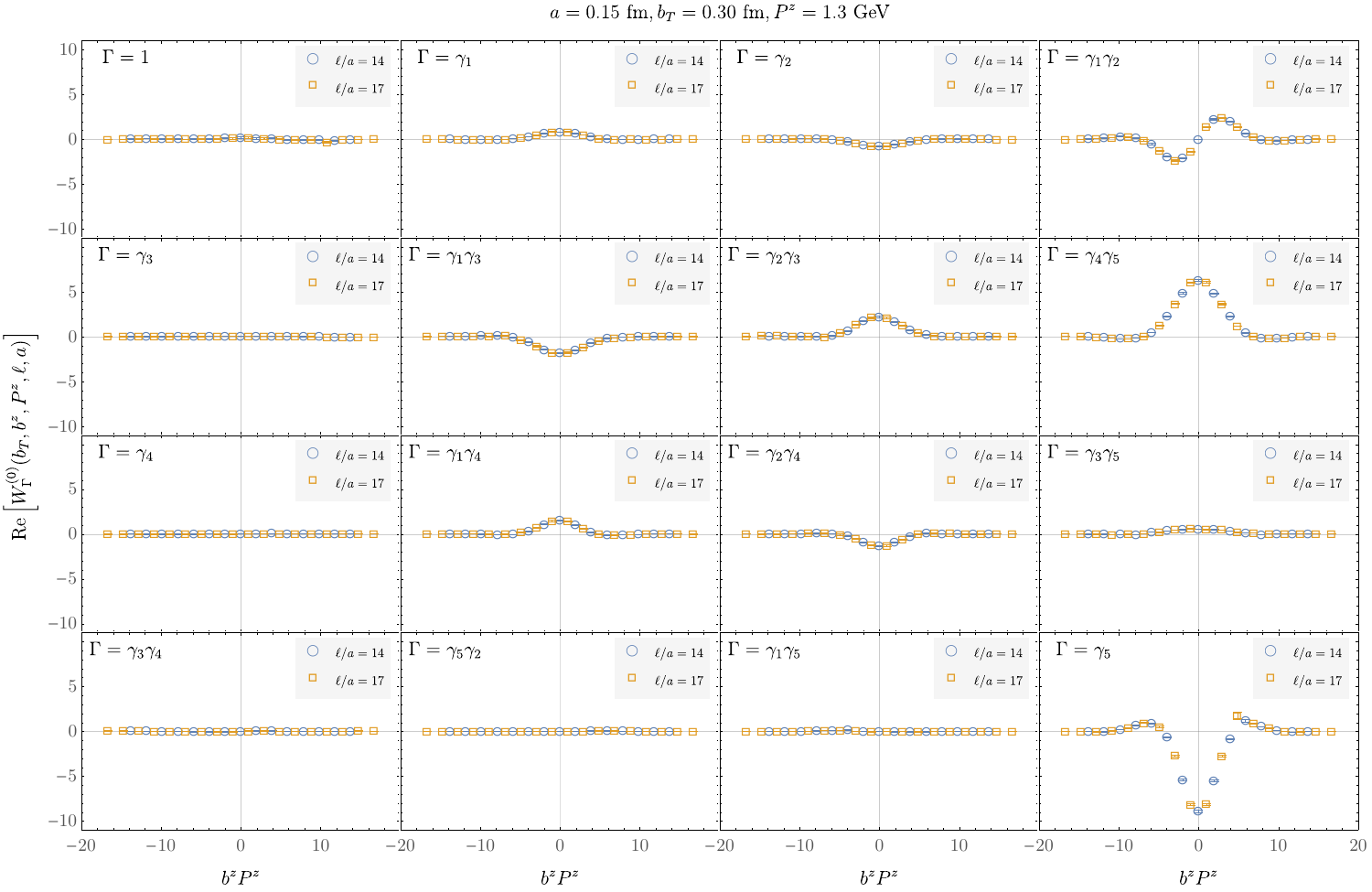} 
        \hspace{20pt}
        \includegraphics[width=0.98\textwidth]{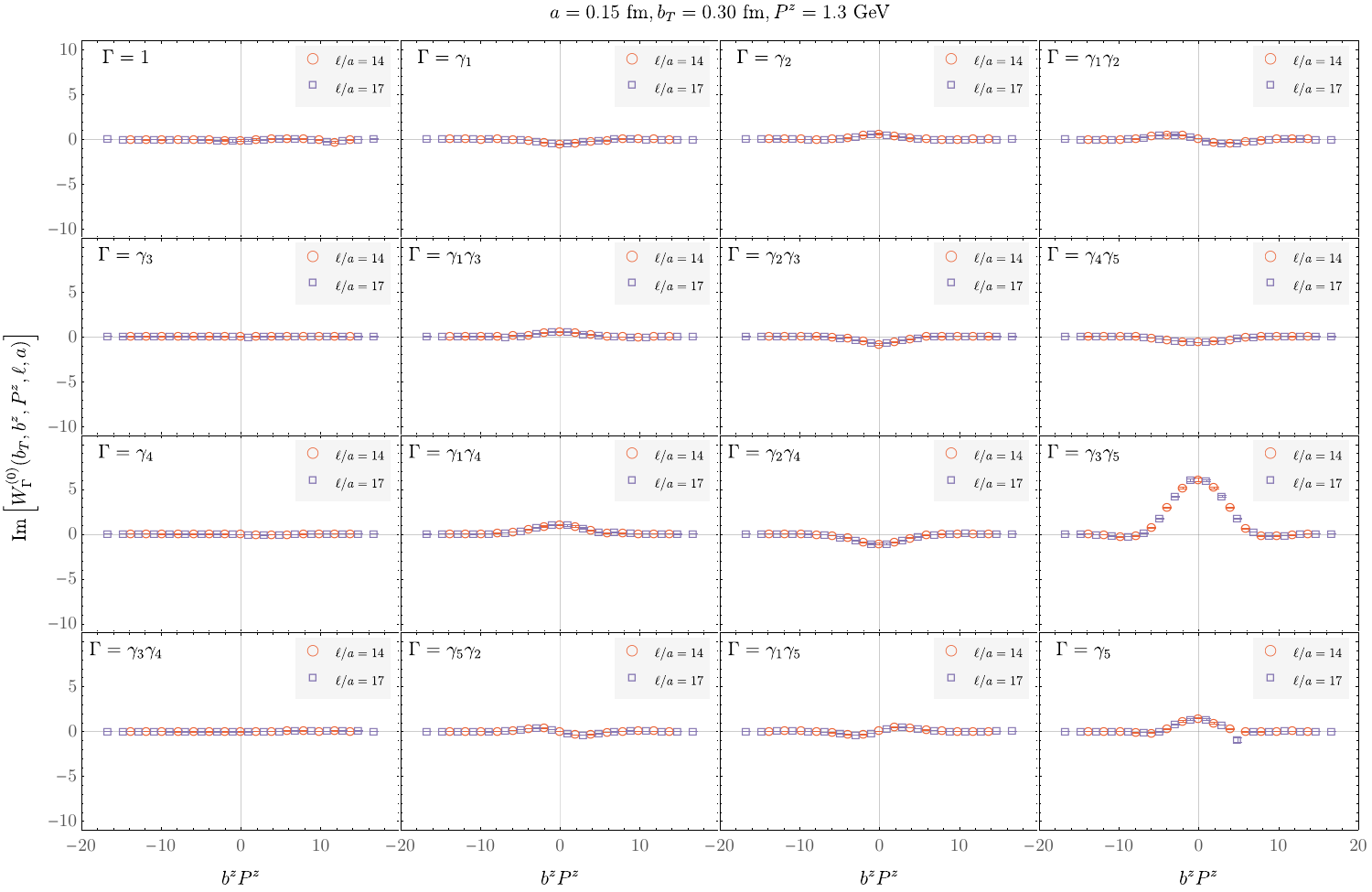} 
        \caption{Examples of real and imaginary parts of the bare quasi-TMD WF ratios $W^{(0)}_{\Gamma}(b_\tran, b^z, P^{z}, \ell,a)$ computed on the $a = 0.15~\text{fm}$ ensemble for all $\Gamma$, $\ell$, and $b^z$ with $b_T/a = 2$ and $n^z = 5$.
        \label{fig:L32_WF_bare_allgamma_comp}
        }
\end{figure*}

\begin{figure*}[t]
    \centering
        \includegraphics[width=0.98\textwidth]{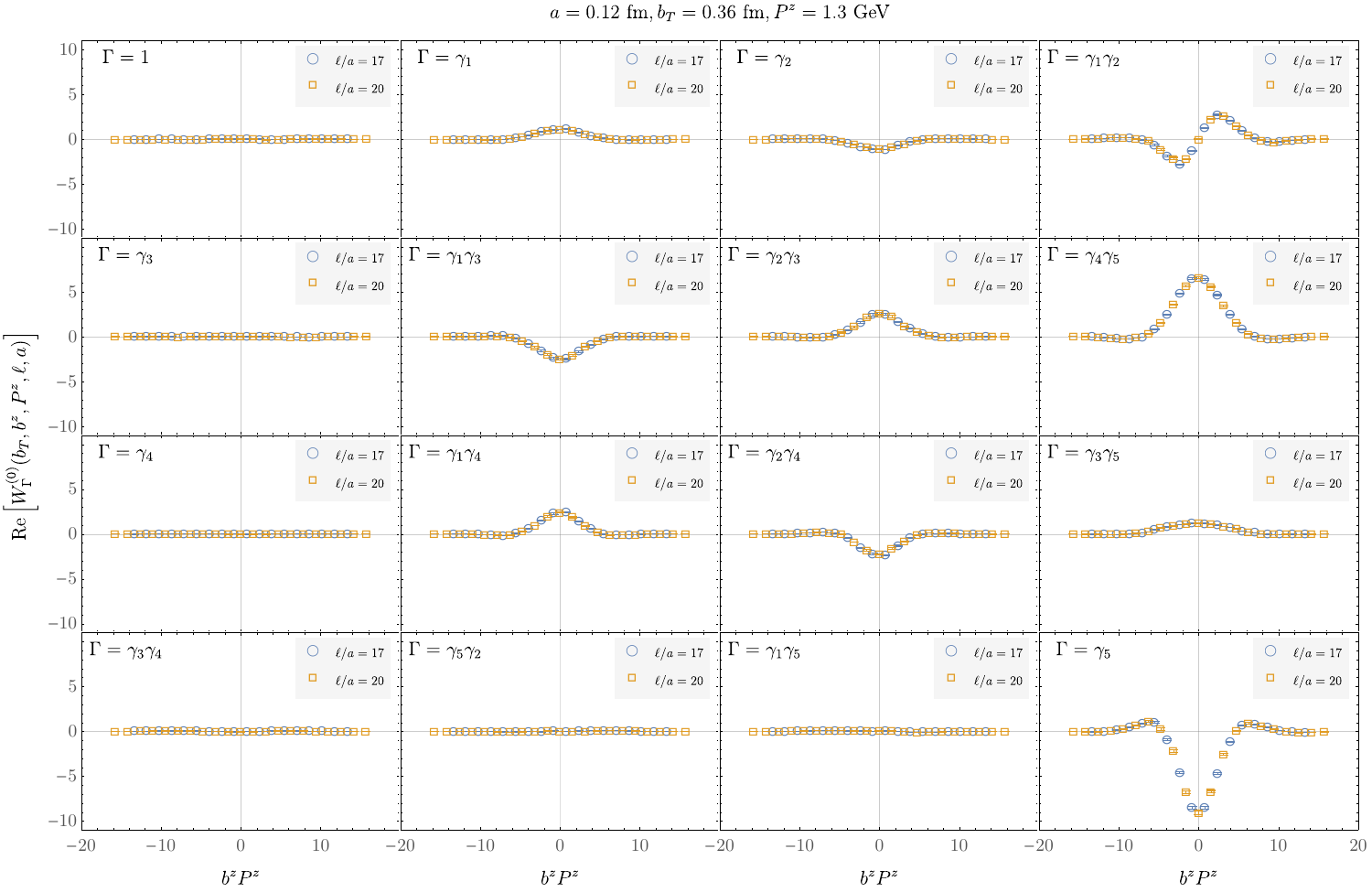} 
        \hspace{20pt}
        \includegraphics[width=0.98\textwidth]{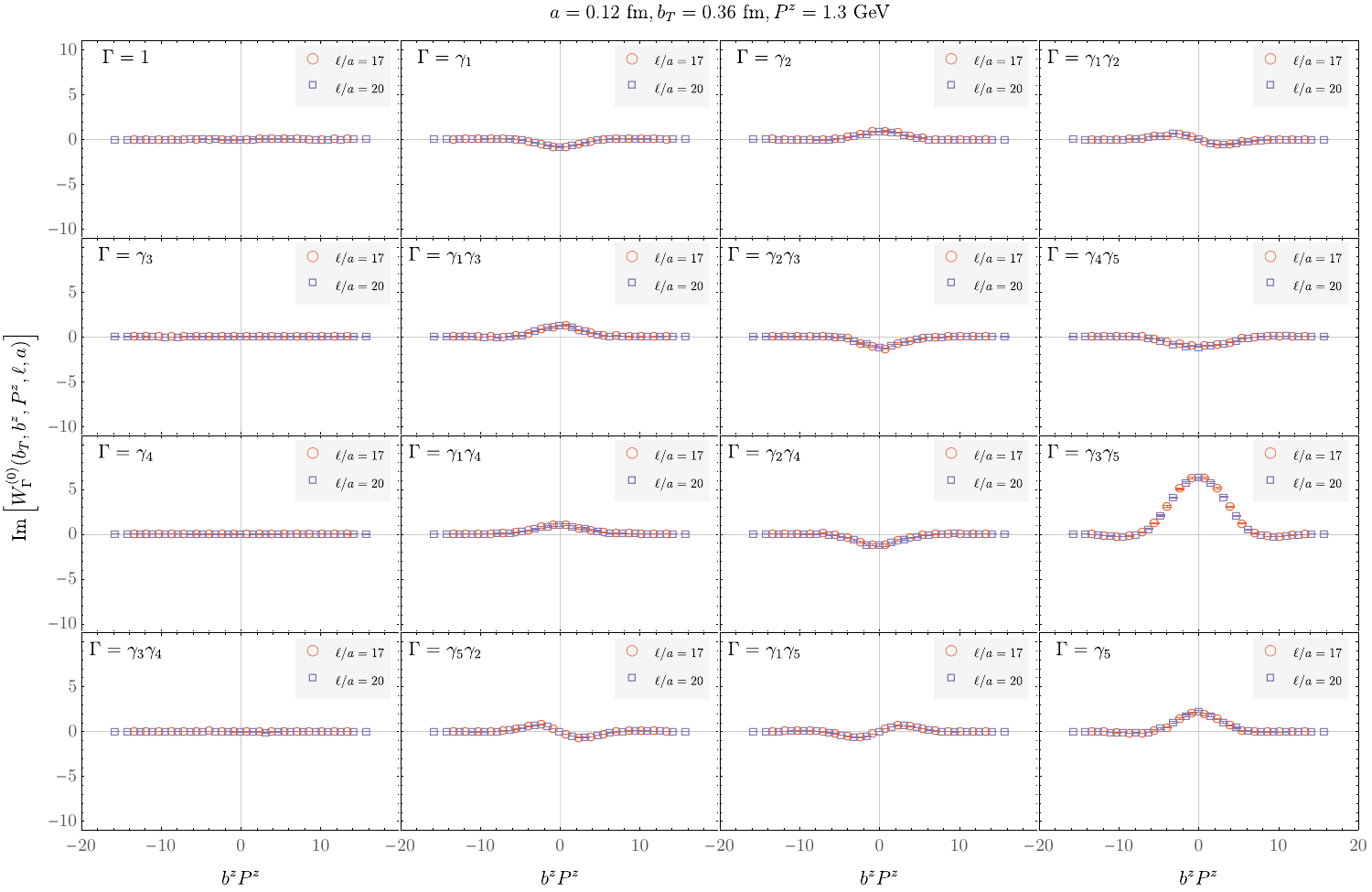} 
        \caption{Examples of real and imaginary parts of the bare quasi-TMD WF ratios $W^{(0)}_{\Gamma}(b_\tran, b^z, P^{z}, \ell,a)$ computed on the $a = 0.12~\text{fm}$ ensemble for all $\Gamma$, $\ell$, and $b^z$ with $b_T/a = 3$ and $n^z = 6$.
        \label{fig:L48_WF_bare_allgamma_comp}
        }
\end{figure*}

\begin{figure*}[t]
    \centering
        \includegraphics[width=0.98\textwidth]{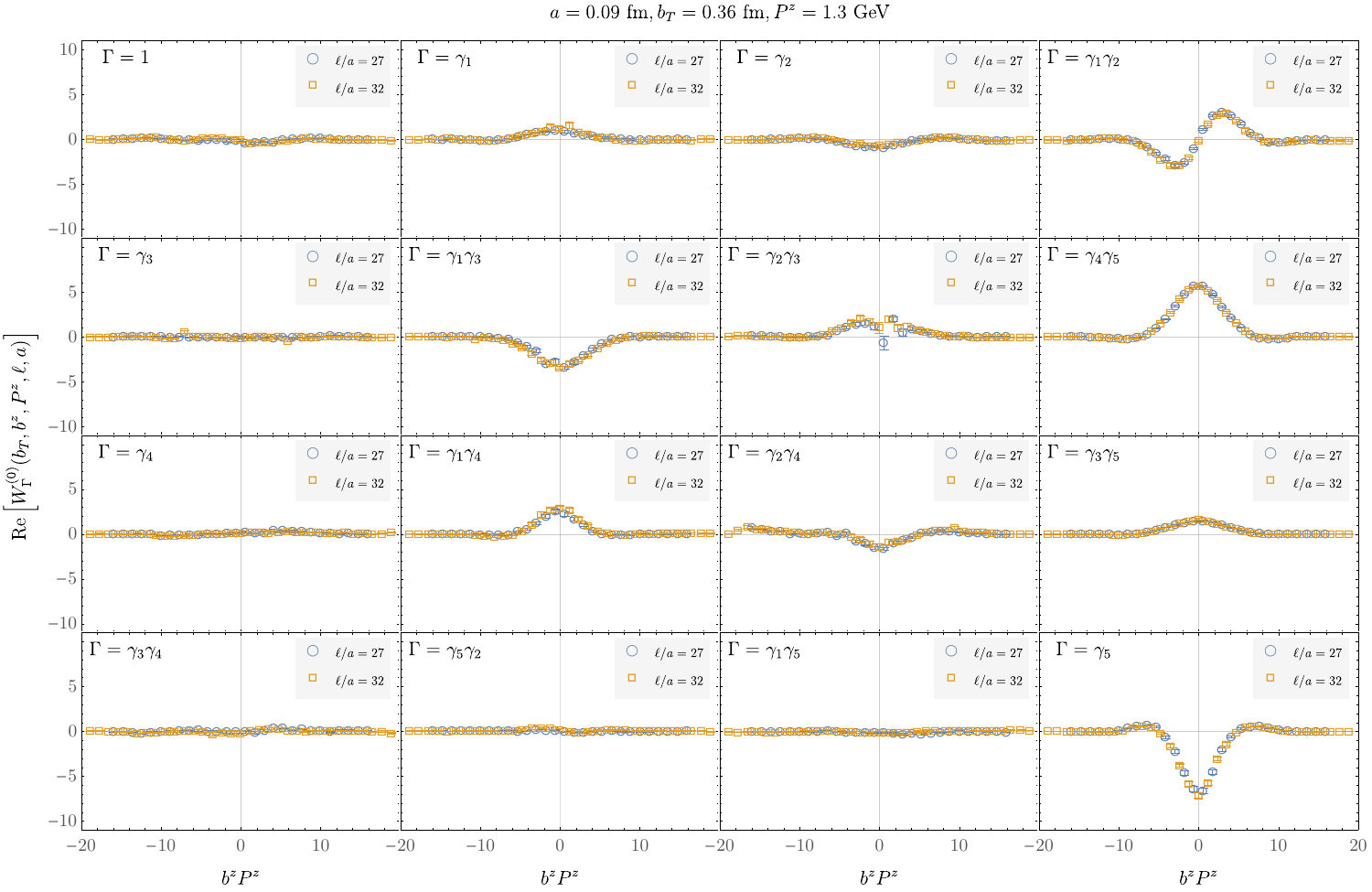} 
        \hspace{20pt}
        \includegraphics[width=0.98\textwidth]{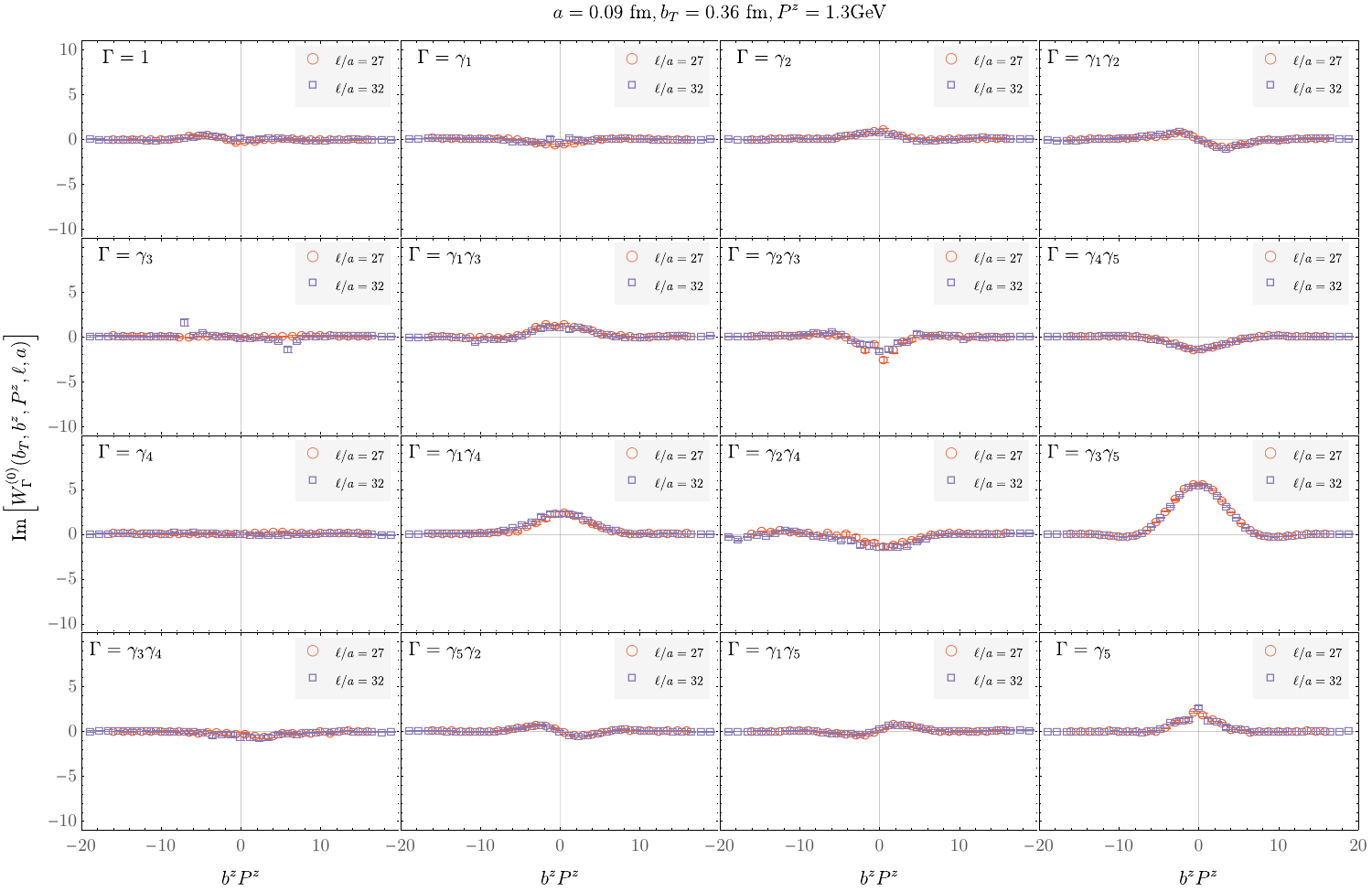} 
        \caption{Examples of real and imaginary parts of the bare quasi-TMD WF ratios $W^{(0)}_{\Gamma}(b_\tran, b^z, P^{z}, \ell,a)$ computed on the $a = 0.09~\text{fm}$ ensemble for all $\Gamma$, $\ell$, and $b^z$ with $b_T/a = 4$ and $n^z = 6$.
        \label{fig:L64_WF_bare_allgamma_comp}
        }
\end{figure*}

Examples of bare quasi-TMD WF ratios $W^{(0)}_{\Gamma}(b_\tran, b^z, P^{z}, \ell,a)$ (Eq.~\eqref{eq:quasi-wf-ratio}) for all $\Gamma$  and all relevant $\ell$ and $b^z$ are shown for representative values of $b_T$ and $P^z$ for all ensembles in Figs.~\ref{fig:L32_WF_bare_allgamma_comp}--\ref{fig:L64_WF_bare_allgamma_comp}. 

\begin{figure*}[t]
    \centering
        \includegraphics[width=0.46\textwidth]{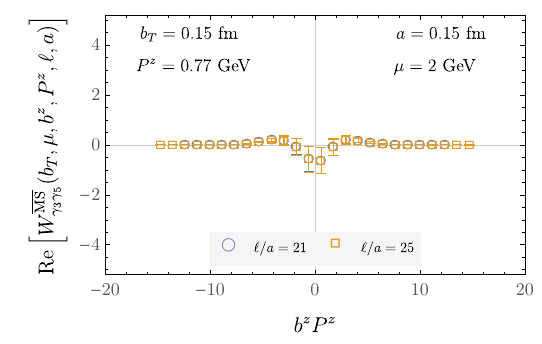}   
        \hspace{20pt}
        \includegraphics[width=0.46\textwidth]{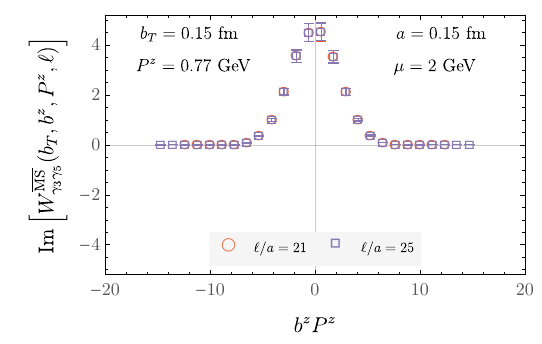}
        \includegraphics[width=0.46\textwidth]{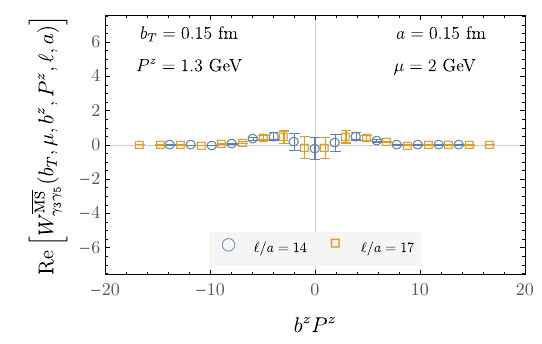} 
        \hspace{20pt}
        \includegraphics[width=0.46\textwidth]{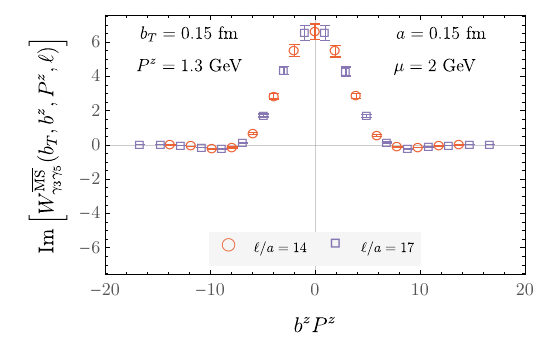} 
        \includegraphics[width=0.46\textwidth]{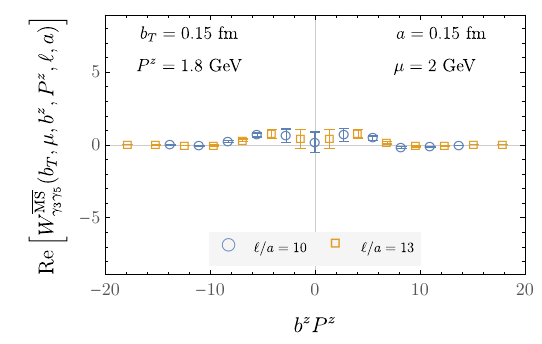} 
        \hspace{20pt}
         \includegraphics[width=0.46\textwidth]{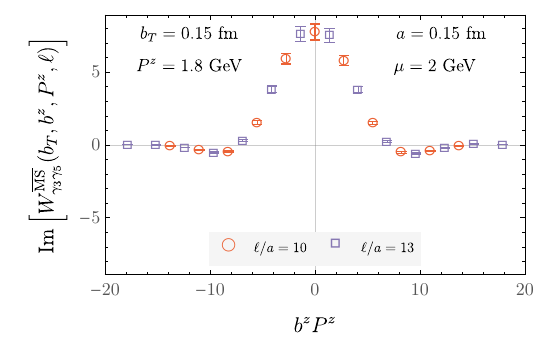} 
        \includegraphics[width=0.46\textwidth]{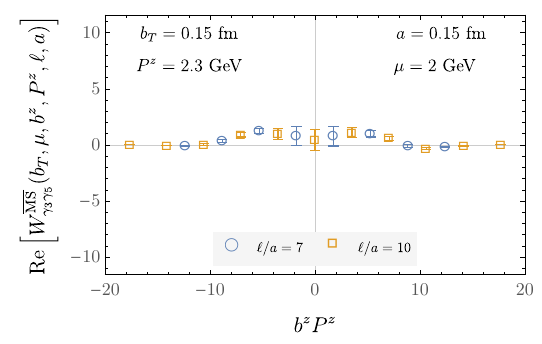} 
        \hspace{20pt}
        \includegraphics[width=0.46\textwidth]{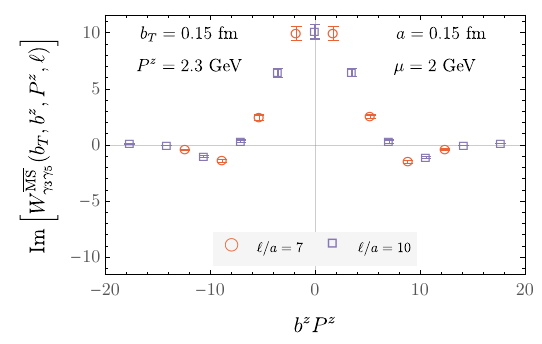} 
        \caption{Real and imaginary parts of the $\MSbar$-renormalized quasi-TMD WF ratios $W^{\MSbar}_{\Gamma}(b_T, \mu, b^z, P^z, \ell, a)$ computed on the $a = 0.15~\text{fm}$ ensemble for $\Gamma = \gamma_3 \gamma_5$ and $b_T/a = 1$.
        \label{fig:wf_ms_L32_gamma11_bT1}
        }
\end{figure*}

\begin{figure*}[t]
    \centering
        \includegraphics[width=0.46\textwidth]{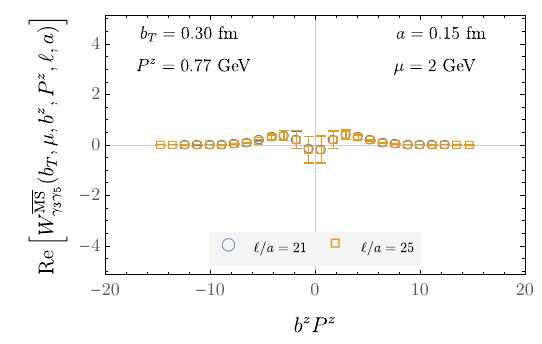}   
        \hspace{20pt}
        \includegraphics[width=0.46\textwidth]{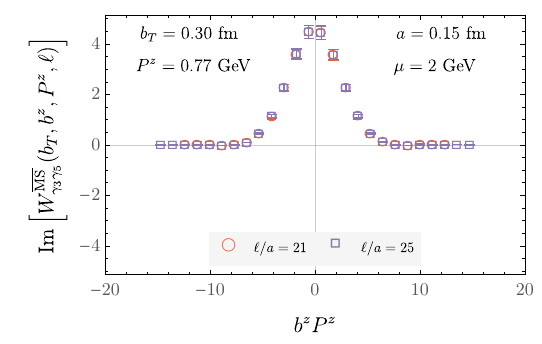}
        \includegraphics[width=0.46\textwidth]{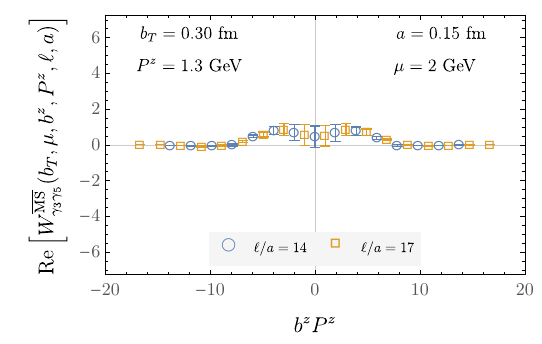} 
        \hspace{20pt}
        \includegraphics[width=0.46\textwidth]{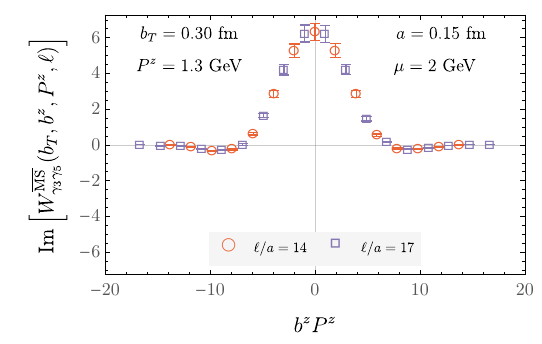} 
        \includegraphics[width=0.46\textwidth]{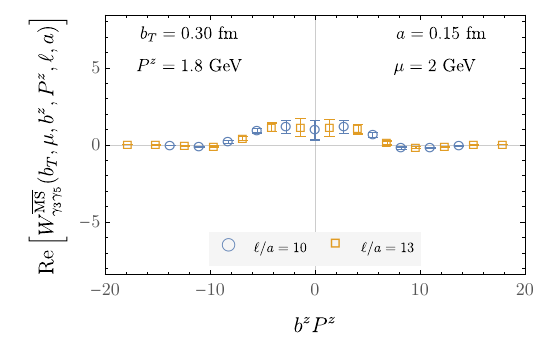} 
        \hspace{20pt}
         \includegraphics[width=0.46\textwidth]{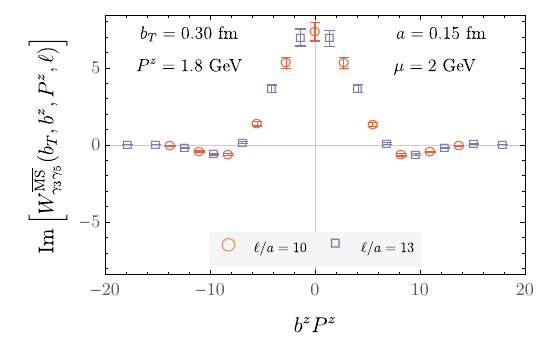} 
        \includegraphics[width=0.46\textwidth]{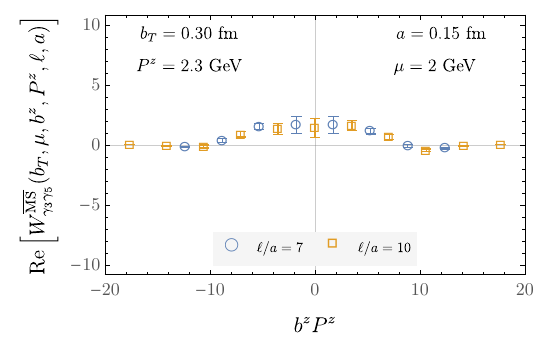} 
        \hspace{20pt}
        \includegraphics[width=0.46\textwidth]{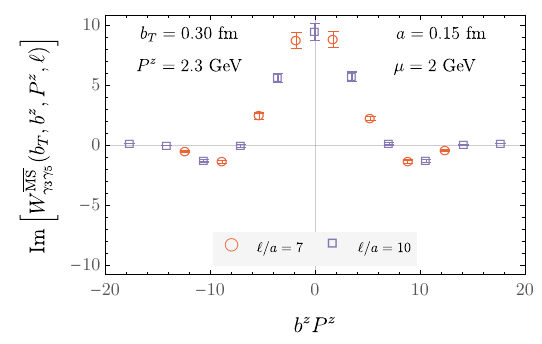} 
        \caption{Real and imaginary parts of the $\MSbar$-renormalized quasi-TMD WF ratios $W^{\MSbar}_{\Gamma}(b_T, \mu, b^z, P^z, \ell, a)$ computed on the $a = 0.15~\text{fm}$ ensemble for $\Gamma = \gamma_3 \gamma_5$ and $b_T/a = 2$.
        \label{fig:wf_ms_L32_gamma11_bT2}
        }
\end{figure*}

\begin{figure*}[t]
    \centering
        \includegraphics[width=0.46\textwidth]{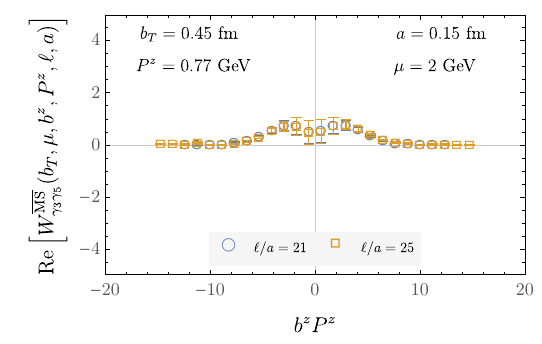}   
        \hspace{20pt}
        \includegraphics[width=0.46\textwidth]{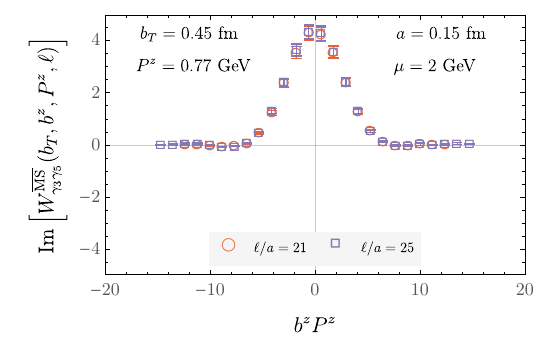}
        \includegraphics[width=0.46\textwidth]{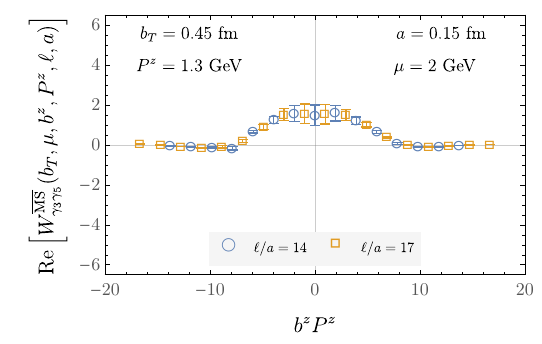} 
        \hspace{20pt}
        \includegraphics[width=0.46\textwidth]{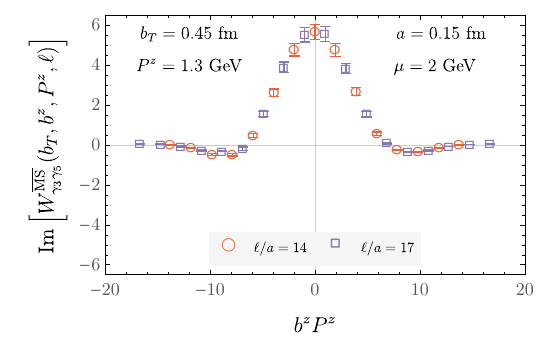} 
        \includegraphics[width=0.46\textwidth]{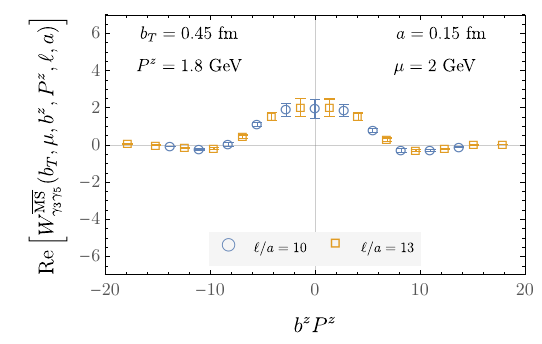} 
        \hspace{20pt}
         \includegraphics[width=0.46\textwidth]{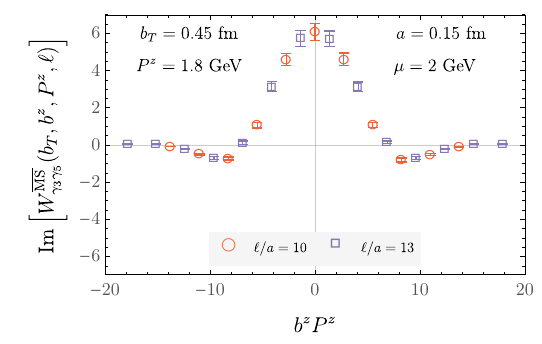} 
        \includegraphics[width=0.46\textwidth]{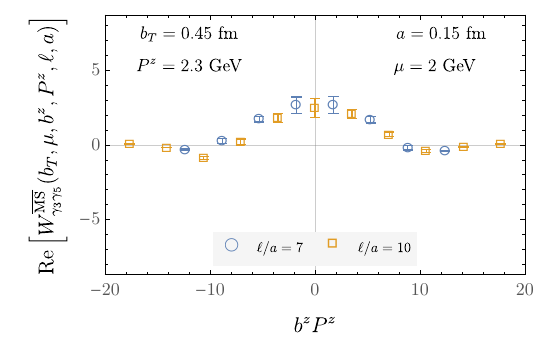} 
        \hspace{20pt}
        \includegraphics[width=0.46\textwidth]{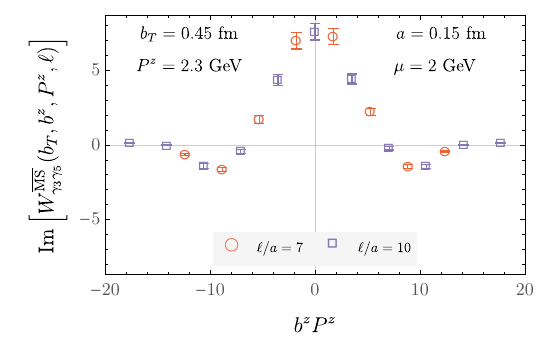} 
        \caption{Real and imaginary parts of the $\MSbar$-renormalized quasi-TMD WF ratios $W^{\MSbar}_{\Gamma}(b_T, \mu, b^z, P^z, \ell, a)$ computed on the $a = 0.15~\text{fm}$ ensemble for $\Gamma = \gamma_3 \gamma_5$ and $b_T/a = 3$.
        \label{fig:wf_ms_L32_gamma11_bT3}
        }
\end{figure*}

\begin{figure*}[t]
    \centering
        \includegraphics[width=0.46\textwidth]{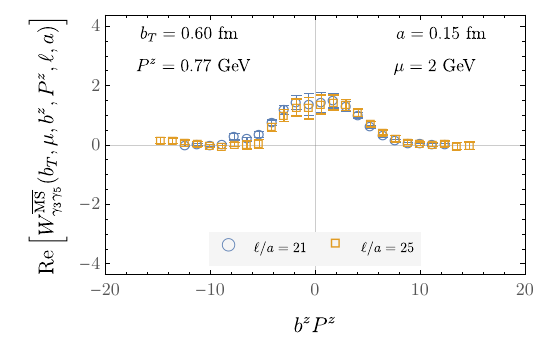}   
        \hspace{20pt}
        \includegraphics[width=0.46\textwidth]{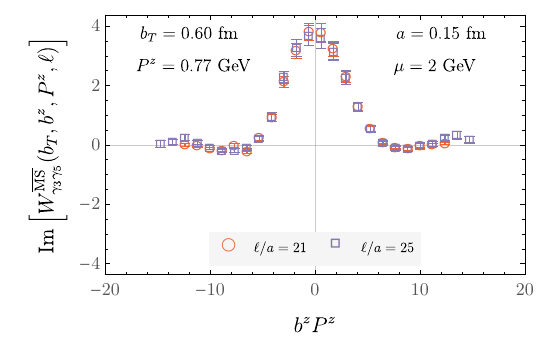}
        \includegraphics[width=0.46\textwidth]{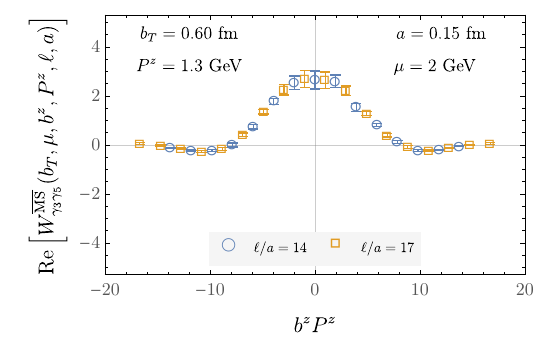} 
        \hspace{20pt}
        \includegraphics[width=0.46\textwidth]{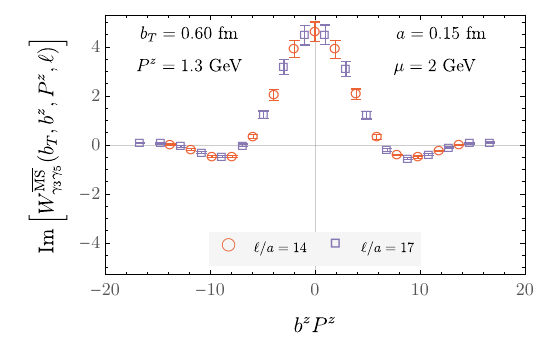} 
        \includegraphics[width=0.46\textwidth]{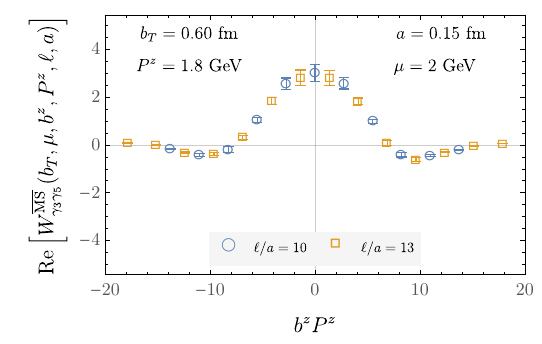} 
        \hspace{20pt}
         \includegraphics[width=0.46\textwidth]{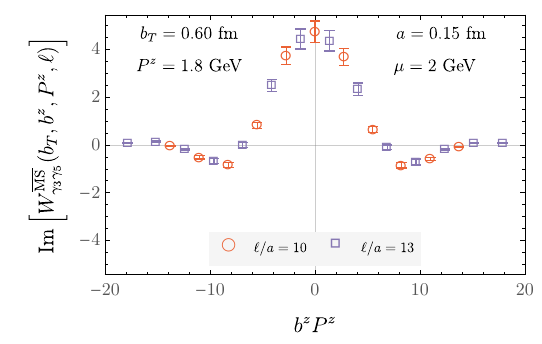} 
        \includegraphics[width=0.46\textwidth]{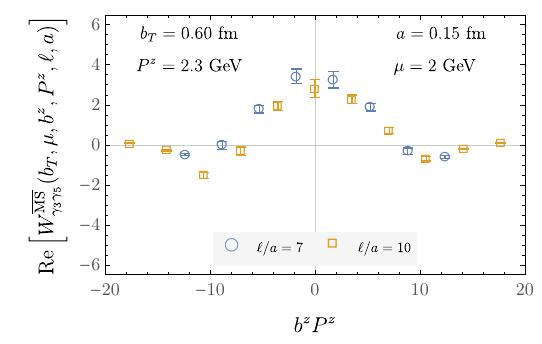} 
        \hspace{20pt}
        \includegraphics[width=0.46\textwidth]{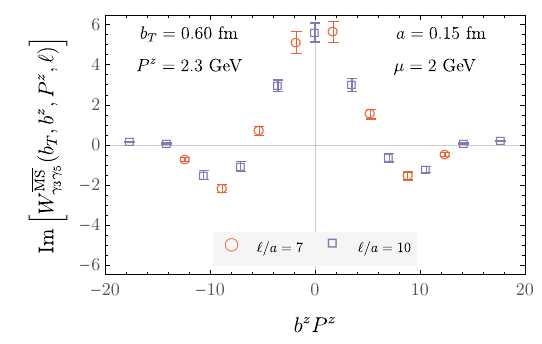} 
        \caption{Real and imaginary parts of the $\MSbar$-renormalized quasi-TMD WF ratios $W^{\MSbar}_{\Gamma}(b_T, \mu, b^z, P^z, \ell, a)$ computed on the $a = 0.15~\text{fm}$ ensemble for $\Gamma = \gamma_3 \gamma_5$ and $b_T/a = 4$.
        \label{fig:wf_ms_L32_gamma11_bT4}
        }
\end{figure*}

\begin{figure*}[t]
    \centering
        \includegraphics[width=0.46\textwidth]{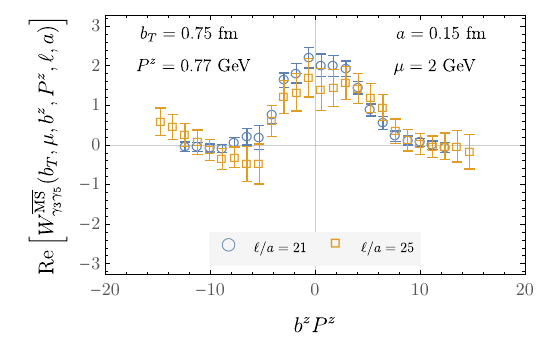}   
        \hspace{20pt}
        \includegraphics[width=0.46\textwidth]{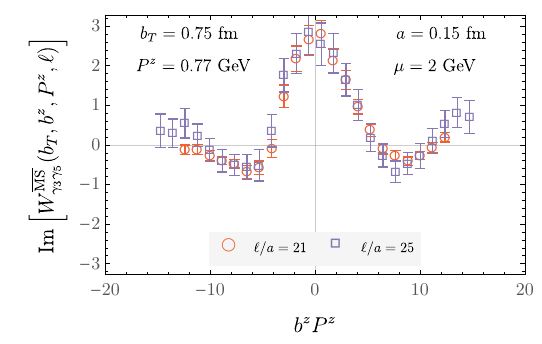}
        \includegraphics[width=0.46\textwidth]{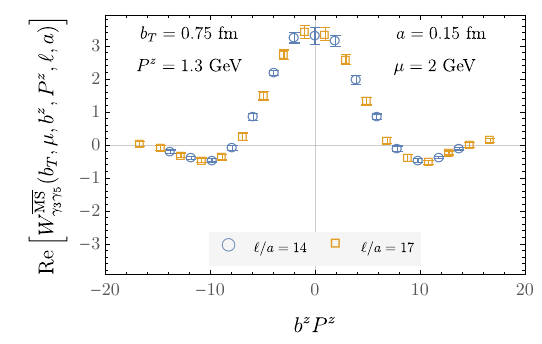} 
        \hspace{20pt}
        \includegraphics[width=0.46\textwidth]{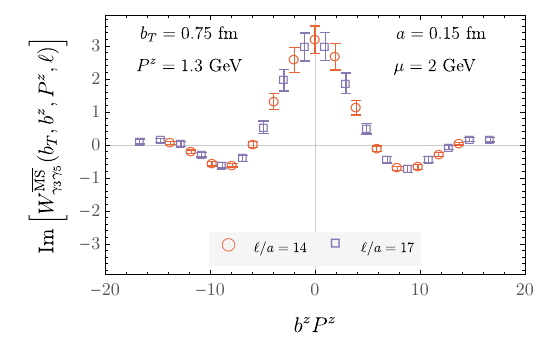} 
        \includegraphics[width=0.46\textwidth]{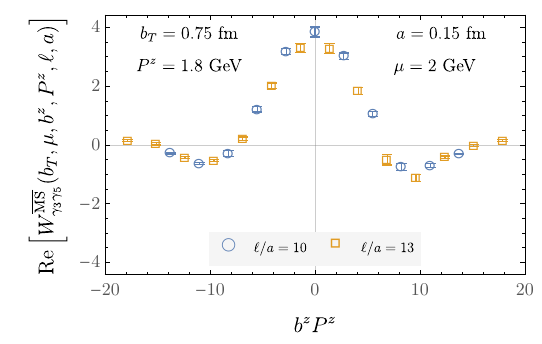} 
        \hspace{20pt}
         \includegraphics[width=0.46\textwidth]{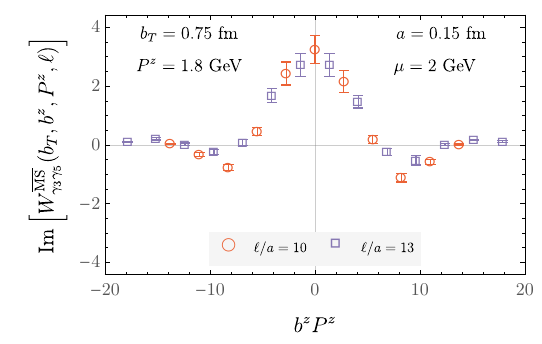} 
        \includegraphics[width=0.46\textwidth]{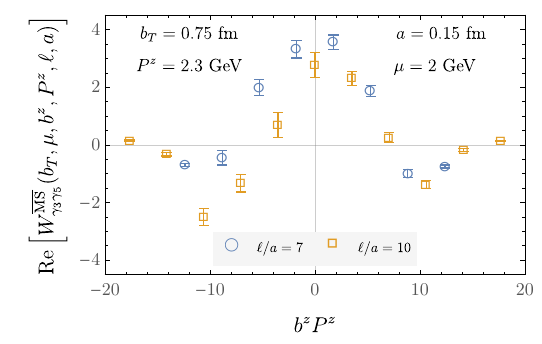} 
        \hspace{20pt}
        \includegraphics[width=0.46\textwidth]{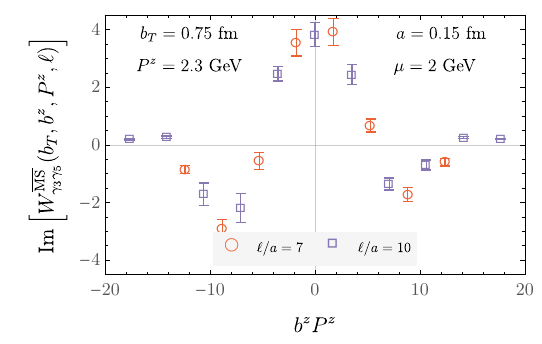} 
        \caption{Real and imaginary parts of the $\MSbar$-renormalized quasi-TMD WF ratios $W^{\MSbar}_{\Gamma}(b_T, \mu, b^z, P^z, \ell, a)$ computed on the $a = 0.15~\text{fm}$ ensemble for $\Gamma = \gamma_3 \gamma_5$ and $b_T/a = 5$.
        \label{fig:wf_ms_L32_gamma11_bT5}
        }
\end{figure*}

\begin{figure*}[t]
    \centering
        \includegraphics[width=0.46\textwidth]{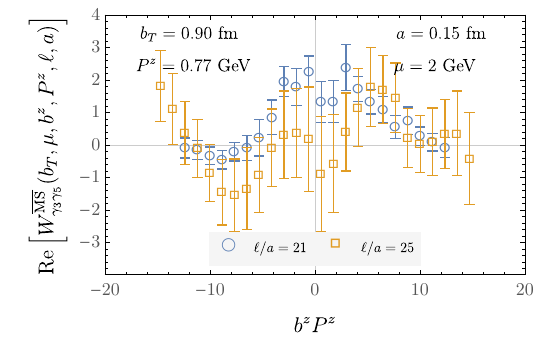}   
        \hspace{20pt}
        \includegraphics[width=0.46\textwidth]{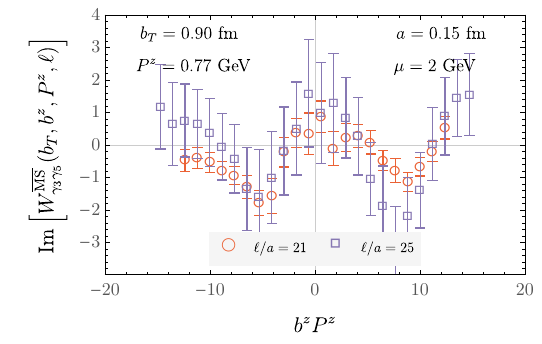}
        \includegraphics[width=0.46\textwidth]{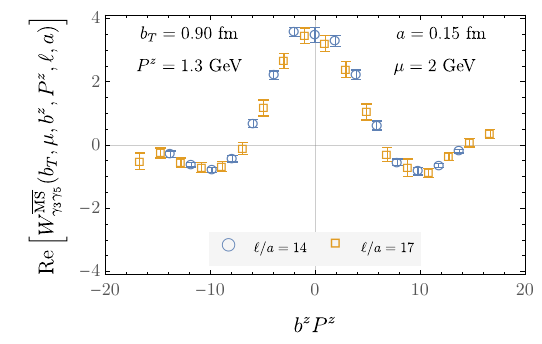} 
        \hspace{20pt}
        \includegraphics[width=0.46\textwidth]{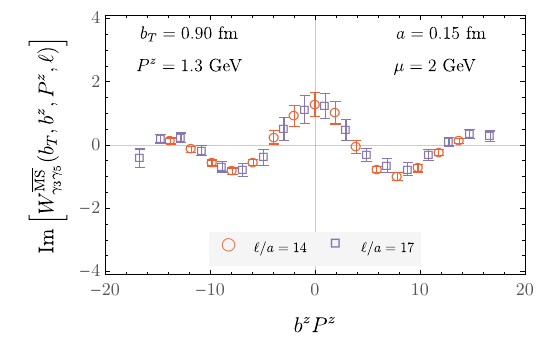} 
        \includegraphics[width=0.46\textwidth]{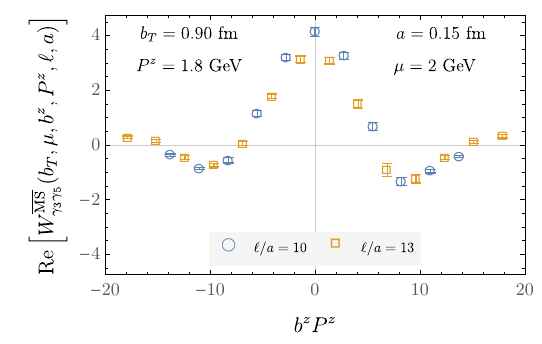} 
        \hspace{20pt}
         \includegraphics[width=0.46\textwidth]{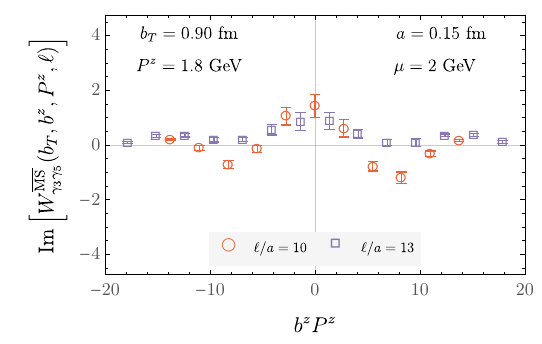} 
        \includegraphics[width=0.46\textwidth]{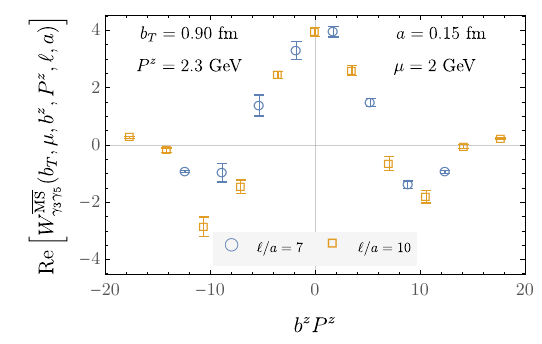} 
        \hspace{20pt}
        \includegraphics[width=0.46\textwidth]{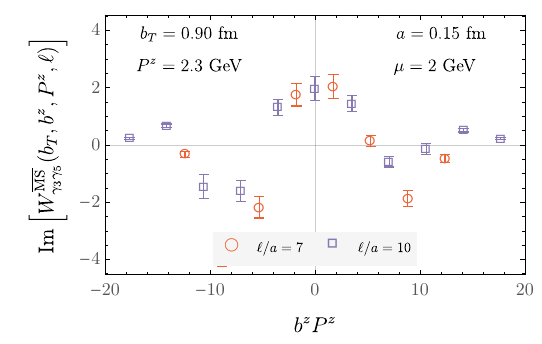} 
        \caption{Real and imaginary parts of the $\MSbar$-renormalized quasi-TMD WF ratios $W^{\MSbar}_{\Gamma}(b_T, \mu, b^z, P^z, \ell, a)$ computed on the $a = 0.15~\text{fm}$ ensemble for $\Gamma = \gamma_3 \gamma_5$ and $b_T/a = 6$.
        \label{fig:wf_ms_L32_gamma11_bT6}
        }
\end{figure*}

\begin{figure*}[t]
    \centering
        \includegraphics[width=0.46\textwidth]{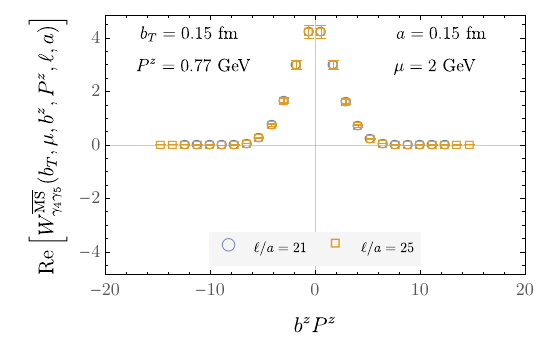}   
        \hspace{20pt}
        \includegraphics[width=0.46\textwidth]{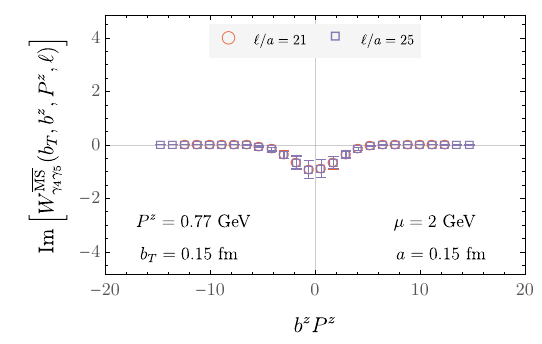}
        \includegraphics[width=0.46\textwidth]{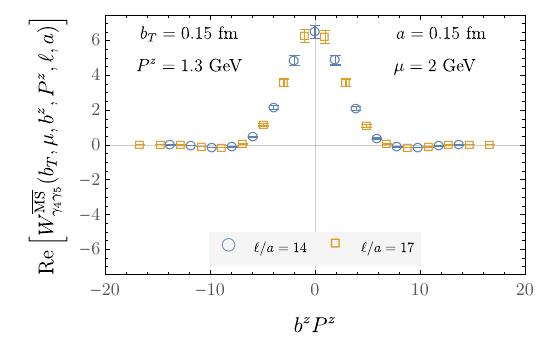} 
        \hspace{20pt}
        \includegraphics[width=0.46\textwidth]{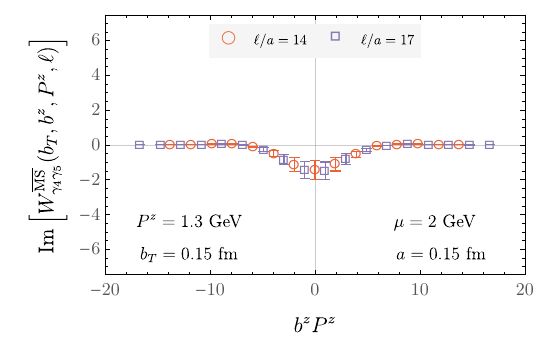} 
        \includegraphics[width=0.46\textwidth]{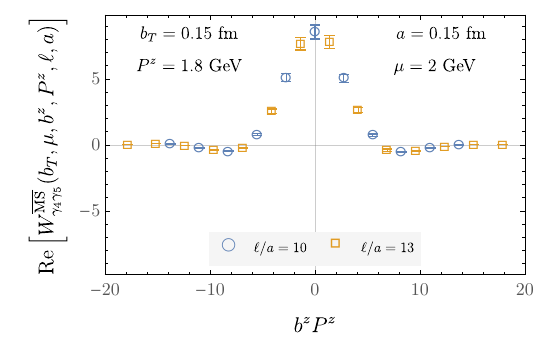} 
        \hspace{20pt}
         \includegraphics[width=0.46\textwidth]{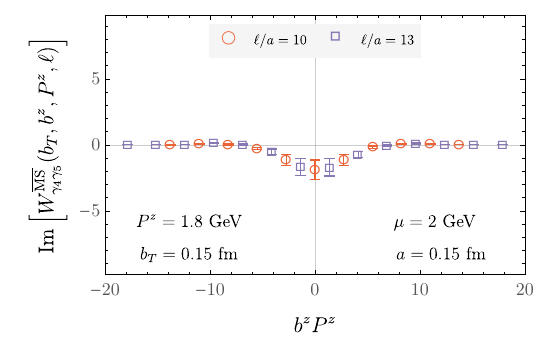} 
        \includegraphics[width=0.46\textwidth]{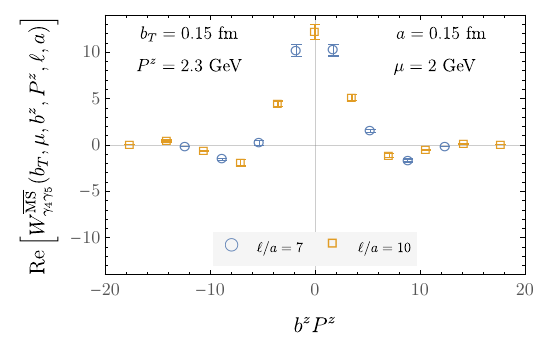} 
        \hspace{20pt}
        \includegraphics[width=0.46\textwidth]{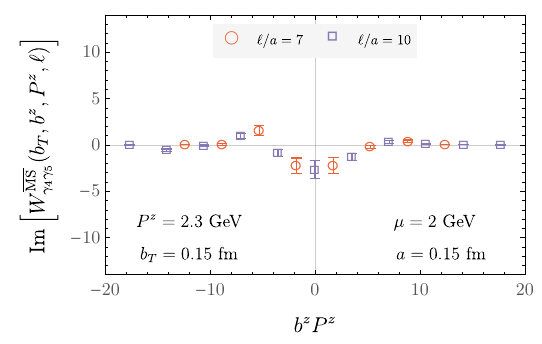} 
        \caption{Real and imaginary parts of the $\MSbar$-renormalized quasi-TMD WF ratios $W^{\MSbar}_{\Gamma}(b_T, \mu, b^z, P^z, \ell, a)$ computed on the $a = 0.15~\text{fm}$ ensemble for $\Gamma = \gamma_4 \gamma_5$ and $b_T/a = 1$.
        \label{fig:wf_ms_L32_gamma7_bT1}
        }
\end{figure*}

\begin{figure*}[t]
    \centering
        \includegraphics[width=0.46\textwidth]{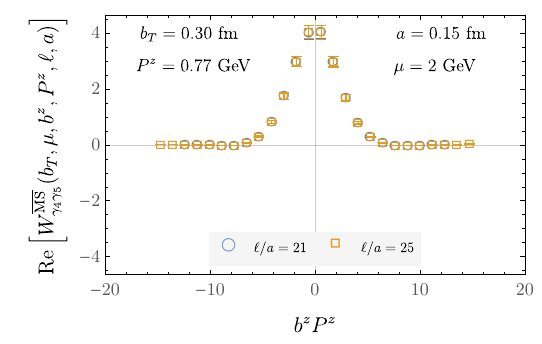}   
        \hspace{20pt}
        \includegraphics[width=0.46\textwidth]{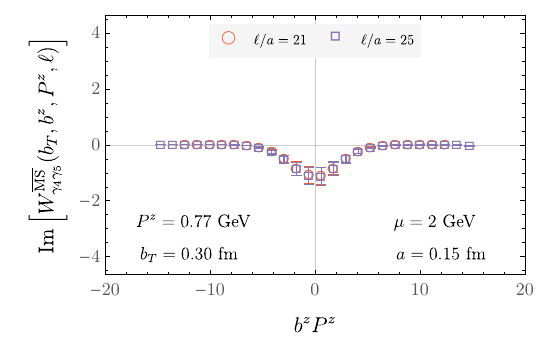}
        \includegraphics[width=0.46\textwidth]{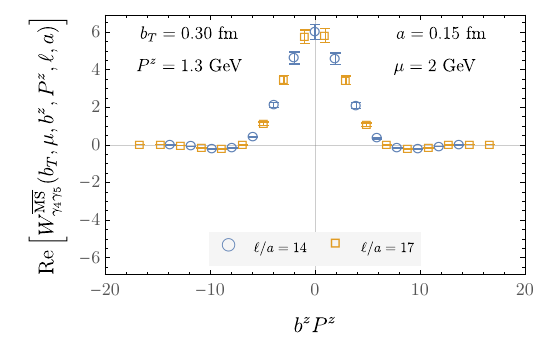} 
        \hspace{20pt}
        \includegraphics[width=0.46\textwidth]{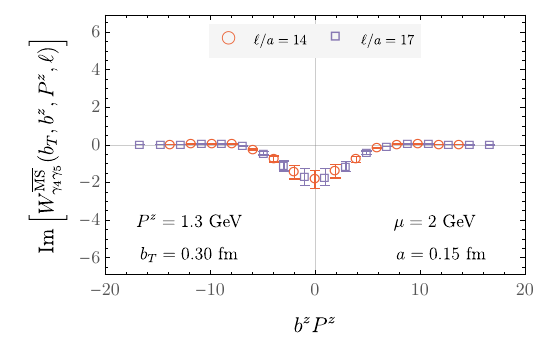} 
        \includegraphics[width=0.46\textwidth]{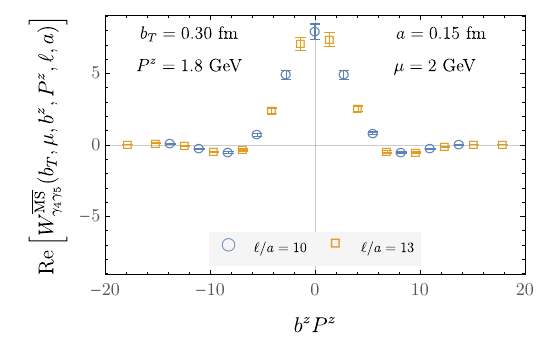} 
        \hspace{20pt}
         \includegraphics[width=0.46\textwidth]{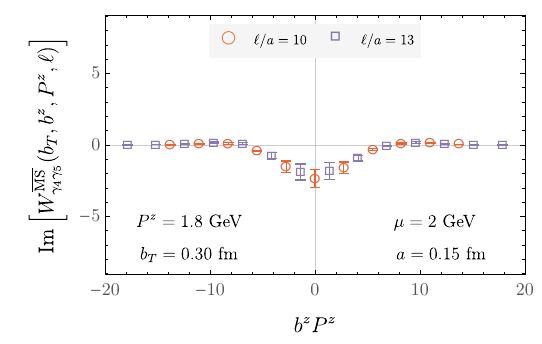} 
        \includegraphics[width=0.46\textwidth]{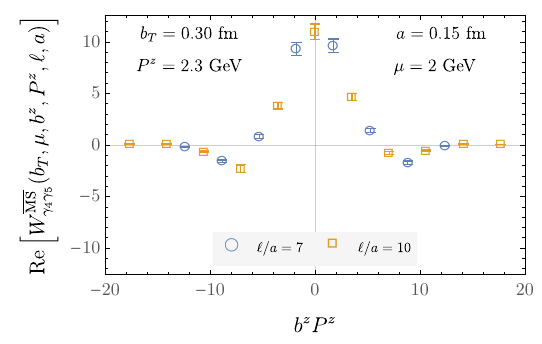} 
        \hspace{20pt}
        \includegraphics[width=0.46\textwidth]{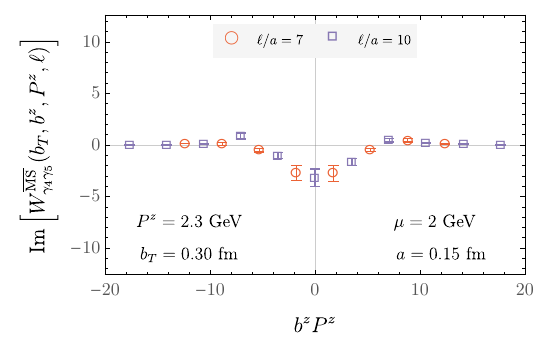} 
        \caption{Real and imaginary parts of the $\MSbar$-renormalized quasi-TMD WF ratios $W^{\MSbar}_{\Gamma}(b_T, \mu, b^z, P^z, \ell, a)$ computed on the $a = 0.15~\text{fm}$ ensemble for $\Gamma = \gamma_4 \gamma_5$ and $b_T/a = 2$.
        \label{fig:wf_ms_L32_gamma7_bT2}
        }
\end{figure*}

\begin{figure*}[t]
    \centering
        \includegraphics[width=0.46\textwidth]{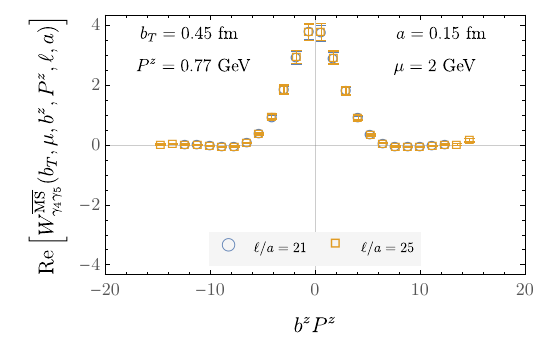}   
        \hspace{20pt}
        \includegraphics[width=0.46\textwidth]{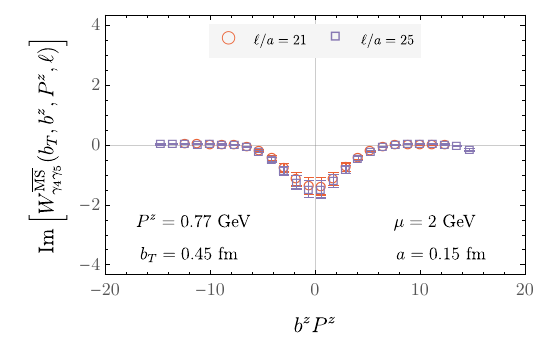}
        \includegraphics[width=0.46\textwidth]{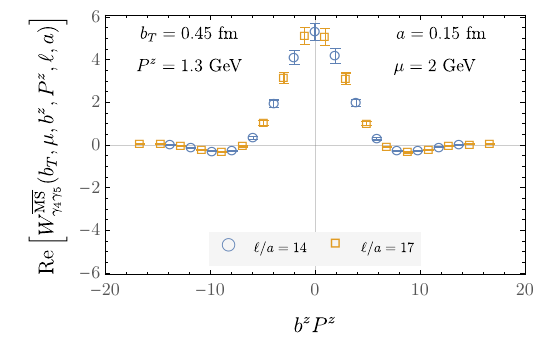} 
        \hspace{20pt}
        \includegraphics[width=0.46\textwidth]{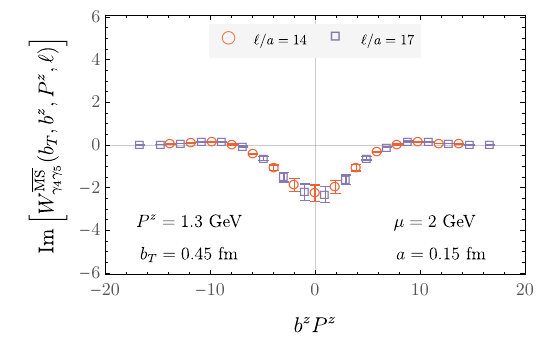} 
        \includegraphics[width=0.46\textwidth]{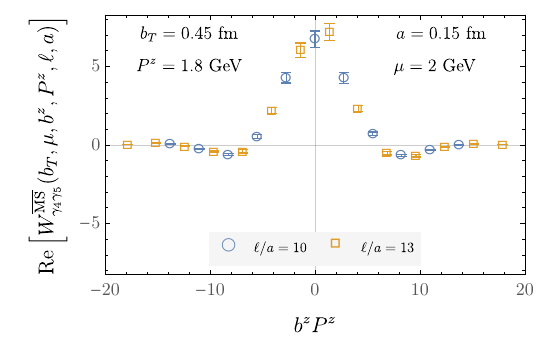} 
        \hspace{20pt}
         \includegraphics[width=0.46\textwidth]{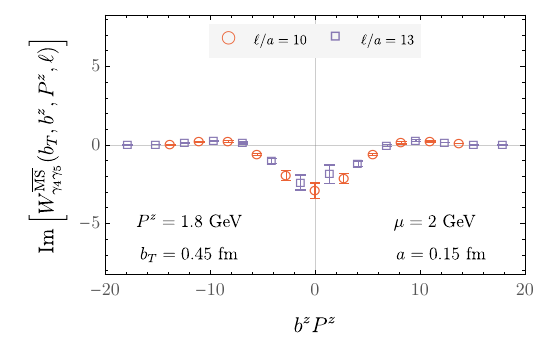} 
        \includegraphics[width=0.46\textwidth]{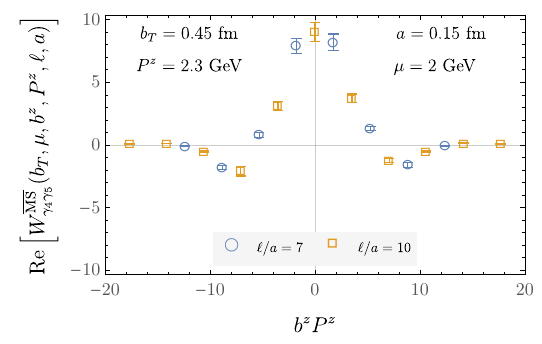} 
        \hspace{20pt}
        \includegraphics[width=0.46\textwidth]{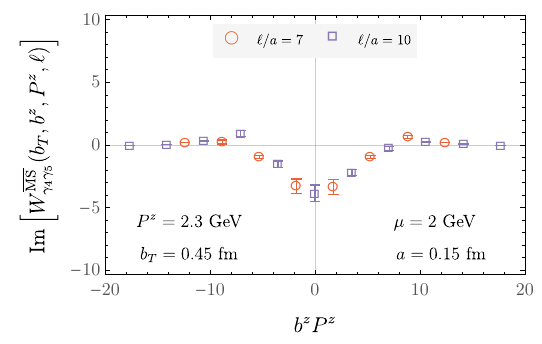} 
        \caption{Real and imaginary parts of the $\MSbar$-renormalized quasi-TMD WF ratios $W^{\MSbar}_{\Gamma}(b_T, \mu, b^z, P^z, \ell, a)$ computed on the $a = 0.15~\text{fm}$ ensemble for $\Gamma = \gamma_4 \gamma_5$ and $b_T/a = 3$.
        \label{fig:wf_ms_L32_gamma7_bT3}
        }
\end{figure*}

\begin{figure*}[t]
    \centering
        \includegraphics[width=0.46\textwidth]{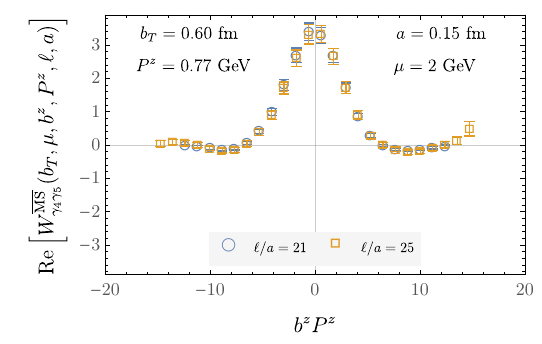}   
        \hspace{20pt}
        \includegraphics[width=0.46\textwidth]{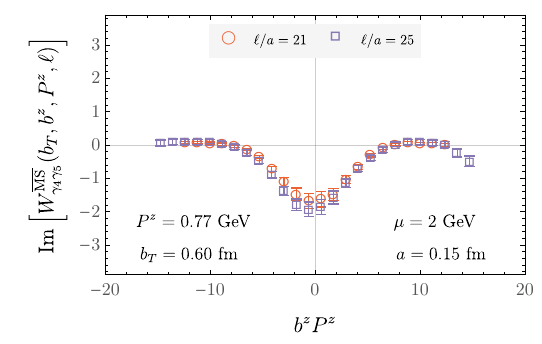}
        \includegraphics[width=0.46\textwidth]{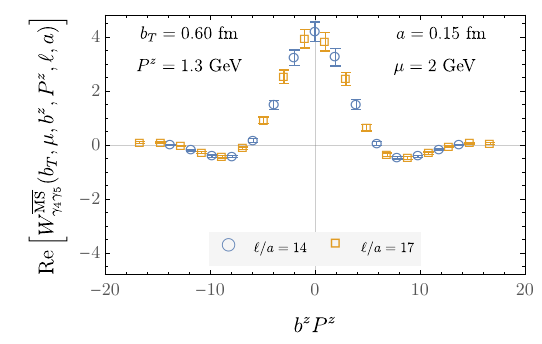} 
        \hspace{20pt}
        \includegraphics[width=0.46\textwidth]{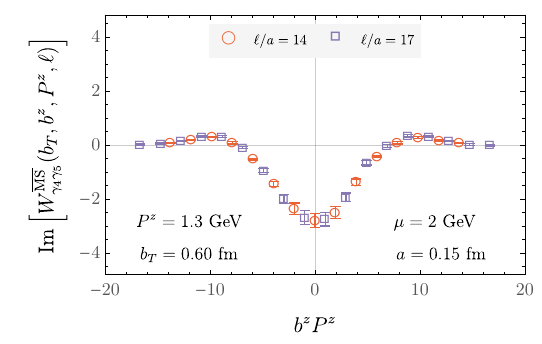} 
        \includegraphics[width=0.46\textwidth]{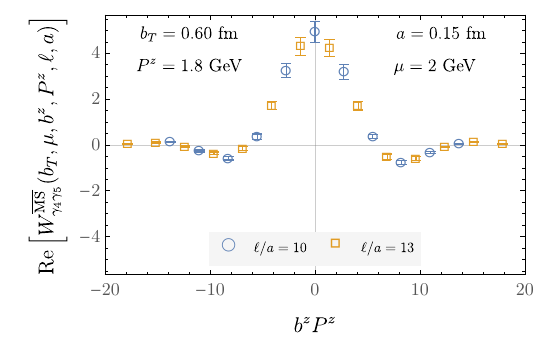} 
        \hspace{20pt}
         \includegraphics[width=0.46\textwidth]{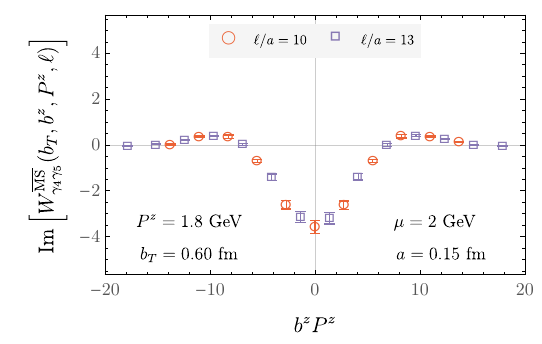} 
        \includegraphics[width=0.46\textwidth]{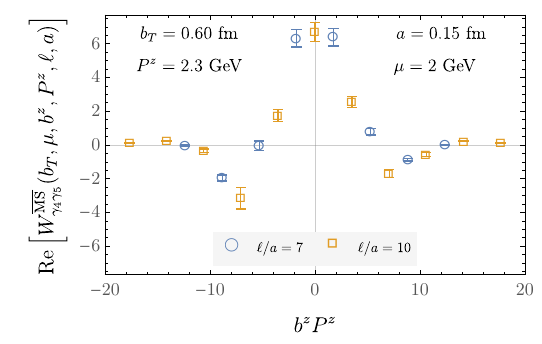} 
        \hspace{20pt}
        \includegraphics[width=0.46\textwidth]{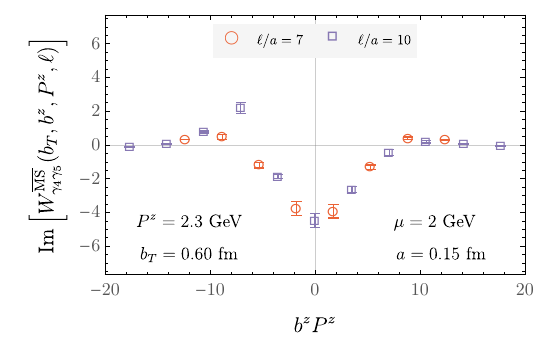} 
        \caption{Real and imaginary parts of the $\MSbar$-renormalized quasi-TMD WF ratios $W^{\MSbar}_{\Gamma}(b_T, \mu, b^z, P^z, \ell, a)$ computed on the $a = 0.15~\text{fm}$ ensemble for $\Gamma = \gamma_4 \gamma_5$ and $b_T/a = 4$.
        \label{fig:wf_ms_L32_gamma7_bT4}
        }
\end{figure*}

\begin{figure*}[t]
    \centering
        \includegraphics[width=0.46\textwidth]{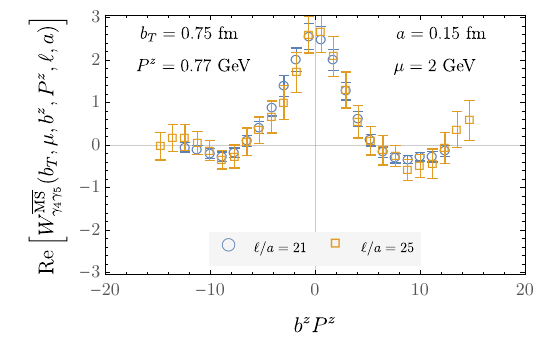}   
        \hspace{20pt}
        \includegraphics[width=0.46\textwidth]{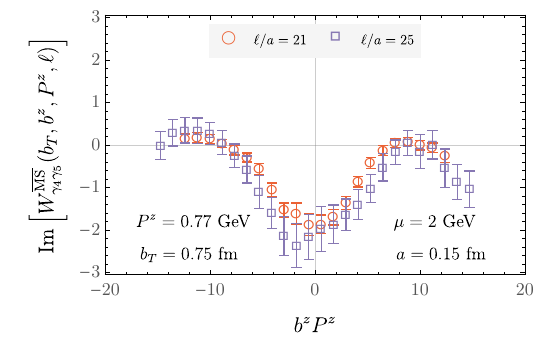}
        \includegraphics[width=0.46\textwidth]{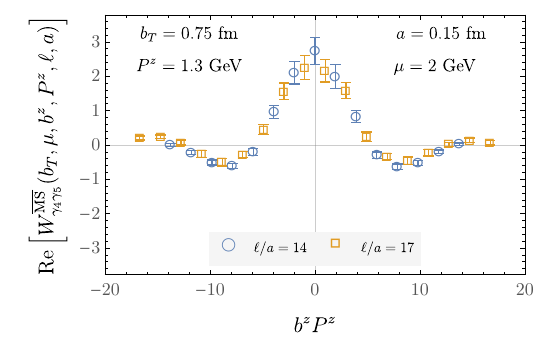} 
        \hspace{20pt}
        \includegraphics[width=0.46\textwidth]{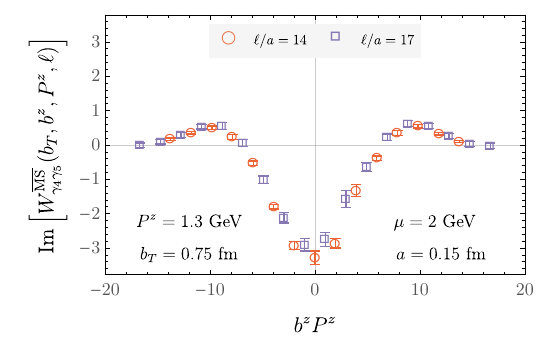} 
        \includegraphics[width=0.46\textwidth]{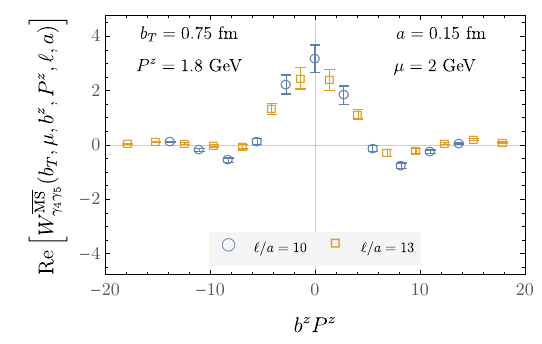} 
        \hspace{20pt}
         \includegraphics[width=0.46\textwidth]{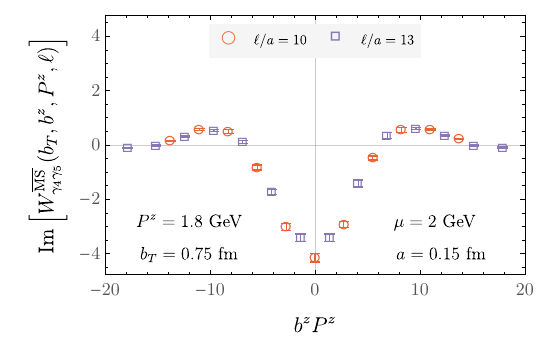} 
        \includegraphics[width=0.46\textwidth]{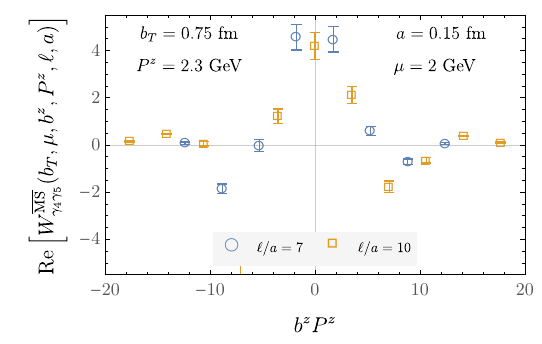} 
        \hspace{20pt}
        \includegraphics[width=0.46\textwidth]{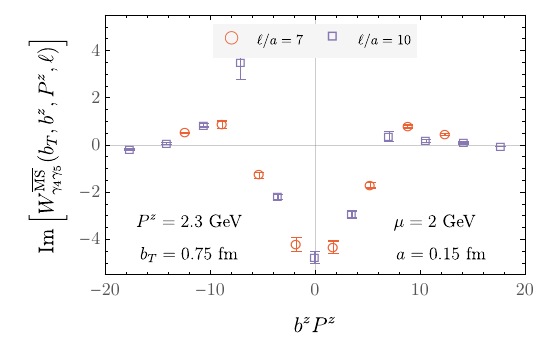} 
        \caption{Real and imaginary parts of the $\MSbar$-renormalized quasi-TMD WF ratios $W^{\MSbar}_{\Gamma}(b_T, \mu, b^z, P^z, \ell, a)$ computed on the $a = 0.15~\text{fm}$ ensemble for $\Gamma = \gamma_4 \gamma_5$ and $b_T/a = 5$.
        \label{fig:wf_ms_L32_gamma7_bT5}
        }
\end{figure*}

\begin{figure*}[t]
    \centering
        \includegraphics[width=0.46\textwidth]{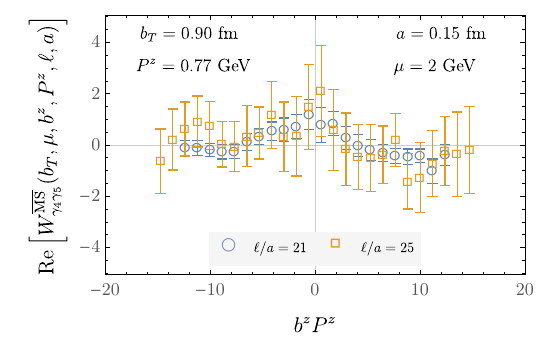}   
        \hspace{20pt}
        \includegraphics[width=0.46\textwidth]{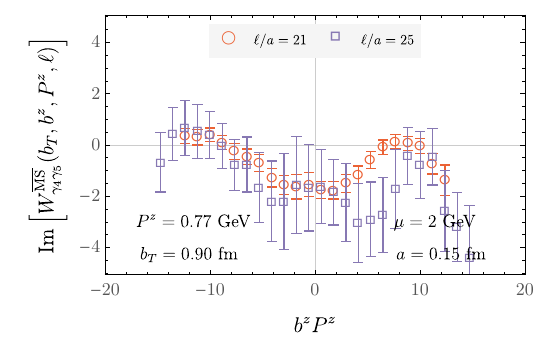}
        \includegraphics[width=0.46\textwidth]{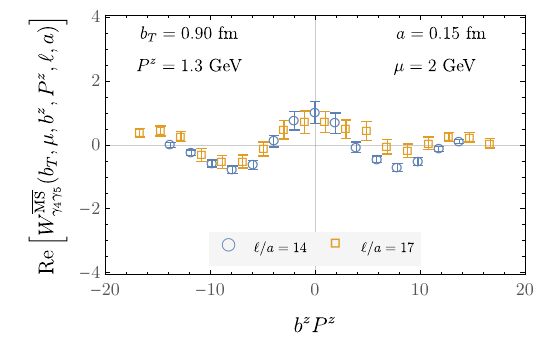} 
        \hspace{20pt}
        \includegraphics[width=0.46\textwidth]{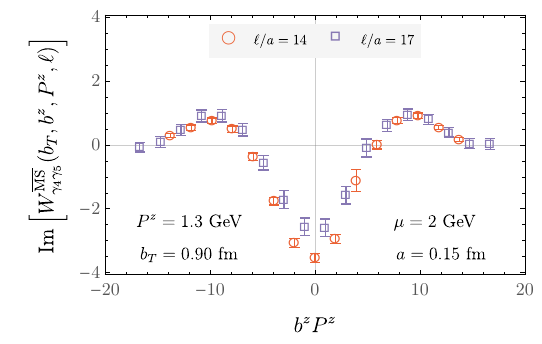} 
        \includegraphics[width=0.46\textwidth]{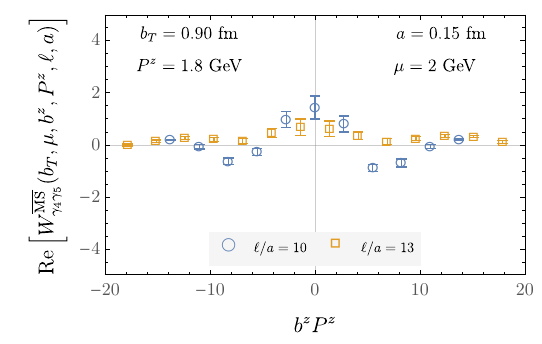} 
        \hspace{20pt}
         \includegraphics[width=0.46\textwidth]{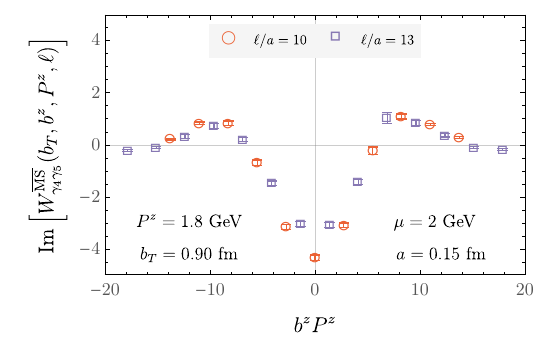} 
        \includegraphics[width=0.46\textwidth]{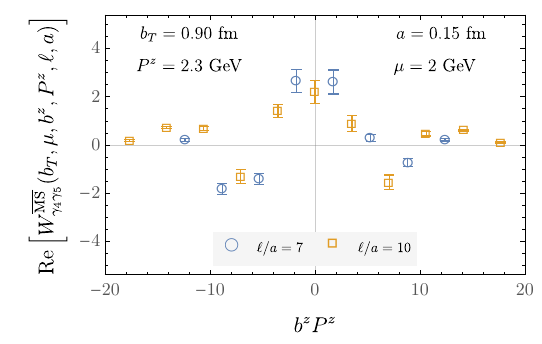} 
        \hspace{20pt}
        \includegraphics[width=0.46\textwidth]{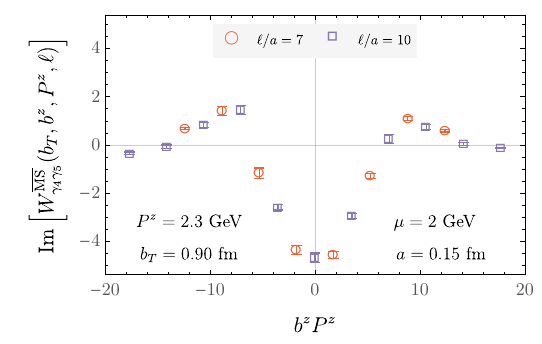} 
        \caption{Real and imaginary parts of the $\MSbar$-renormalized quasi-TMD WF ratios $W^{\MSbar}_{\Gamma}(b_T, \mu, b^z, P^z, \ell, a)$ computed on the $a = 0.15~\text{fm}$ ensemble for $\Gamma = \gamma_4 \gamma_5$ and $b_T/a = 6$.
        \label{fig:wf_ms_L32_gamma7_bT6}
        }
\end{figure*}

\begin{figure*}[t]
    \centering
        \includegraphics[width=0.46\textwidth]{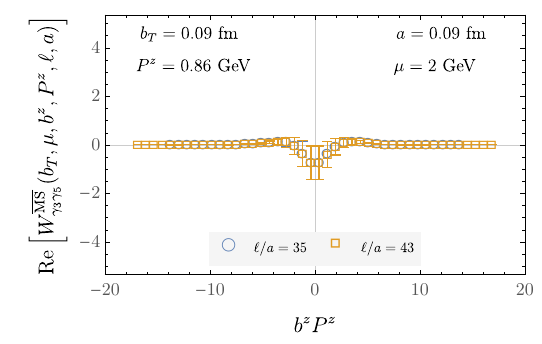}   
        \hspace{20pt}
        \includegraphics[width=0.46\textwidth]{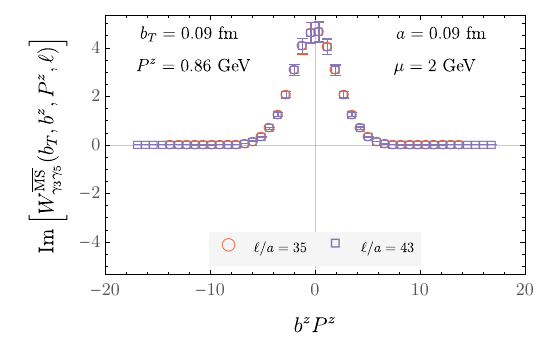}
        \includegraphics[width=0.46\textwidth]{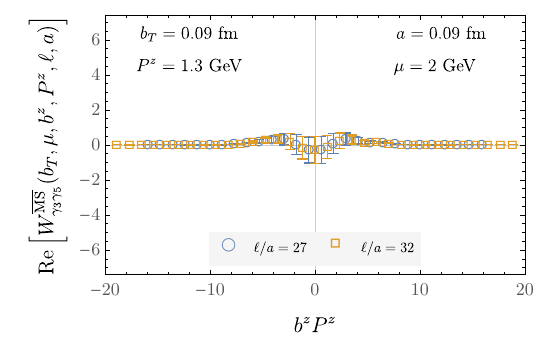} 
        \hspace{20pt}
        \includegraphics[width=0.46\textwidth]{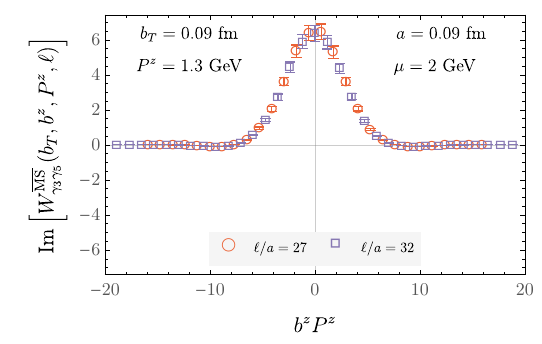} 
        \includegraphics[width=0.46\textwidth]{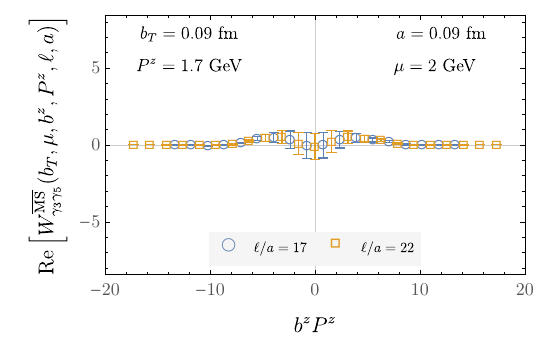} 
        \hspace{20pt}
         \includegraphics[width=0.46\textwidth]{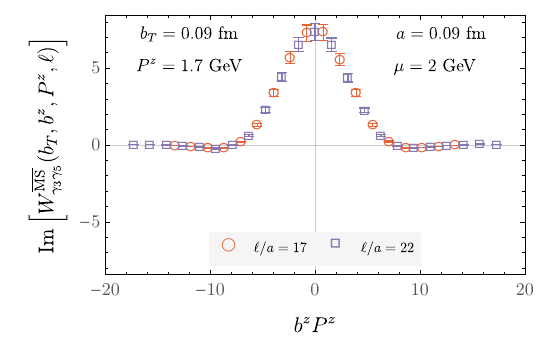} 
        \includegraphics[width=0.46\textwidth]{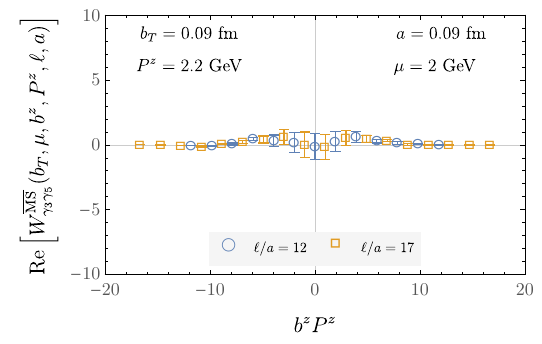} 
        \hspace{20pt}
        \includegraphics[width=0.46\textwidth]{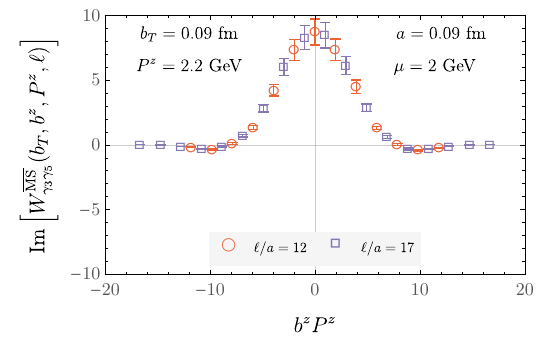} 
        \caption{Real and imaginary parts of the $\MSbar$-renormalized quasi-TMD WF ratios $W^{\MSbar}_{\Gamma}(b_T, \mu, b^z, P^z, \ell, a)$ computed on the $a = 0.09~\text{fm}$ ensemble for $\Gamma = \gamma_3 \gamma_5$ and $b_T/a = 1$.
        \label{fig:wf_ms_L64_gamma11_bT1}
        }
\end{figure*}

\begin{figure*}[t]
    \centering
        \includegraphics[width=0.46\textwidth]{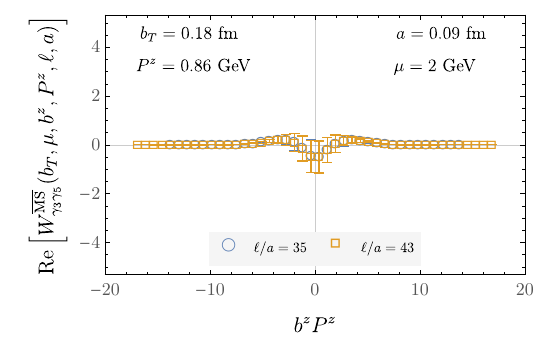}   
        \hspace{20pt}
        \includegraphics[width=0.46\textwidth]{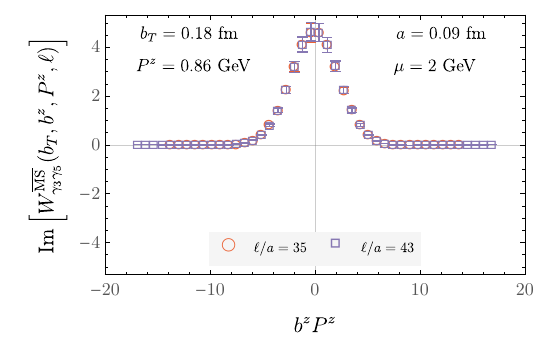}
        \includegraphics[width=0.46\textwidth]{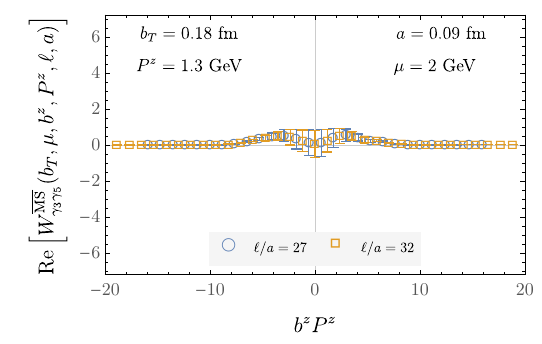} 
        \hspace{20pt}
        \includegraphics[width=0.46\textwidth]{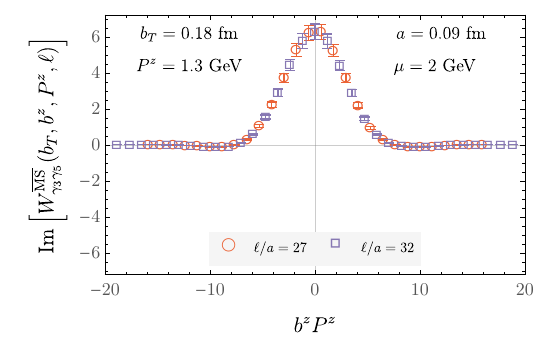} 
        \includegraphics[width=0.46\textwidth]{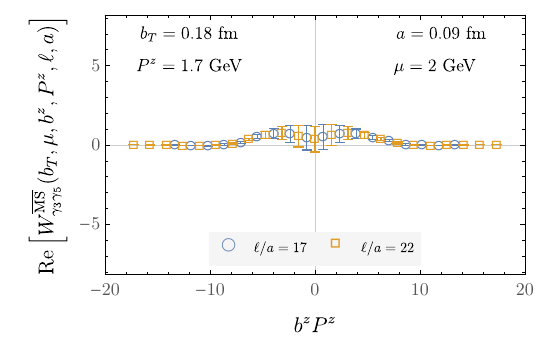} 
        \hspace{20pt}
         \includegraphics[width=0.46\textwidth]{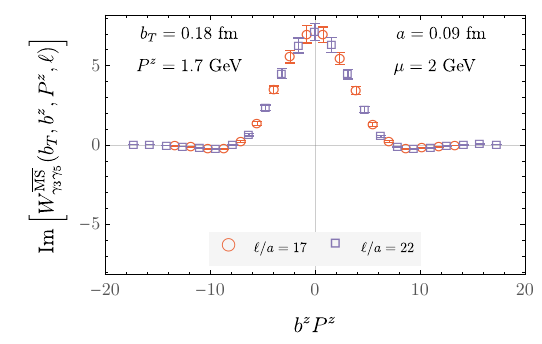} 
        \includegraphics[width=0.46\textwidth]{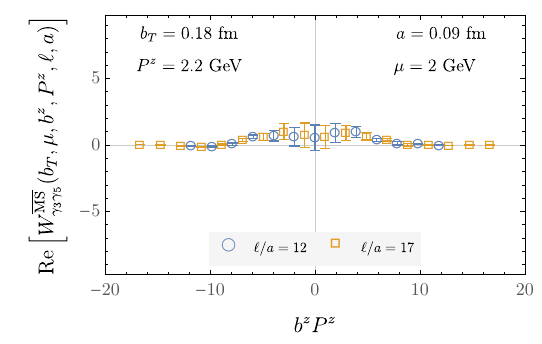} 
        \hspace{20pt}
        \includegraphics[width=0.46\textwidth]{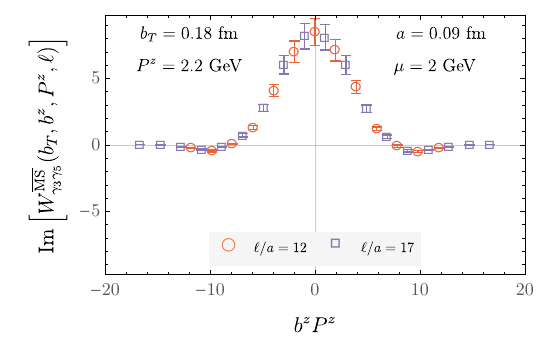} 
        \caption{Real and imaginary parts of the $\MSbar$-renormalized quasi-TMD WF ratios $W^{\MSbar}_{\Gamma}(b_T, \mu, b^z, P^z, \ell, a)$ computed on the $a = 0.09~\text{fm}$ ensemble for $\Gamma = \gamma_3 \gamma_5$ and $b_T/a = 2$.
        \label{fig:wf_ms_L64_gamma11_bT2}
        }
\end{figure*}

\begin{figure*}[t]
    \centering
        \includegraphics[width=0.46\textwidth]{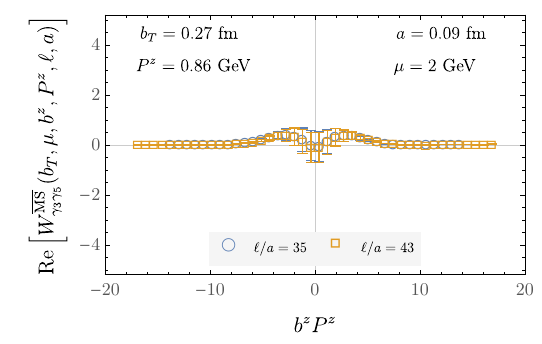}   
        \hspace{20pt}
        \includegraphics[width=0.46\textwidth]{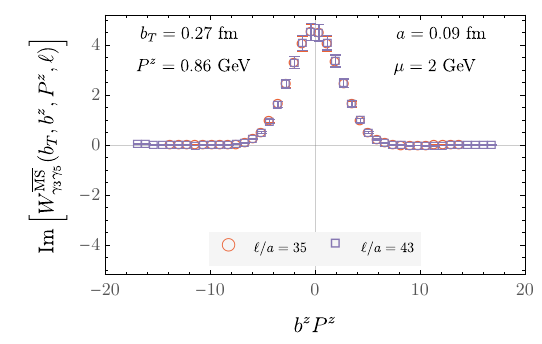}
        \includegraphics[width=0.46\textwidth]{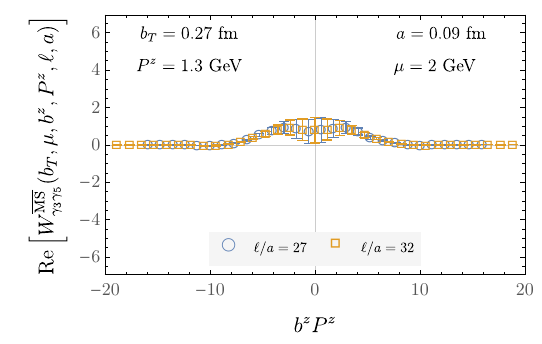} 
        \hspace{20pt}
        \includegraphics[width=0.46\textwidth]{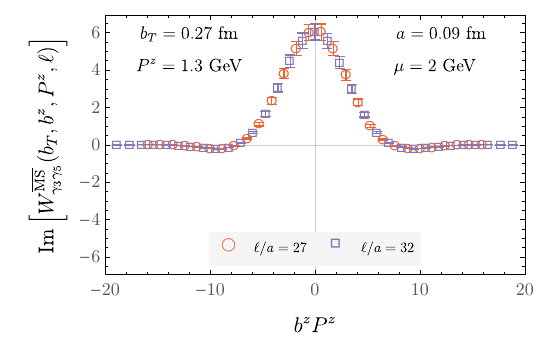} 
        \includegraphics[width=0.46\textwidth]{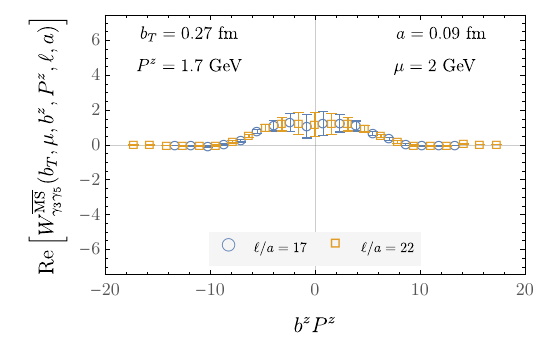} 
        \hspace{20pt}
         \includegraphics[width=0.46\textwidth]{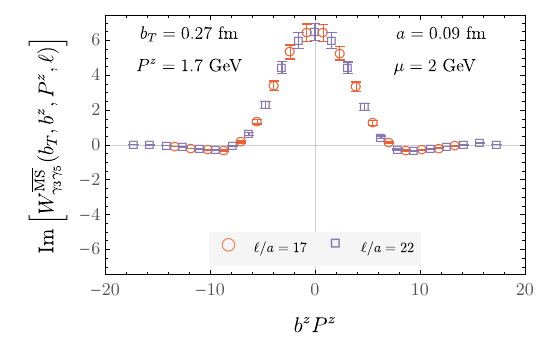} 
        \includegraphics[width=0.46\textwidth]{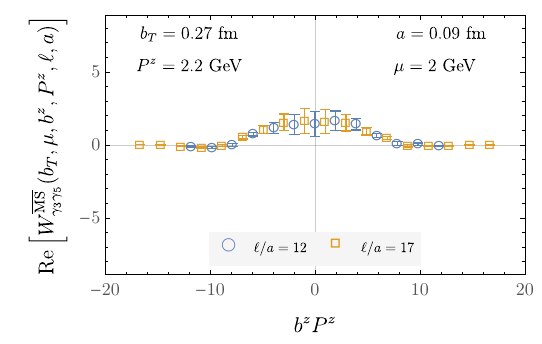} 
        \hspace{20pt}
        \includegraphics[width=0.46\textwidth]{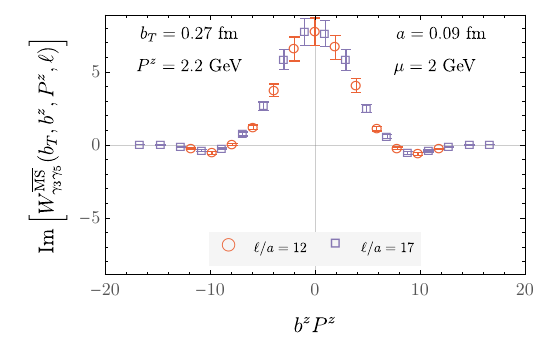} 
        \caption{Real and imaginary parts of the $\MSbar$-renormalized quasi-TMD WF ratios $W^{\MSbar}_{\Gamma}(b_T, \mu, b^z, P^z, \ell, a)$ computed on the $a = 0.09~\text{fm}$ ensemble for $\Gamma = \gamma_3 \gamma_5$ and $b_T/a = 3$.
        \label{fig:wf_ms_L64_gamma11_bT3}
        }
\end{figure*}

\begin{figure*}[t]
    \centering
        \includegraphics[width=0.46\textwidth]{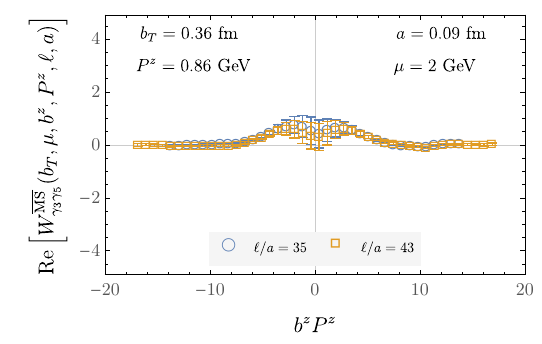}   
        \hspace{20pt}
        \includegraphics[width=0.46\textwidth]{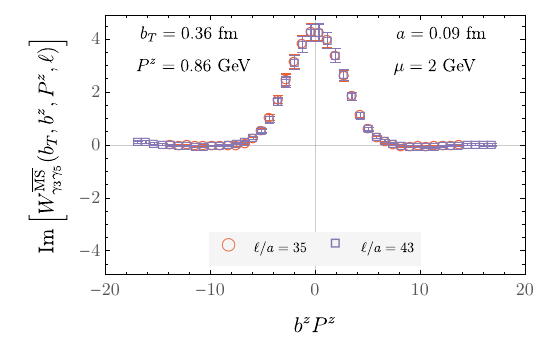}
        \includegraphics[width=0.46\textwidth]{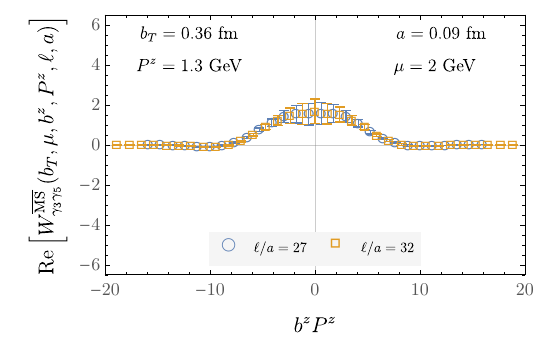} 
        \hspace{20pt}
        \includegraphics[width=0.46\textwidth]{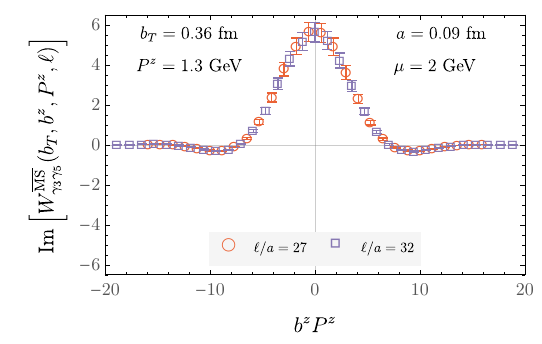} 
        \includegraphics[width=0.46\textwidth]{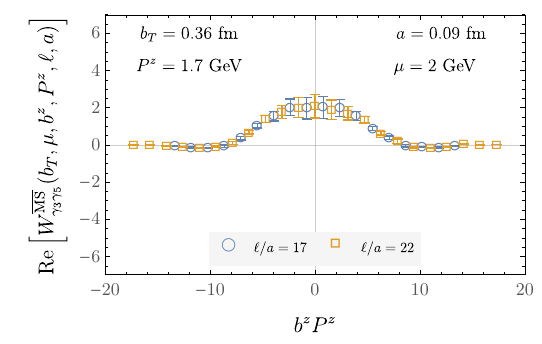} 
        \hspace{20pt}
         \includegraphics[width=0.46\textwidth]{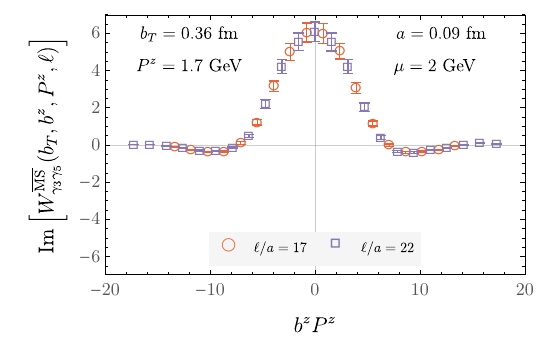} 
        \includegraphics[width=0.46\textwidth]{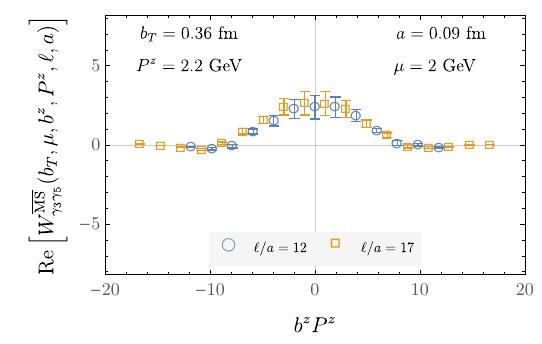} 
        \hspace{20pt}
        \includegraphics[width=0.46\textwidth]{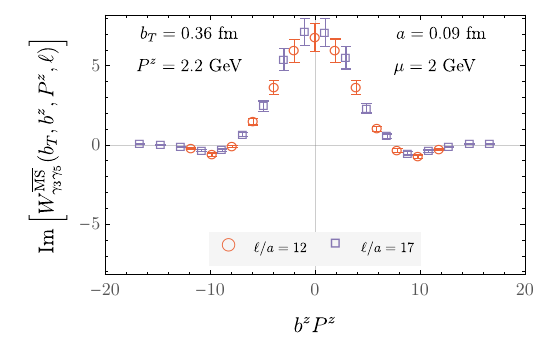} 
        \caption{Real and imaginary parts of the $\MSbar$-renormalized quasi-TMD WF ratios $W^{\MSbar}_{\Gamma}(b_T, \mu, b^z, P^z, \ell, a)$ computed on the $a = 0.09~\text{fm}$ ensemble for $\Gamma = \gamma_3 \gamma_5$ and $b_T/a = 4$.
        \label{fig:wf_ms_L64_gamma11_bT4}
        }
\end{figure*}

\begin{figure*}[t]
    \centering
        \includegraphics[width=0.46\textwidth]{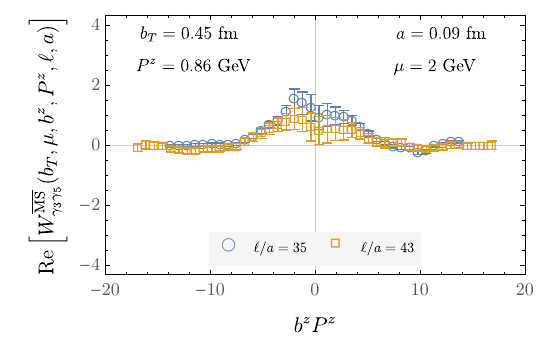}   
        \hspace{20pt}
        \includegraphics[width=0.46\textwidth]{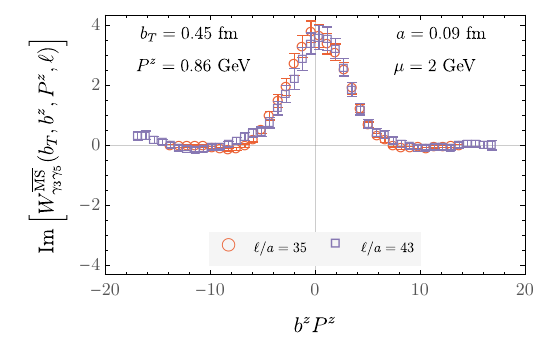}
        \includegraphics[width=0.46\textwidth]{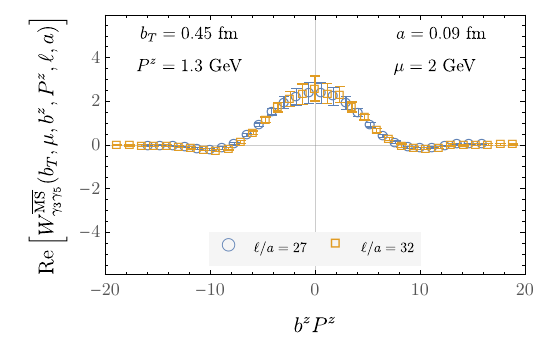} 
        \hspace{20pt}
        \includegraphics[width=0.46\textwidth]{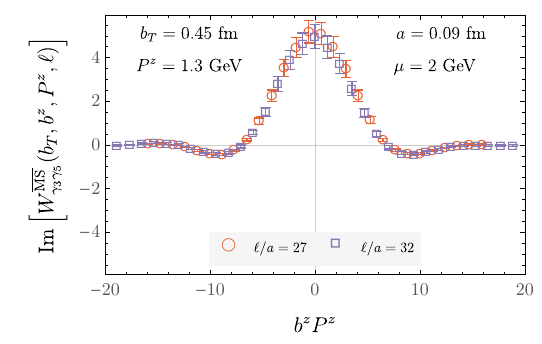} 
        \includegraphics[width=0.46\textwidth]{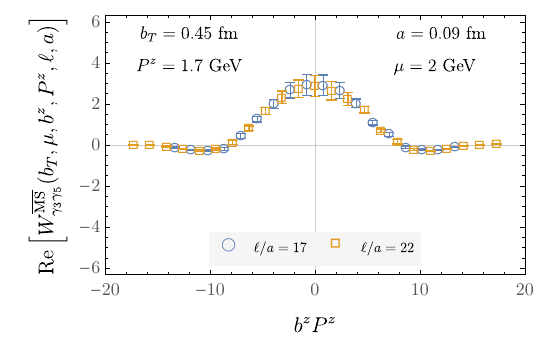} 
        \hspace{20pt}
         \includegraphics[width=0.46\textwidth]{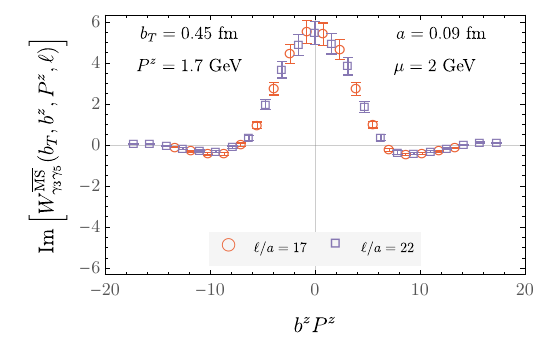} 
        \includegraphics[width=0.46\textwidth]{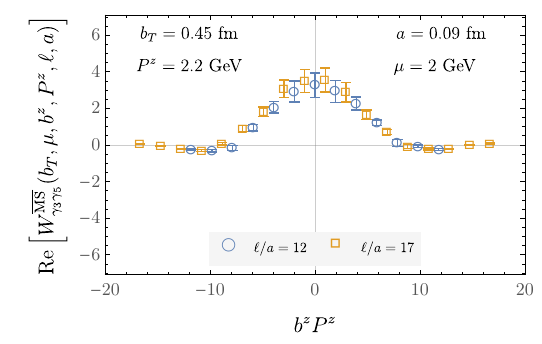} 
        \hspace{20pt}
        \includegraphics[width=0.46\textwidth]{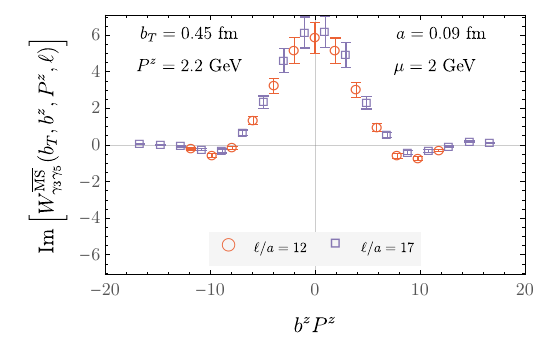} 
        \caption{Real and imaginary parts of the $\MSbar$-renormalized quasi-TMD WF ratios $W^{\MSbar}_{\Gamma}(b_T, \mu, b^z, P^z, \ell, a)$ computed on the $a = 0.09~\text{fm}$ ensemble for $\Gamma = \gamma_3 \gamma_5$ and $b_T/a = 5$.
        \label{fig:wf_ms_L64_gamma11_bT5}
        }
\end{figure*}

\begin{figure*}[t]
    \centering
        \includegraphics[width=0.46\textwidth]{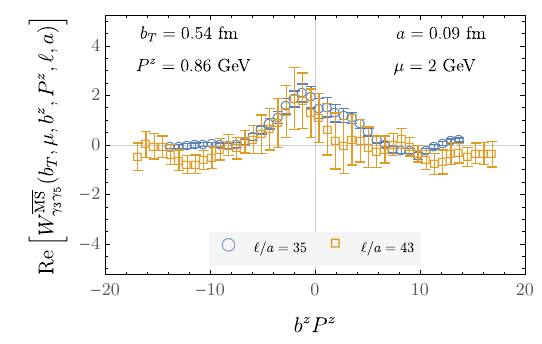}   
        \hspace{20pt}
        \includegraphics[width=0.46\textwidth]{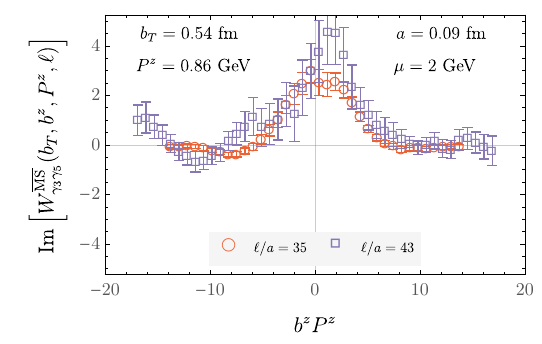}
        \includegraphics[width=0.46\textwidth]{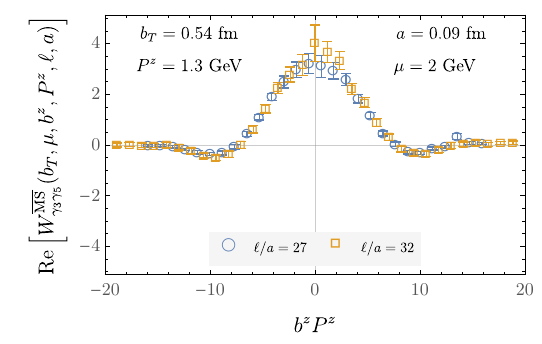} 
        \hspace{20pt}
        \includegraphics[width=0.46\textwidth]{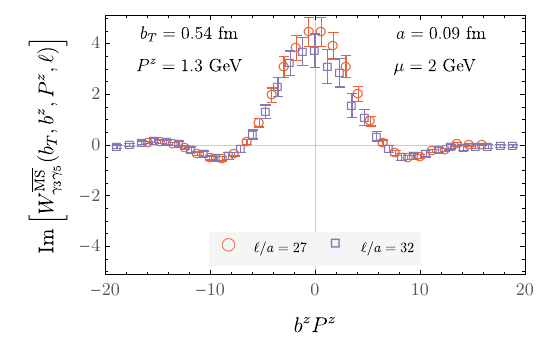} 
        \includegraphics[width=0.46\textwidth]{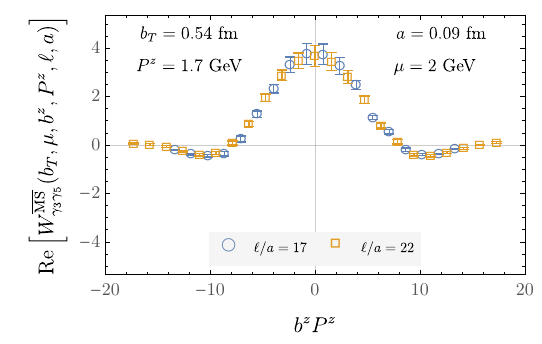} 
        \hspace{20pt}
         \includegraphics[width=0.46\textwidth]{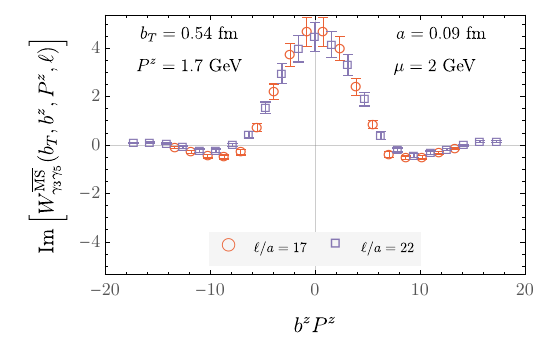} 
        \includegraphics[width=0.46\textwidth]{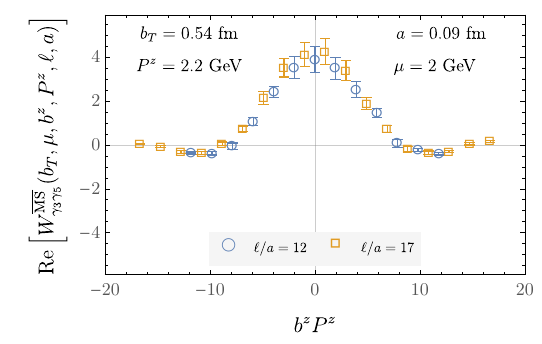} 
        \hspace{20pt}
        \includegraphics[width=0.46\textwidth]{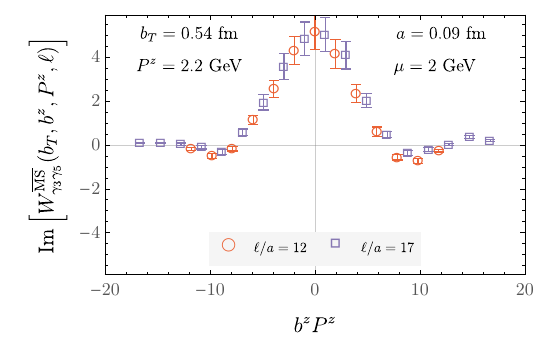} 
        \caption{Real and imaginary parts of the $\MSbar$-renormalized quasi-TMD WF ratios $W^{\MSbar}_{\Gamma}(b_T, \mu, b^z, P^z, \ell, a)$ computed on the $a = 0.09~\text{fm}$ ensemble for $\Gamma = \gamma_3 \gamma_5$ and $b_T/a = 6$.
        \label{fig:wf_ms_L64_gamma11_bT6}
        }
\end{figure*}

\begin{figure*}[t]
    \centering
        \includegraphics[width=0.46\textwidth]{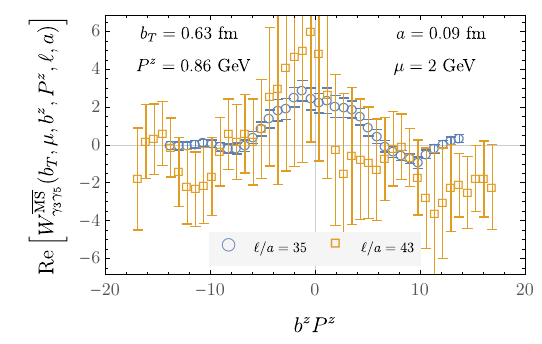}   
        \hspace{20pt}
        \includegraphics[width=0.46\textwidth]{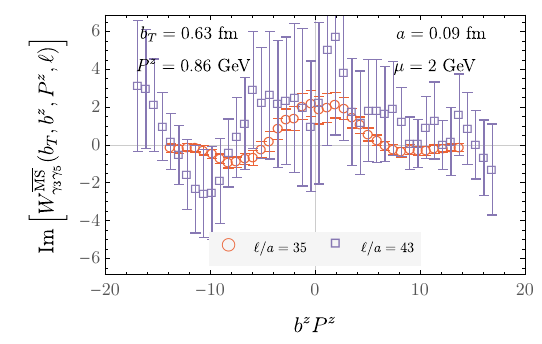}
        \includegraphics[width=0.46\textwidth]{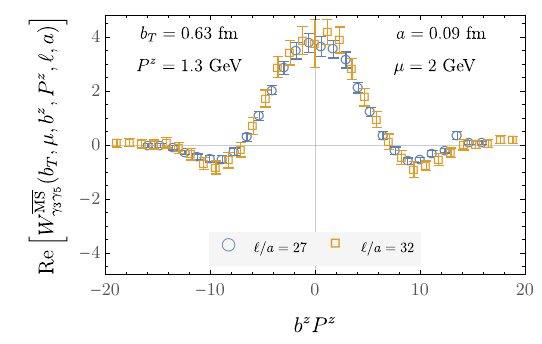} 
        \hspace{20pt}
        \includegraphics[width=0.46\textwidth]{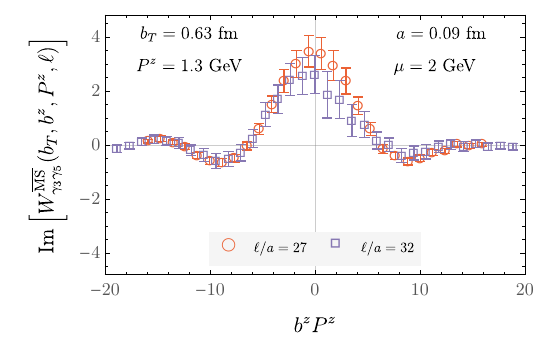} 
        \includegraphics[width=0.46\textwidth]{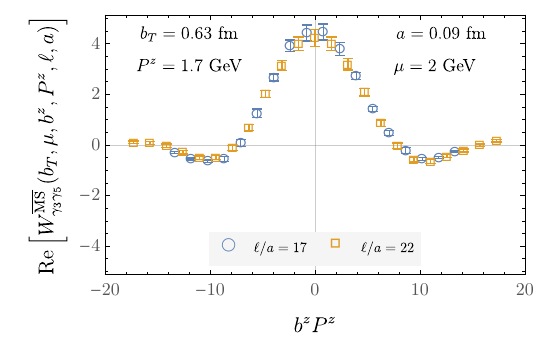} 
        \hspace{20pt}
         \includegraphics[width=0.46\textwidth]{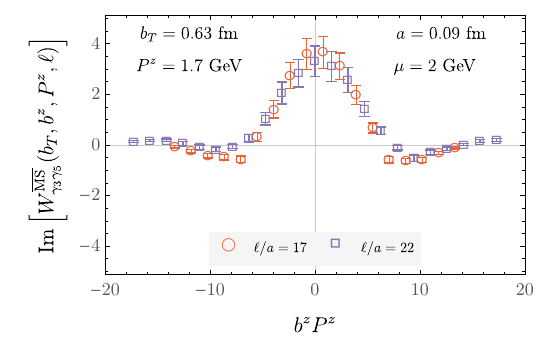} 
        \includegraphics[width=0.46\textwidth]{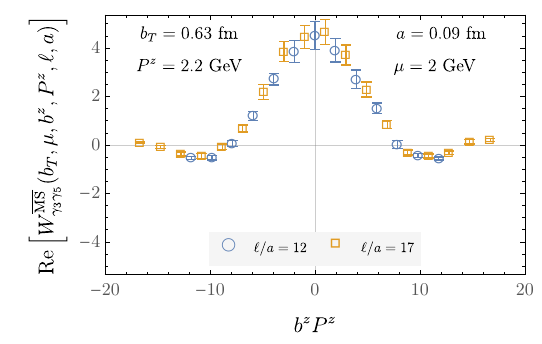} 
        \hspace{20pt}
        \includegraphics[width=0.46\textwidth]{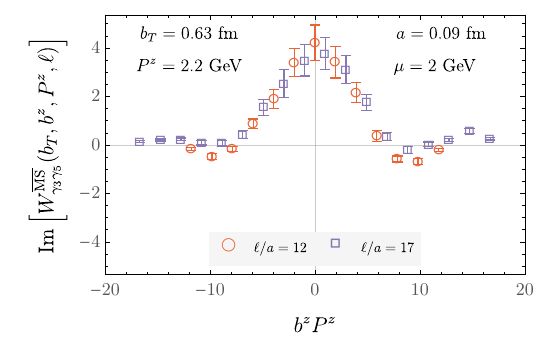} 
        \caption{Real and imaginary parts of the $\MSbar$-renormalized quasi-TMD WF ratios $W^{\MSbar}_{\Gamma}(b_T, \mu, b^z, P^z, \ell, a)$ computed on the $a = 0.09~\text{fm}$ ensemble for $\Gamma = \gamma_3 \gamma_5$ and $b_T/a = 7$.
        \label{fig:wf_ms_L64_gamma11_bT7}
        }
\end{figure*}

\begin{figure*}[t]
    \centering
        \includegraphics[width=0.46\textwidth]{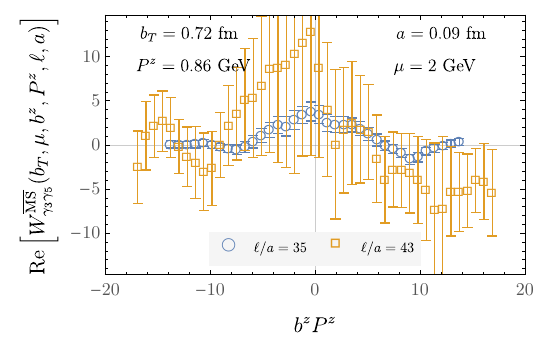}   
        \hspace{20pt}
        \includegraphics[width=0.46\textwidth]{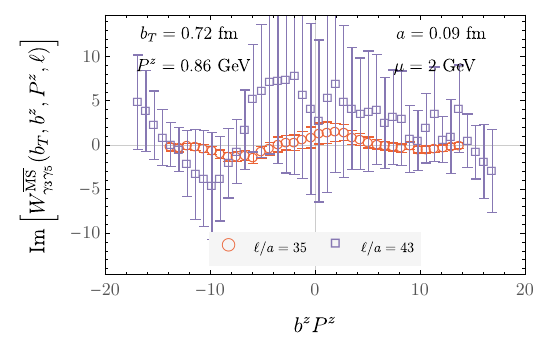}
        \includegraphics[width=0.46\textwidth]{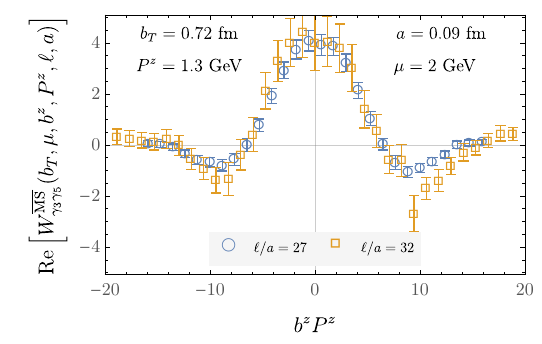} 
        \hspace{20pt}
        \includegraphics[width=0.46\textwidth]{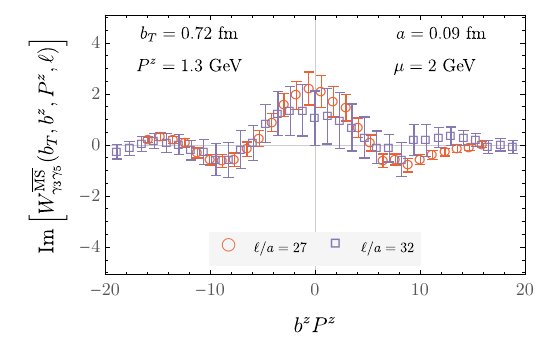} 
        \includegraphics[width=0.46\textwidth]{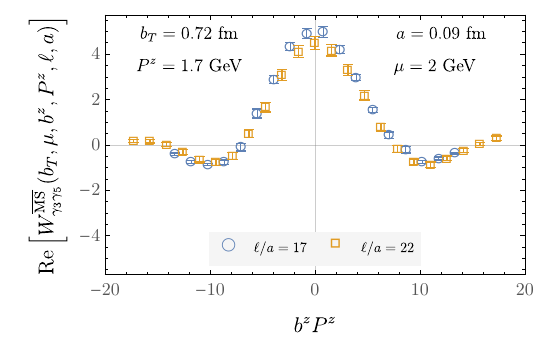} 
        \hspace{20pt}
         \includegraphics[width=0.46\textwidth]{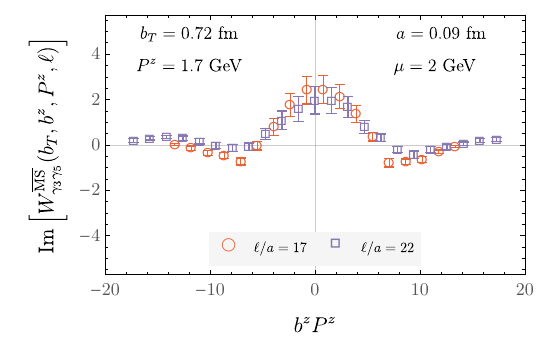} 
        \includegraphics[width=0.46\textwidth]{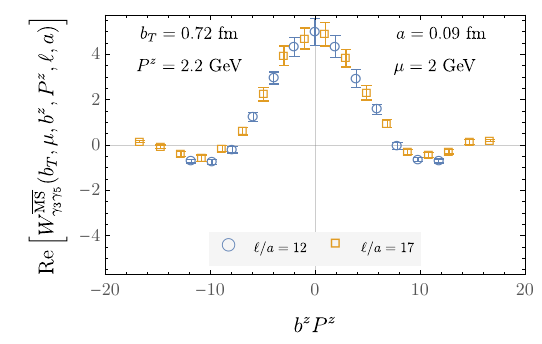} 
        \hspace{20pt}
        \includegraphics[width=0.46\textwidth]{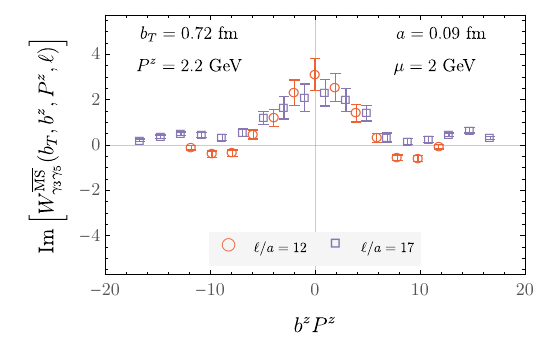} 
        \caption{Real and imaginary parts of the $\MSbar$-renormalized quasi-TMD WF ratios $W^{\MSbar}_{\Gamma}(b_T, \mu, b^z, P^z, \ell, a)$ computed on the $a = 0.09~\text{fm}$ ensemble for $\Gamma = \gamma_3 \gamma_5$ and $b_T/a = 8$.
        \label{fig:wf_ms_L64_gamma11_bT8}
        }
\end{figure*}

\begin{figure*}[t]
    \centering
        \includegraphics[width=0.46\textwidth]{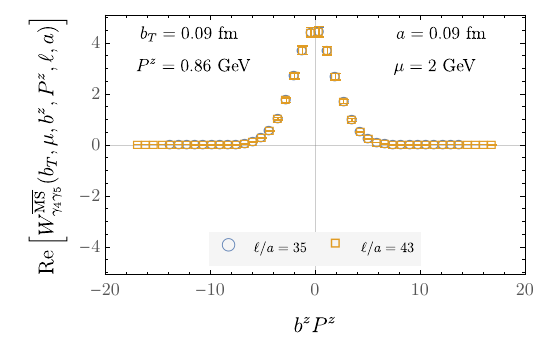}   
        \hspace{20pt}
        \includegraphics[width=0.46\textwidth]{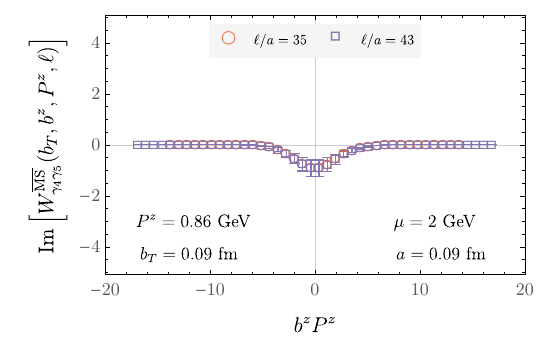}
        \includegraphics[width=0.46\textwidth]{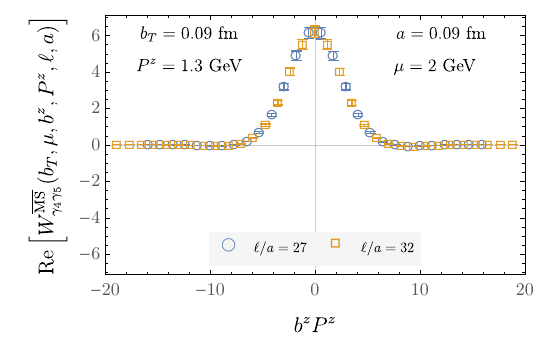} 
        \hspace{20pt}
        \includegraphics[width=0.46\textwidth]{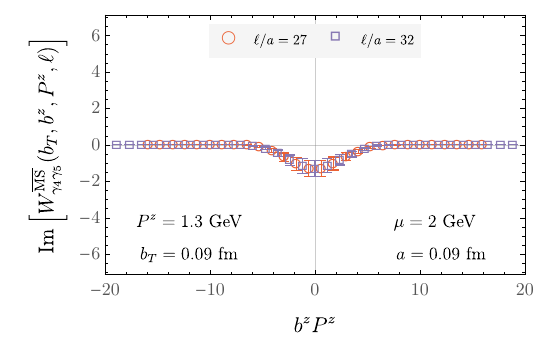} 
        \includegraphics[width=0.46\textwidth]{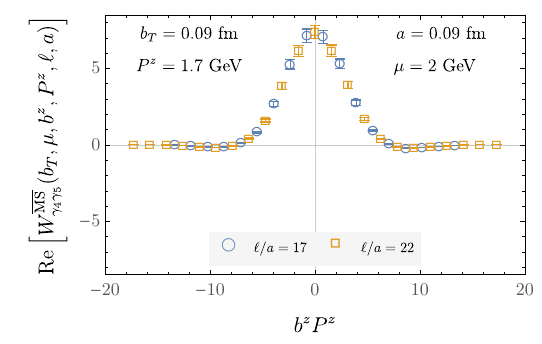} 
        \hspace{20pt}
         \includegraphics[width=0.46\textwidth]{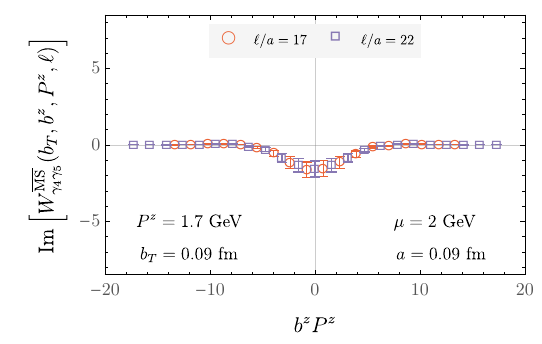} 
        \includegraphics[width=0.46\textwidth]{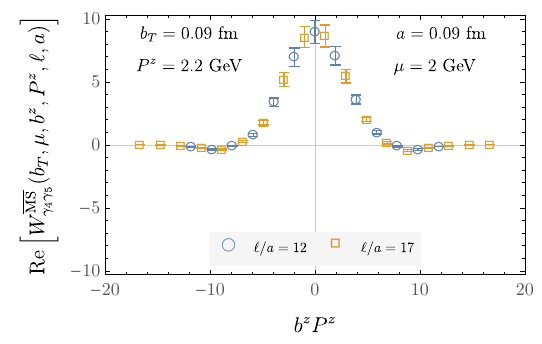} 
        \hspace{20pt}
        \includegraphics[width=0.46\textwidth]{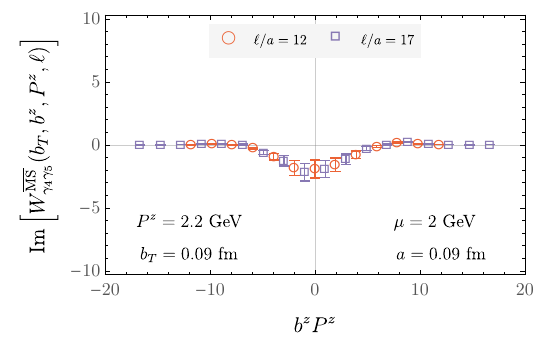} 
        \caption{Real and imaginary parts of the $\MSbar$-renormalized quasi-TMD WF ratios $W^{\MSbar}_{\Gamma}(b_T, \mu, b^z, P^z, \ell, a)$ computed on the $a = 0.09~\text{fm}$ ensemble for $\Gamma = \gamma_4 \gamma_5$ and $b_T/a = 1$.
        \label{fig:wf_ms_L64_gamma7_bT1}
        }
\end{figure*}

\begin{figure*}[t]
    \centering
        \includegraphics[width=0.46\textwidth]{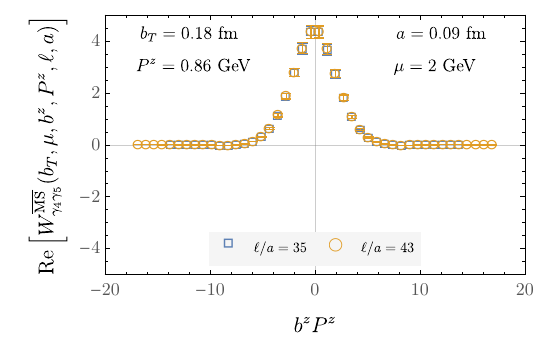}   
        \hspace{20pt}
        \includegraphics[width=0.46\textwidth]{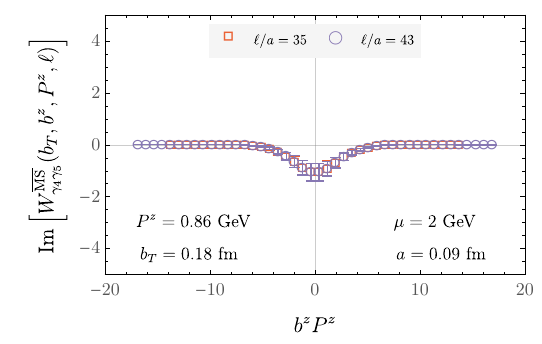}
        \includegraphics[width=0.46\textwidth]{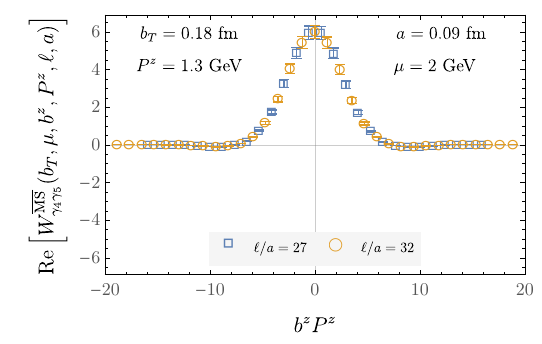} 
        \hspace{20pt}
        \includegraphics[width=0.46\textwidth]{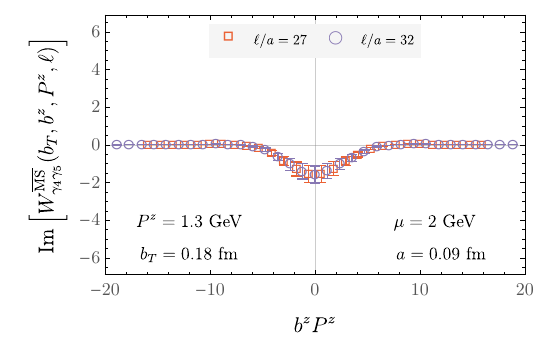} 
        \includegraphics[width=0.46\textwidth]{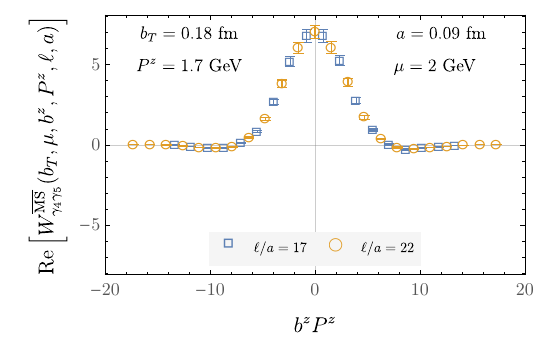} 
        \hspace{20pt}
         \includegraphics[width=0.46\textwidth]{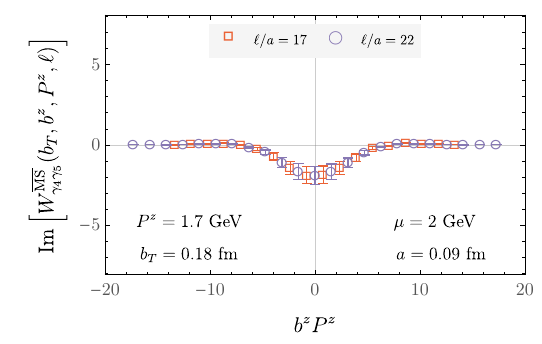} 
        \includegraphics[width=0.46\textwidth]{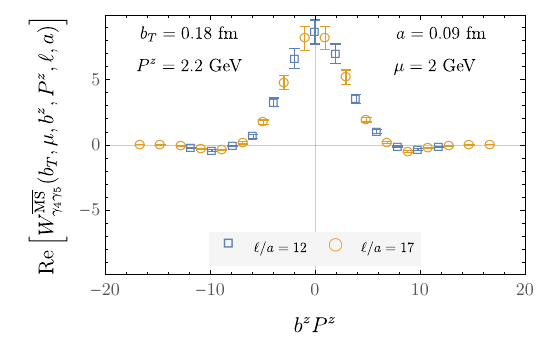} 
        \hspace{20pt}
        \includegraphics[width=0.46\textwidth]{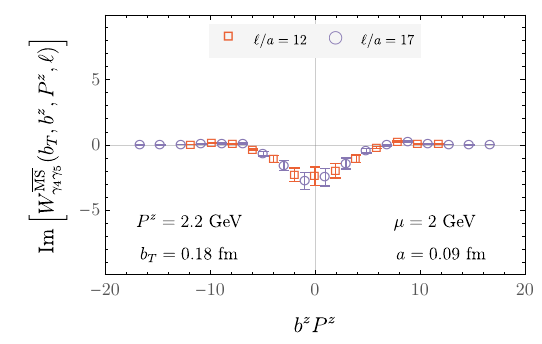} 
        \caption{Real and imaginary parts of the $\MSbar$-renormalized quasi-TMD WF ratios $W^{\MSbar}_{\Gamma}(b_T, \mu, b^z, P^z, \ell, a)$ computed on the $a = 0.09~\text{fm}$ ensemble for $\Gamma = \gamma_4 \gamma_5$ and $b_T/a = 2$.
        \label{fig:wf_ms_L64_gamma7_bT2}
        }
\end{figure*}

\begin{figure*}[t]
    \centering
        \includegraphics[width=0.46\textwidth]{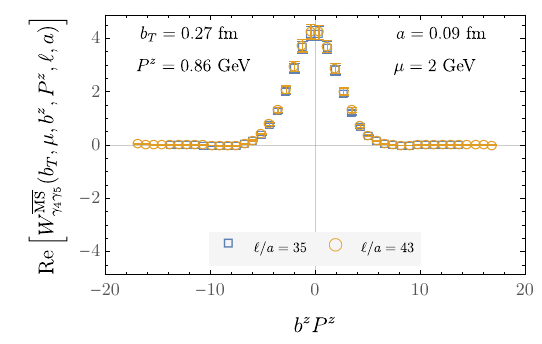}   
        \hspace{20pt}
        \includegraphics[width=0.46\textwidth]{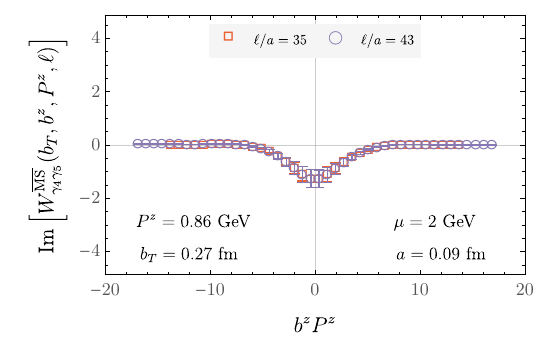}
        \includegraphics[width=0.46\textwidth]{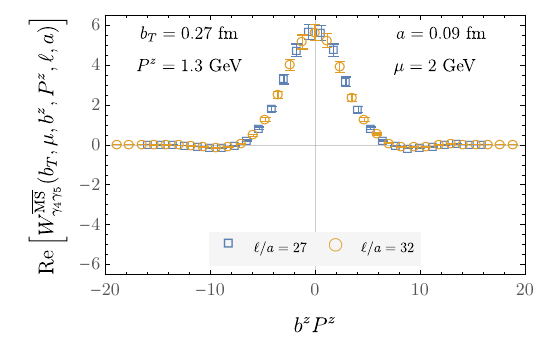} 
        \hspace{20pt}
        \includegraphics[width=0.46\textwidth]{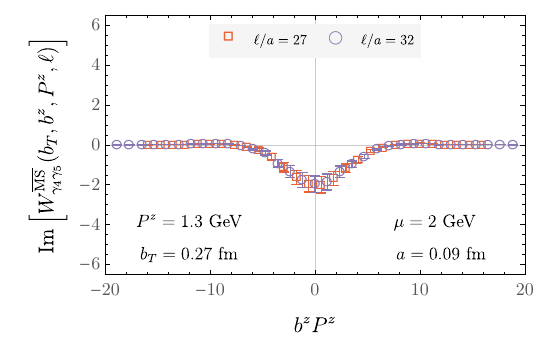} 
        \includegraphics[width=0.46\textwidth]{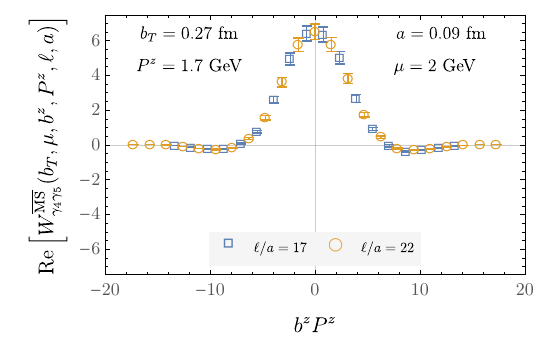} 
        \hspace{20pt}
         \includegraphics[width=0.46\textwidth]{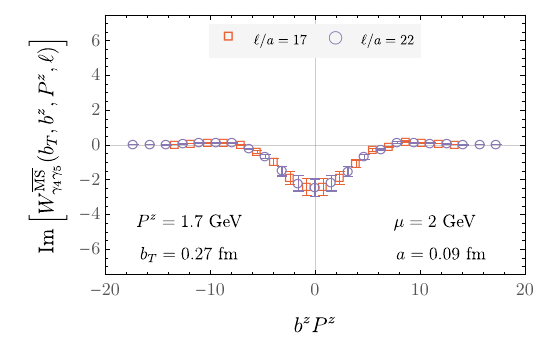} 
        \includegraphics[width=0.46\textwidth]{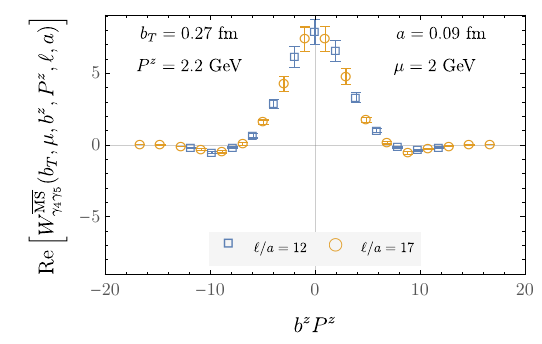} 
        \hspace{20pt}
        \includegraphics[width=0.46\textwidth]{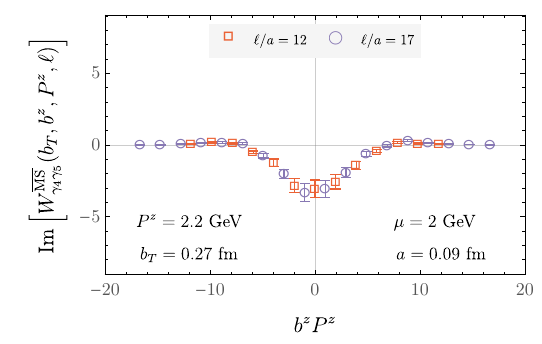} 
        \caption{Real and imaginary parts of the $\MSbar$-renormalized quasi-TMD WF ratios $W^{\MSbar}_{\Gamma}(b_T, \mu, b^z, P^z, \ell, a)$ computed on the $a = 0.09~\text{fm}$ ensemble for $\Gamma = \gamma_4 \gamma_5$ and $b_T/a = 3$.
        \label{fig:wf_ms_L64_gamma7_bT3}
        }
\end{figure*}

\begin{figure*}[t]
    \centering
        \includegraphics[width=0.46\textwidth]{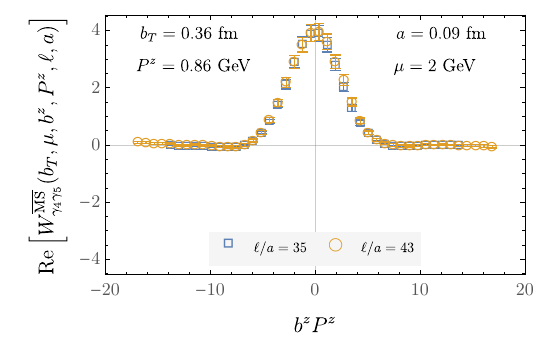}   
        \hspace{20pt}
        \includegraphics[width=0.46\textwidth]{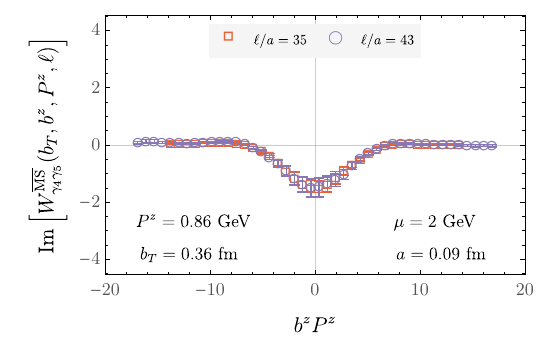}
        \includegraphics[width=0.46\textwidth]{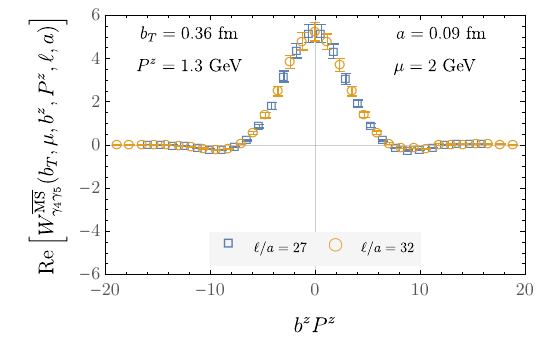} 
        \hspace{20pt}
        \includegraphics[width=0.46\textwidth]{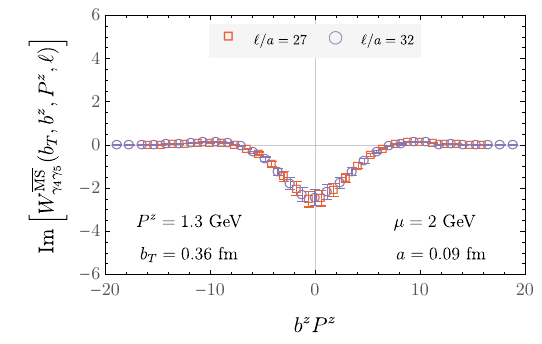} 
        \includegraphics[width=0.46\textwidth]{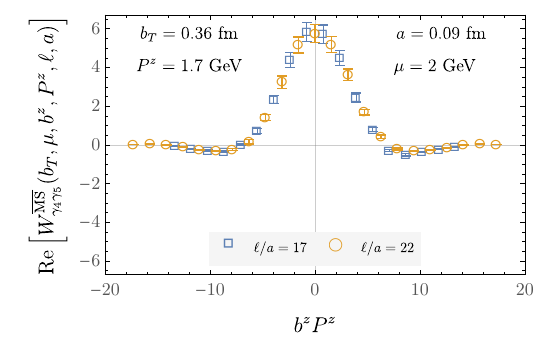} 
        \hspace{20pt}
         \includegraphics[width=0.46\textwidth]{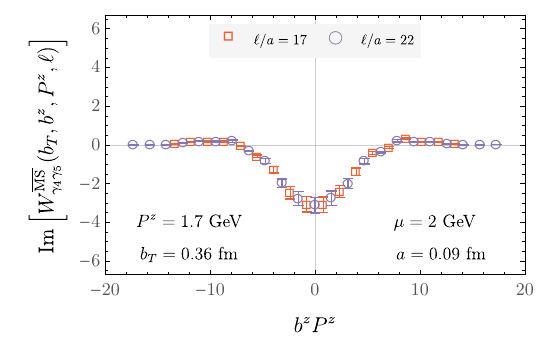} 
        \includegraphics[width=0.46\textwidth]{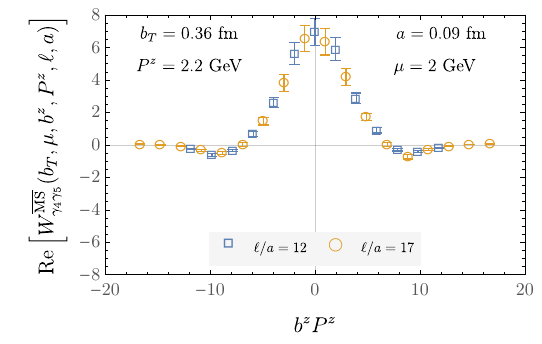} 
        \hspace{20pt}
        \includegraphics[width=0.46\textwidth]{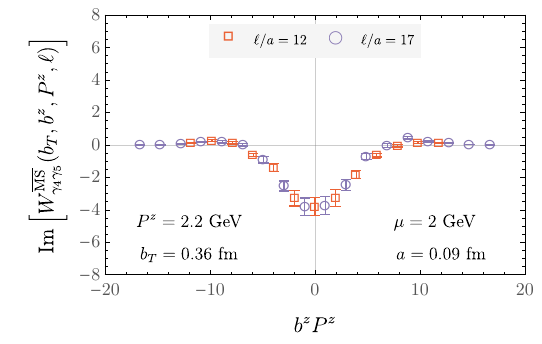} 
        \caption{Real and imaginary parts of the $\MSbar$-renormalized quasi-TMD WF ratios $W^{\MSbar}_{\Gamma}(b_T, \mu, b^z, P^z, \ell, a)$ computed on the $a = 0.09~\text{fm}$ ensemble for $\Gamma = \gamma_4 \gamma_5$ and $b_T/a = 4$.
        \label{fig:wf_ms_L64_gamma7_bT4}
        }
\end{figure*}

\begin{figure*}[t]
    \centering
        \includegraphics[width=0.46\textwidth]{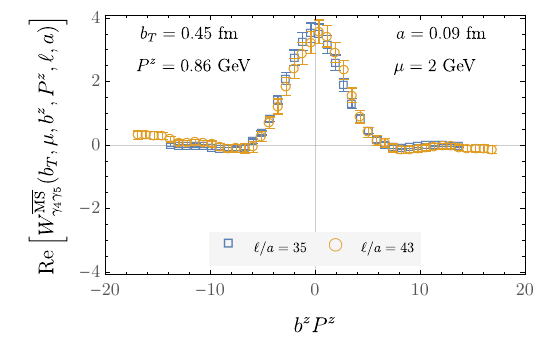}   
        \hspace{20pt}
        \includegraphics[width=0.46\textwidth]{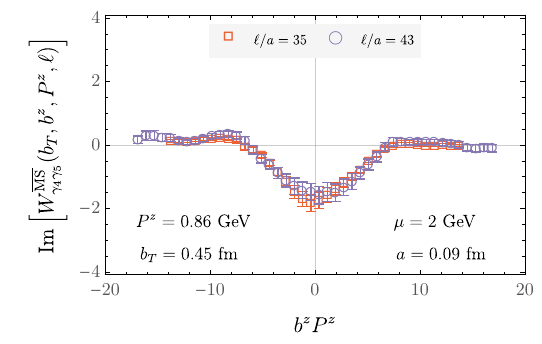}
        \includegraphics[width=0.46\textwidth]{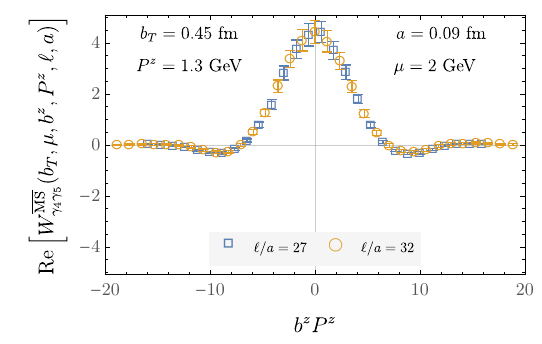} 
        \hspace{20pt}
        \includegraphics[width=0.46\textwidth]{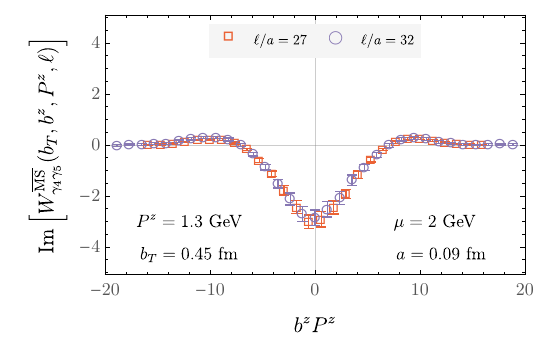} 
        \includegraphics[width=0.46\textwidth]{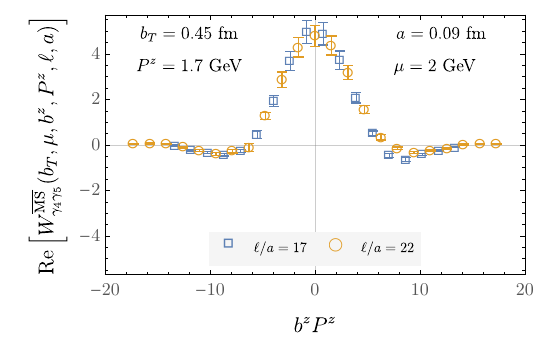} 
        \hspace{20pt}
         \includegraphics[width=0.46\textwidth]{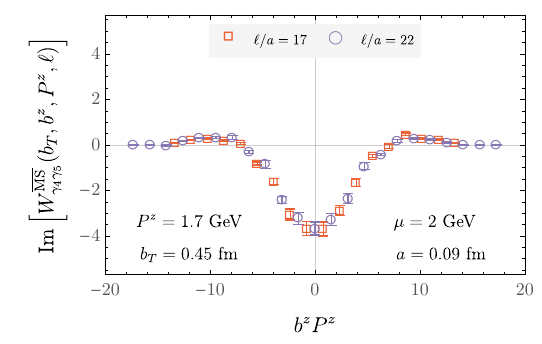} 
        \includegraphics[width=0.46\textwidth]{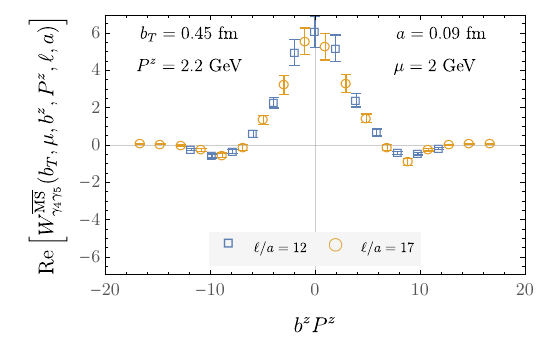} 
        \hspace{20pt}
        \includegraphics[width=0.46\textwidth]{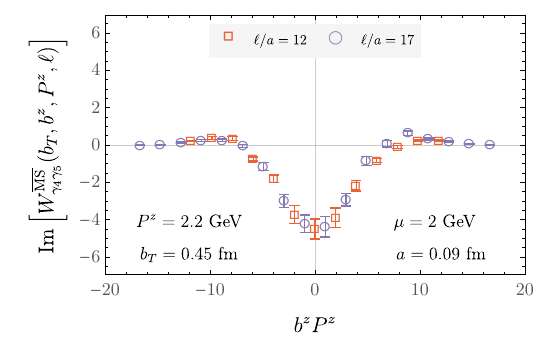} 
        \caption{Real and imaginary parts of the $\MSbar$-renormalized quasi-TMD WF ratios $W^{\MSbar}_{\Gamma}(b_T, \mu, b^z, P^z, \ell, a)$ computed on the $a = 0.09~\text{fm}$ ensemble for $\Gamma = \gamma_4 \gamma_5$ and $b_T/a = 5$.
        \label{fig:wf_ms_L64_gamma7_bT5}
        }
\end{figure*}

\begin{figure*}[t]
    \centering
        \includegraphics[width=0.46\textwidth]{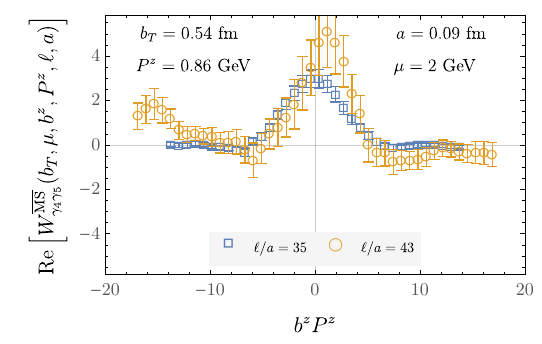}   
        \hspace{20pt}
        \includegraphics[width=0.46\textwidth]{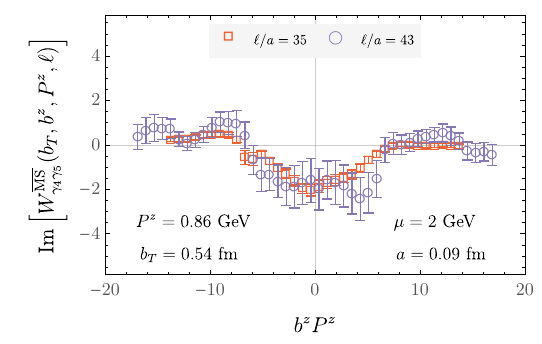}
        \includegraphics[width=0.46\textwidth]{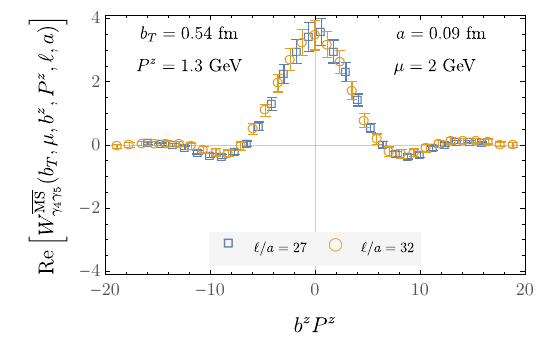} 
        \hspace{20pt}
        \includegraphics[width=0.46\textwidth]{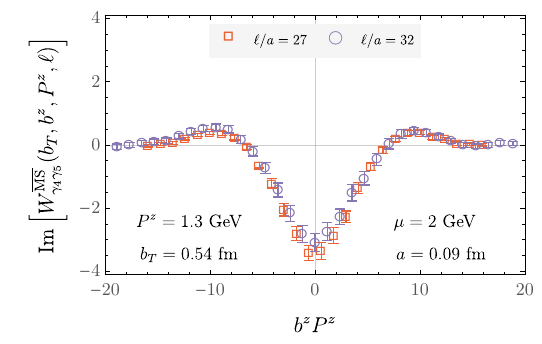} 
        \includegraphics[width=0.46\textwidth]{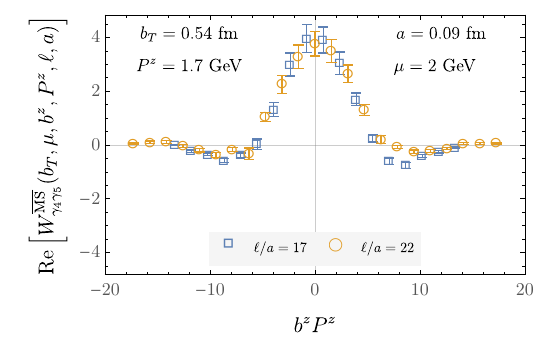} 
        \hspace{20pt}
         \includegraphics[width=0.46\textwidth]{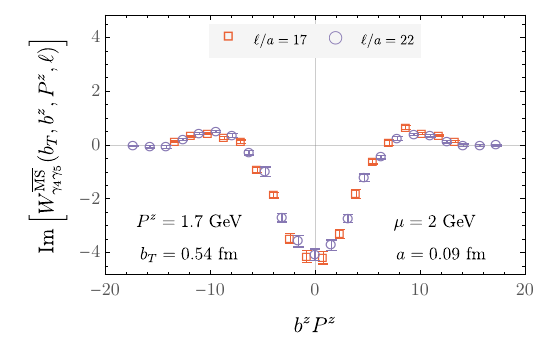} 
        \includegraphics[width=0.46\textwidth]{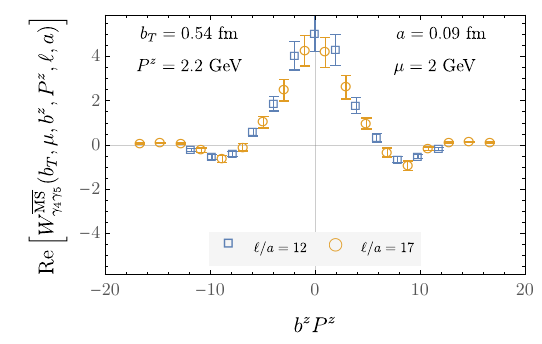} 
        \hspace{20pt}
        \includegraphics[width=0.46\textwidth]{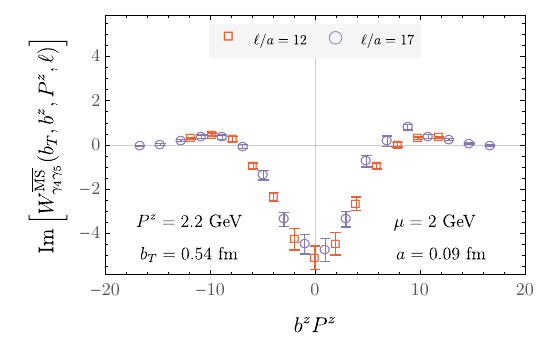} 
        \caption{Real and imaginary parts of the $\MSbar$-renormalized quasi-TMD WF ratios $W^{\MSbar}_{\Gamma}(b_T, \mu, b^z, P^z, \ell, a)$ computed on the $a = 0.09~\text{fm}$ ensemble for $\Gamma = \gamma_4 \gamma_5$ and $b_T/a = 6$.
        \label{fig:wf_ms_L64_gamma7_bT6}
        }
\end{figure*}

\begin{figure*}[t]
    \centering
        \includegraphics[width=0.46\textwidth]{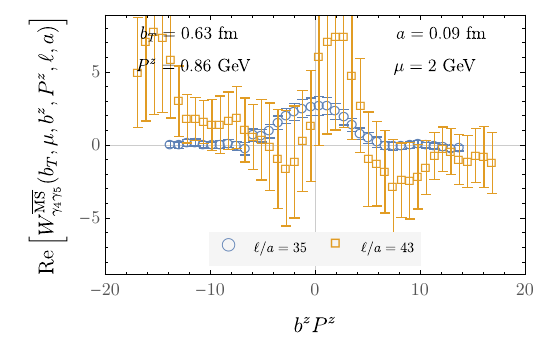}   
        \hspace{20pt}
        \includegraphics[width=0.46\textwidth]{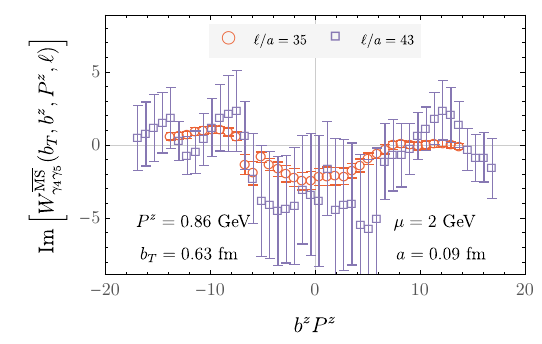}
        \includegraphics[width=0.46\textwidth]{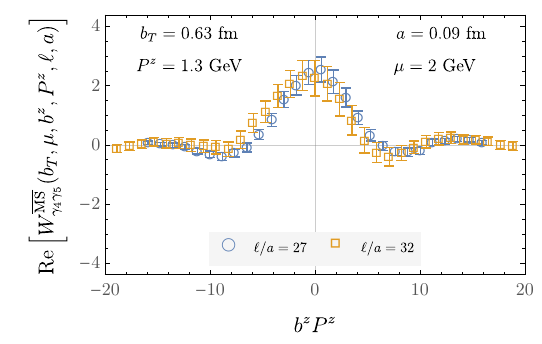} 
        \hspace{20pt}
        \includegraphics[width=0.46\textwidth]{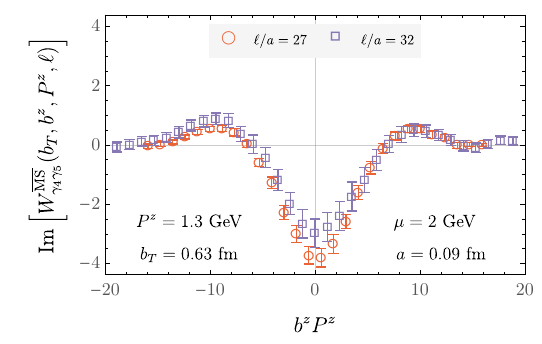} 
        \includegraphics[width=0.46\textwidth]{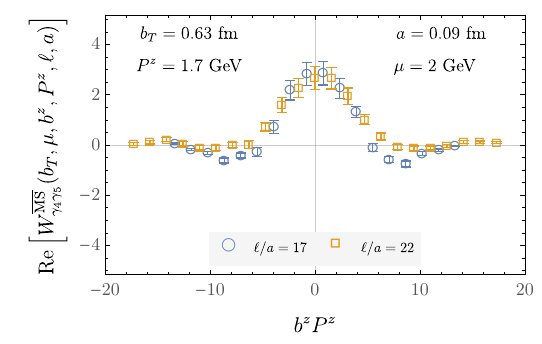} 
        \hspace{20pt}
         \includegraphics[width=0.46\textwidth]{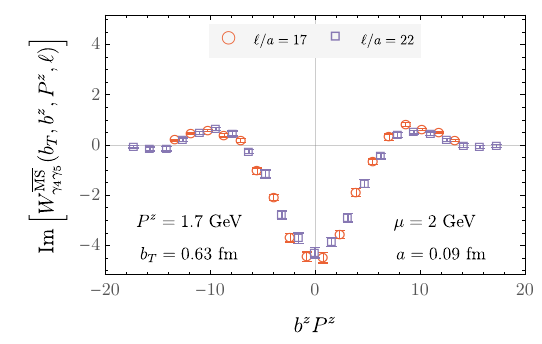} 
        \includegraphics[width=0.46\textwidth]{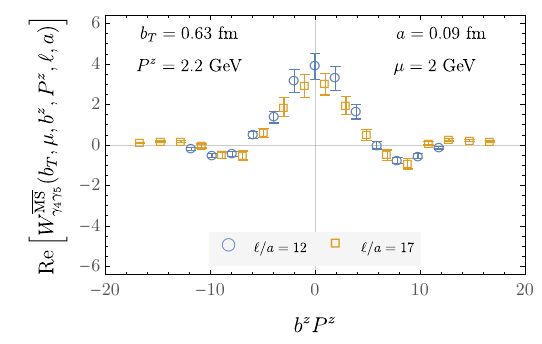} 
        \hspace{20pt}
        \includegraphics[width=0.46\textwidth]{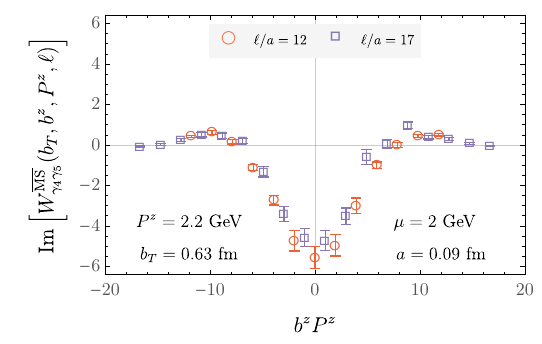} 
        \caption{Real and imaginary parts of the $\MSbar$-renormalized quasi-TMD WF ratios $W^{\MSbar}_{\Gamma}(b_T, \mu, b^z, P^z, \ell, a)$ computed on the $a = 0.09~\text{fm}$ ensemble for $\Gamma = \gamma_4 \gamma_5$ and $b_T/a = 7$.
        \label{fig:wf_ms_L64_gamma7_bT7}
        }
\end{figure*}

\begin{figure*}[t]
    \centering
        \includegraphics[width=0.46\textwidth]{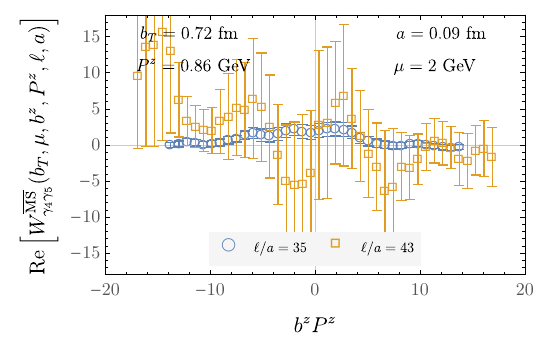}   
        \hspace{20pt}
        \includegraphics[width=0.46\textwidth]{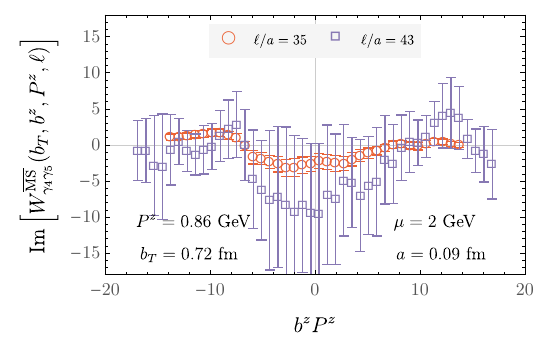}
        \includegraphics[width=0.46\textwidth]{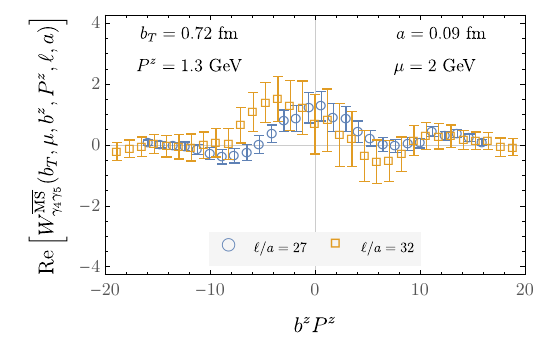} 
        \hspace{20pt}
        \includegraphics[width=0.46\textwidth]{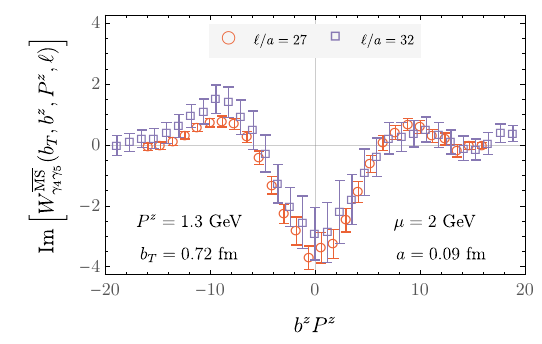} 
        \includegraphics[width=0.46\textwidth]{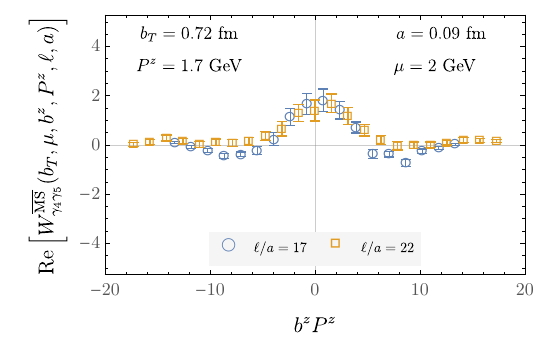} 
        \hspace{20pt}
         \includegraphics[width=0.46\textwidth]{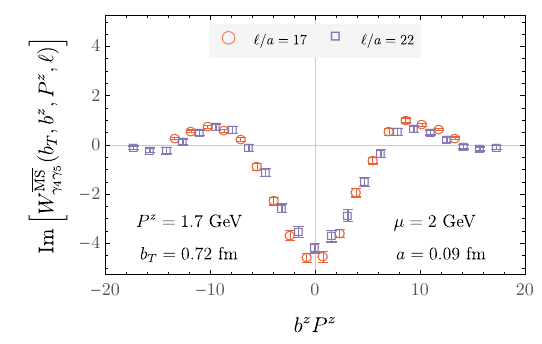} 
        \includegraphics[width=0.46\textwidth]{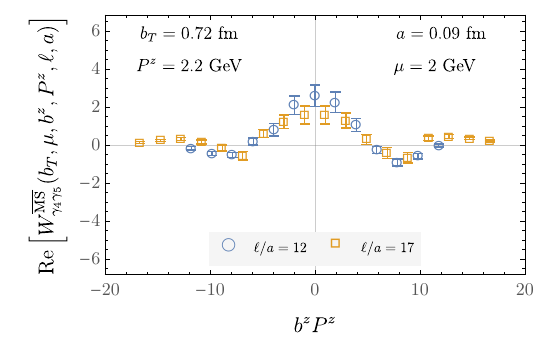} 
        \hspace{20pt}
        \includegraphics[width=0.46\textwidth]{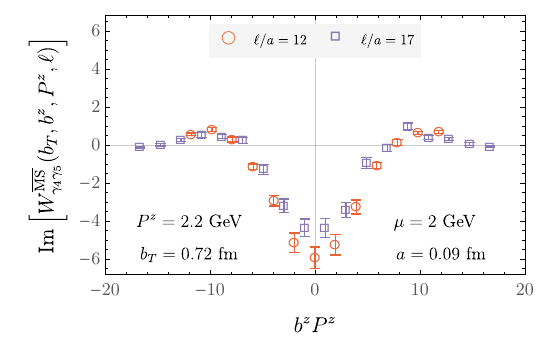} 
        \caption{Real and imaginary parts of the $\MSbar$-renormalized quasi-TMD WF ratios $W^{\MSbar}_{\Gamma}(b_T, \mu, b^z, P^z, \ell, a)$ computed on the $a = 0.09~\text{fm}$ ensemble for $\Gamma = \gamma_4 \gamma_5$ and $b_T/a = 8$.
        \label{fig:wf_ms_L64_gamma7_bT8}
        }
\end{figure*}

Renormalized quasi-TMD WF ratios $W^{\MSbar}_{\Gamma}(b_\tran, \mu, b^z, P^z, \ell,a)$ for the $\Gamma \in \{\gamma_4 \gamma_5, \gamma_3 \gamma_5\}$ used to constrain the CS kernel are shown for all $b_T$, $b^z$, $P^z$, and $\ell$ for the $a = 0.15~\text{fm}$ and $a = 0.09~\text{fm}$ ensembles in Figs.~\ref{fig:wf_ms_L32_gamma11_bT1}--\ref{fig:wf_ms_L64_gamma7_bT8}.
The corresponding results for the $a = 0.12~\text{fm}$ ensemble are presented in Ref.~\cite{Avkhadiev:2023poz}.

\subsection{CS kernel estimators}\label{app:CS}

\begin{figure*}[t]
    \centering
        \includegraphics[width=0.46\textwidth]{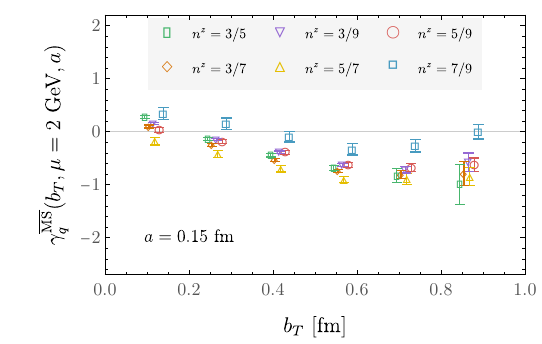}   
        \hspace{20pt}
        \includegraphics[width=0.46\textwidth]{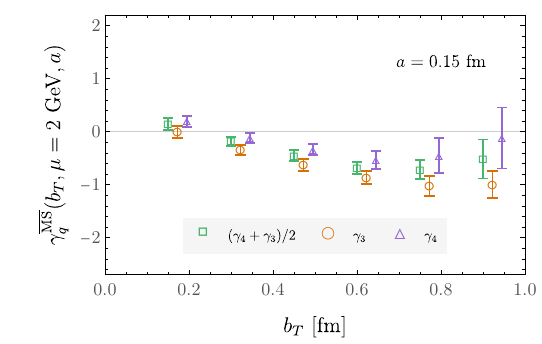}   
        \caption{For the $a = 0.15~\text{fm}$ ensemble. [Left] The $x$- and $\Gamma$-averaged CS kernel constraints $\gamma_q^{\MSbar}(b_T, \mu, a)$ computed separately for each momentum pair $P_1^z, P_2^z$. [Right]  $x$- and $P^z$-averaged CS kernel constraints computed separately for each Dirac structure $\Gamma$. 
        \label{fig:CS_mtm_gamma_comp_L32}
        }
\end{figure*}

\begin{figure*}[t]
    \centering
        \includegraphics[width=0.46\textwidth]{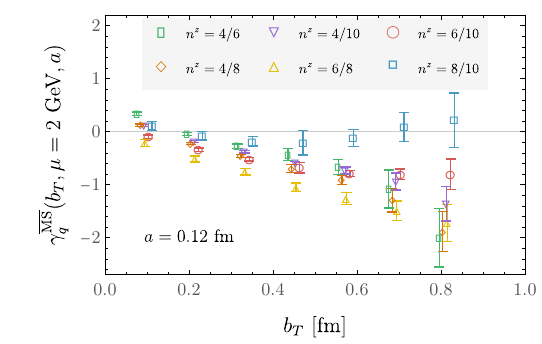}   
        \hspace{20pt}
        \includegraphics[width=0.46\textwidth]{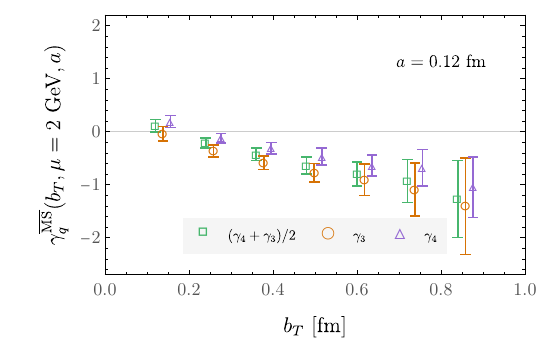}   
        \caption{As in Fig.~\ref{fig:CS_mtm_gamma_comp_L32}, for the $a = 0.12~\text{fm}$ ensemble.
        \label{fig:CS_mtm_gamma_comp_L48}
        }
\end{figure*}

\begin{figure*}[t]
    \centering
        \includegraphics[width=0.46\textwidth]{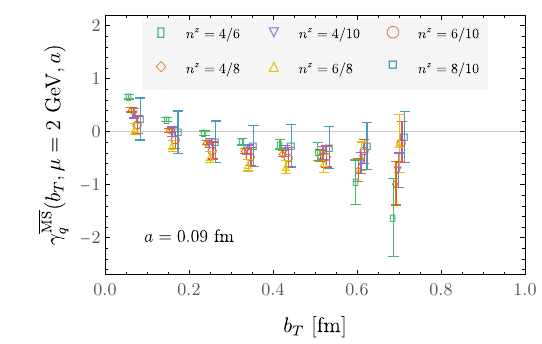}   
        \hspace{20pt}
        \includegraphics[width=0.46\textwidth]{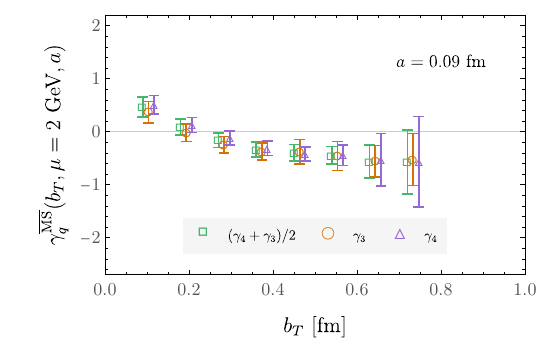}   
        \caption{As in Fig.~\ref{fig:CS_mtm_gamma_comp_L32}, for the $a = 0.09~\text{fm}$ ensemble.
        \label{fig:CS_mtm_gamma_comp_L64}
        }
\end{figure*}

Results for the CS kernel estimator $\hat{\gamma}_{\Gamma}^{\MSbar}(b_{\tran}, \mu, x, P_1^z, P_2^z, a)$ (Eq.~\eqref{eq:kernel-wf-estimator}) averaged over $x \in [0.3, 0.7]$ and $\Gamma \in \{\gamma_4 \gamma_5, \gamma_3 \gamma_5\}$ but shown separately for each momentum pair $P_1^z, P_2^z$ are presented for all three ensembles in Figs.~\ref{fig:CS_mtm_gamma_comp_L32}-\ref{fig:CS_mtm_gamma_comp_L64}.
Analogous results for the CS kernel estimator averaged over $x$ and $P_1^z$, $P_2^z$ but shown separately for each $\Gamma$ are also presented in Figs.~\ref{fig:CS_mtm_gamma_comp_L32}-\ref{fig:CS_mtm_gamma_comp_L64} 
The results shown for the $a = 0.12~\text{fm}$ ensemble are reproduced from Ref.~\cite{Avkhadiev:2023poz} to enable comparison with the results from the other two ensembles.

\subsection{Imaginary parts of CS kernel estimators}\label{app:imag}

Previous analyses of lattice QCD results for the CS kernel in Refs.~\cite{LPC:2022ibr,Avkhadiev:2023poz} yielded nonzero values of the imaginary part. Without explanation, this would indicate the presence of systematic uncertainties unaccounted for in the calculations. However, these observations can be understood in the context of a recent analysis~\cite{Liu:2023onm} which reveals that the imaginary part of the matching coefficient 
for the quasi-TMD WF includes a linear infrared renormalon. To mitigate the nonconvergence of the perturbative series, it was proposed in Ref.~\cite{Liu:2023onm} to subtract the leading asymptotic series from the matching coefficient, i.e., a leading-renormalon subtraction (LRR) scheme, with the uncertainties in the perturbative series absorbed into a linear power correction.

The matching correction to the CS kernel in \Eq{kernel-wf-estimator} can be expressed as
\begin{align}
    \delta \gamma_q^{\MSbar}(\mu, x, P_1^z, P_2^z) &= - \dfrac{1}{\ln(P_{1}^{z}/P_{2}^{z})} \ln\frac{H(xP_1^z, \bar{x}P_1^z,\mu)}{H(xP_2^z, \bar{x}P_2^z,\mu)}\,,    
\end{align}
where $\bar{x}=1-x$, and the matching coefficient $H$ is the product of two matching kernels for the `heavy-light' current,
\begin{align}
	H(xP^z, \bar{x}P^z,\mu) &= C(xP^z, \mu) C(\bar{x}P^z,\mu)\,,
\end{align}
which are defined in Eqs.~(C2-C8) in Ref.~\cite{Avkhadiev:2023poz}. Here the $b_\tran$-dependence in $\delta \gamma_q^{\MSbar}$ is suppressed as it exists only in the $b_\tran$-unexpanded matching coefficient in Eq.~(C43) of Ref.~\cite{Avkhadiev:2023poz}.

The asymptotic series in $C$ is given by
\begin{align}
	R(p^z, \mu) &=i N_m{\mu \over p^z} \sum_{n=0}^\infty \beta_0^n \alpha_s^{n+1}(\mu)n!\,,
\end{align}
from Eqs.~(2.48-52) of Ref.~\cite{Liu:2023onm}, where $\beta_0$ is the lowest order $\beta$-function, and $N_m= 0.552$ for $n_f=4$.
Therefore, the LRR subtracted matching coefficient is simply
\begin{align}
	C^{\rm LRR}(p^z, \mu) &= C(p^z, \mu) - R(p^z, \mu) \,,
\end{align}
which is shown to converge quickly in the perturbative order~\cite{Liu:2023onm}. For the analysis in this work, $C$ is thus replaced by $C^{\rm LRR}$, including the $b_\tran$-unexpanded uNNLL matching coefficient, defined in Eq.~(27) and Eq.~(C43) of Ref.~\cite{Avkhadiev:2023poz}, to obtain the final quoted results~\cite{Avkhadiev:2023poz}.

\begin{figure*}[t]
    \centering
        \includegraphics[width=0.46\textwidth]{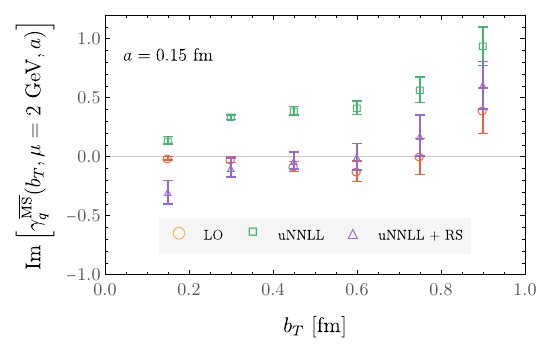}  
        \hspace{20pt}
        \includegraphics[width=0.46\textwidth]{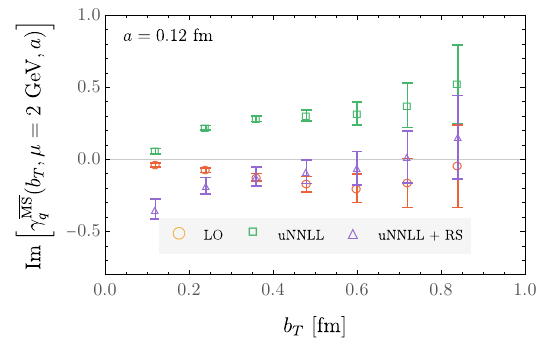}   \\
        \includegraphics[width=0.46\textwidth]{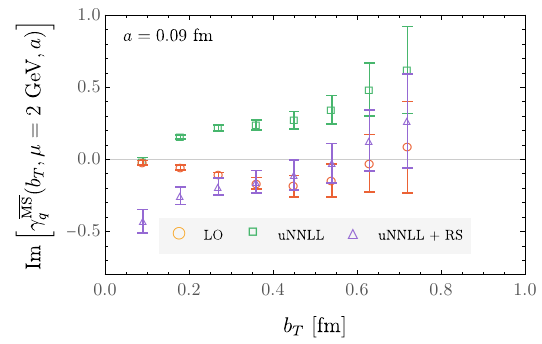}   
        \caption{
        Imaginary part of the CS kernel computed on each of the three lattice ensembles used in the numerical calculations, analogous to the results for the real part of the CS kernel shown in Fig.~\ref{fig:CSkerneldata} of the main text. Labels `LO', `uNNLL', and `uNNLL+RS' denote the order of matching used, as described in Sec.~\ref{app:imag}. 
        \label{fig:CS_imag_comp}
        }
\end{figure*}

Since the linear renormalon is a feature in the TMD factorization of the quasi-TMD WF~\cite{Liu:2023onm}, the LRR is only expected to improve the perturbative convergence in the large $b_\tran$, or large $P^z b_\tran$, region. When $P^z b_\tran \lesssim 1$, the $b_\tran$-dependent uNNLL matching kernel takes care of the perturbative power corrections, and has been demonstrated to yield an imaginary part of the CS kernel which is consistent with zero~\cite{Avkhadiev:2023poz}. As shown in Fig.~\ref{fig:CS_imag_comp}, in the numerical calculation of this work results with both LRR and uNNLL matching indeed display the expected behavior in the relevant parameter regions. Thus, the results do not imply additional uncontrolled systematic uncertainties beyond those discussed in the main text.

\subsection{Perturbative part of the CS kernel}\label{app:param}

The perturbative part of the CS kernel $\DP$ in \Eq{CSkernelparam} is given by
\begin{align}\label{eq:dres}
	\DP(b^*,\mu)&={1\over2}K(\mu,\mu_{b^*}) +  d[\alpha_s(\mu_{b^*})]\nn\\
	&\equiv \int_{\mu_{b^*}}^\mu {d\mu'\over \mu'} \Gamma_{\rm cusp}[\alpha_s(\mu')] +  d[\alpha_s(\mu_{b^*})]\,,
\end{align}
where $\mu_{b^*}=2e^{-\gamma_E}/b^*$, $ \Gamma_{\rm cusp}$ is the cusp anomalous dimension known to four loops~\cite{Korchemsky:1987wg,Moch:2004pa,Henn:2019swt,vonManteuffel:2020vjv,Moult:2022xzt,Duhr:2022yyp} and approximated at five-loop order~\cite{Herzog:2018kwj}, and $d$ is the non-cusp part of the rapidity anomalous dimension known to four loops~\cite{Li:2016ctv,Vladimirov:2016dll,Duhr:2022yyp,Moult:2022xzt}.

Both anomalous dimensions are perturbative series in the strong coupling $\alpha_s$,
\begin{align}
	    \Gamma_{\rm cusp}[a_s(\mu)] &= \sum_{n=0}^\infty a_s^{n+1}(\mu) \Gamma_n\,, \\
        d[a_s(\mu)] &= \sum_{n=0}^\infty a_s^{n+1}(\mu)d_{n}\,,
\end{align}
where $a_s=\alpha_s/(4\pi)$. For N$^3$LL resummation, the required anomalous dimensions are
\begin{align}\label{eq:cusp-pert}
	\Gamma_0 &= {16\over 3}, \\
    \Gamma_1 &= \frac{1072}{9}-\frac{16 \pi ^2}{3} -\frac{160}{27} n_f, \\
	\Gamma_2 &= 352 \zeta (3)+\frac{176 \pi ^4}{15}-\frac{2144 \pi ^2}{9}+1960\nn\\
	&\quad  + n_f\left( -\frac{832 \zeta (3)}{9}+\frac{320 \pi ^2}{27}-\frac{5104}{27} \right) -\frac{64}{81} n_f^2,\\
	\Gamma_3 &= \left(-1536 \zeta (3)^2-704 \pi ^2 \zeta (3)+28032 \zeta (3)-\frac{34496 \zeta (5)}{3}\right. \nn\\
	&\qquad \left.+\frac{337112}{9}-\frac{178240 \pi ^2}{27}+\frac{3608 \pi ^4}{5}-\frac{32528 \pi ^6}{945}\right)\nn\\
	&\qquad +  n_f \left( \frac{1664 \pi ^2 \zeta (3)}{9}-\frac{616640 \zeta (3)}{81}+\frac{25472 \zeta (5)}{9}\right.\nn\\
	&\qquad \left.-\frac{1377380}{243} + \frac{51680 \pi ^2}{81}-\frac{2464 \pi ^4}{135} \right) \nn\\
	&\qquad + n_f^2 \left( \frac{16640 \zeta (3)}{81} +\frac{71500}{729} -\frac{1216 \pi ^2}{243}-\frac{416 \pi ^4}{405} \right) \nn\\
	&\qquad + n_f^3\left( \frac{256 \zeta (3)}{81}-\frac{128}{243} \right) \,,
\end{align}
and
\begin{align}
	d_0 &= d_1 =0\,,\\
	d_2 &=-56 \zeta (3)+\frac{1616}{27}  -\frac{224 }{81}n_f\,,\\
	d_3&=\left(\frac{176 \pi ^2 \zeta (3)}{3}-\frac{24656 \zeta (3)}{9}+1152 \zeta (5)+\frac{594058}{243}\right.\nn\\
	&\quad \left.-\frac{6392 \pi ^2}{81}-\frac{154 \pi ^4}{45}\right) + n_f\left(\frac{7856 \zeta (3)}{81}-\frac{166316}{729}\right.\nn\\
	&\quad \left.+\frac{824 \pi ^2}{243}+\frac{4 \pi ^4}{405}\right) + n_f^2\left(\frac{64 \zeta (3)}{27}+\frac{3712}{2187} \right)\,.
\end{align}

Additionally, the QCD $\beta$-function is defined as
\begin{align}
		\beta[\alpha_s(\mu)] &\equiv {d\alpha_s(\mu)\over d\ln \mu} = - 2\alpha_s(\mu) \sum_{n=0}^\infty a_s^{n+1}(\mu)b_n\,,
\end{align}
with
\begin{align}
    \label{eq:beta}
	b_0 &= 11-\frac{2 }{3}n_f, \\
	b_1 &= 102-\frac{38 }{3}n_f, \\
        b_2 &=\frac{2857}{2}-\frac{5033 }{18}n_f+ \frac{325 }{54}n_f^2\,, \\ 
	b_3 &=3564 \zeta (3)+\frac{149753}{6} -n_f \left(\frac{6508 \zeta (3)}{27}+\frac{1078361}{162}\right) \nn\\
	&\qquad +n_f^2 \left(\frac{6472 \zeta (3)}{81}+\frac{50065}{162}\right) +  \frac{1093}{729}n_f^3\,.  
    \end{align}
    
The kernel $K$ in \Eq{dres} at N$^3$LL is~\cite{Stewart:2010qs,Hoang:2014wka}
\begin{align}
	K(\mu,\mu_0)
	&= -\frac{\Gamma_0}{b_0}\left\{\ln r +a_s \left(\frac{\Gamma_1}{\Gamma_0}-\frac{b_1}{b_0}\right) (r-1) \right.\nn\\
	&\quad + \frac{1}{2} a_s^2  \left(\frac{b_1^2}{b_0^2}-\frac{b_1 \Gamma_1}{b_0 \Gamma_0}-\frac{b_2}{b_0}+\frac{\Gamma_2}{\Gamma_0}\right) \left(r^2-1\right)\nn\\
	&\quad  + \frac{1}{3} a_s^3  \left[\frac{\Gamma_1 }{\Gamma_0} \left(\frac{b_1^2}{b_0^2}-\frac{b_2}{b_0}\right) -\frac{b_1 }{b_0} \left(\frac{b_1^2}{b_0^2}-\frac{2 b_2}{b_0}+\frac{\Gamma_2}{\Gamma_0}\right) \right.\nn\\
	&\quad\left.\left.-\frac{b_3}{b_0}+\frac{\Gamma_3}{\Gamma_0}\right]\left(r^3-1\right) \right\}\,,
\end{align}
where $r=\alpha_s(\mu)/\alpha_s(\mu_0)$. The four-loop running coupling is needed for N$^3$LL resummation:
\begin{align}
	&{1\over a_s(\mu)} = {X\over a_s(\mu_0)} + \frac{b_1 \ln X}{b_0}  \nn\\
	&\quad + a_s(\mu_0) \left[\frac{b_1^2 }{b_0^2}\left(\frac{1}{X}+\frac{\ln X}{X}-1\right) + \frac{b_2 }{b_0}\left(1-\frac{1}{X}\right)\right]\nn\\
	&\quad + a_s^2(\mu_0) \left[-\frac{b_1^3 }{2 b_0^3} {\ln^2 X \over X^2}+ \frac{b_1 b_2}{b_0^2 }{ \ln X\over X^2}\right.\nn\\
	&\quad  +\left(1-\frac{1}{X}\right) \left(-\frac{b_1 b_2}{b_0^2}+\frac{b_3 }{2 b_0}\left(\frac{1}{X}+1\right)+\frac{1}{2} \left(1-\frac{1}{X}\right)\right) \nn\\
	&\quad \left. - \left(1-\Big(\frac{b_1}{b_0}\Big)^3\right)\frac{(1-X)^2}{2 X^2} \right]\,.
\end{align}
In the numerical analysis, the initial condition of $\alpha_s$ is set to be $\alpha_s(\mu_0=2\ {\rm GeV}) = 0.293$ and $n_f=4$.

\end{document}